\documentclass[]{aa}
\usepackage		[utf8]					{inputenc}
\usepackage		[english]				{babel}
\usepackage								{amsmath}
\usepackage								{amsfonts}
\usepackage								{amssymb}
\usepackage 					  		{graphicx}
\usepackage								{natbib}
\usepackage     [T1]                    {fontenc}
\usepackage                             {color}
\usepackage                             {hyperref}
\usepackage								{gensymb}
\usepackage								{subfigure}
\usepackage								{color}
\usepackage                             {longtable}

\usepackage{soul,xcolor}

\definecolor{darkblue}{rgb}{0.0,0.0,0.4}
\definecolor{darkgreen}{rgb}{0.0,0.4,0.0}
\hypersetup{
    colorlinks,
    linkcolor=black,
    citecolor=darkgreen,
    urlcolor=darkblue
}

\begin{document}

\title{Structure and Fragmentation of a high line-mass filament: Nessie}

\author{M. Mattern\inst{1}
          \and
          J. Kainulainen\inst{2,3}
          \and
          M. Zhang\inst{2}
          \and 
          H. Beuther\inst{2}
          }
\institute{Max-Planck-Institut f\"ur Radioastronomie, Auf dem H\"ugel 69, D-53121 Bonn \email{mmattern@mpifr-bonn.mpg.de}
         \and
         Max-Planck-Institut f\"ur Astronomie, K\"onigstuhl 17, D-69117 Heidelberg
         \and 
         Dept. of Space, Earth and Environment, Chalmers University of Technology, Onsala Space Observatory, 439 92 Onsala, Sweden
           }
\date{Received ...; accepted 29/03/2018}

\abstract{An increasing number of hundred-parsec scale, high line-mass filaments have been detected in the Galaxy. Their evolutionary path, including fragmentation towards star formation, is virtually unknown.
}
         {We  characterize the fragmentation within the hundred-parsec-scale, high line-mass Nessie filament, covering size-scales between $\rm \sim 0.1-100~pc$. We also connect the small-scale fragments to the star-forming potential of the cloud.}
         {We combine near-infrared data from the VISTA Variables in the Via Lactea (VVV) survey with mid-infrared \emph{Spitzer}/GLIMPSE data to derive a high-resolution dust extinction map for Nessie . We then apply a wavelet decomposition technique on the map to analyze the fragmentation characteristics of the cloud. The characteristics are then compared with predictions from gravitational fragmentation models. We compare the detected objects to those identified in $\sim 10$ times coarser resolution from ATLASGAL 870 $\mu$m dust emission data.}
         {We present a high-resolution extinction map of Nessie ($2\arcsec$ full-width-half-max, $FWHM$, corresponding to 0.03 pc). We estimate the mean line mass of Nessie to be $\sim 627~\text{M}_\odot~\text{pc}^{-1}$ and the distance to be $\sim 3.5~\text{kpc}$. We find that Nessie shows fragmentation at multiple size scales. The median nearest-neighbour separations of the fragments at all scales are within a factor of two of the Jeans' length at that scale. However, the relationship between the mean densities of the fragments and their separations is significantly shallower than expected for Jeans' fragmentation. The relationship is similar to the one predicted for a filament that exhibits a Larson-like scaling between size-scale and velocity dispersion; such a scaling may result from turbulent support. Based on the number of young stellar objects (YSOs) in the cloud, we estimate that the star formation rate of Nessie is $\sim \rm 371~ M_\odot \, Myr^{-1}$; similar values result if using the number of dense cores, or the amount of dense gas, as the proxy of star formation. The star formation efficiency is $0.017$. These numbers indicate that by its star-forming content, Nessie is comparable to the Solar neighborhood giant molecular clouds like Orion A.}
         {}

\keywords{Stars: formation --
          Infrared: ISM --
          ISM: clouds --
          dust, extinction
          }

\maketitle

\section{Introduction}

Star formation is an important process in the evolution of galaxies and the Universe. It plays a crucial role in gas-to-stars conversion through parameters such as star-forming rate and -efficiency, and the initial mass function \citep[e.g.,][]{McKee2007a, Hennebelle2012, Padoan2014}. 
Star formation takes place in dense regions of  molecular clouds, which appear to be commonly composed of filamentary structures \citep[][see \citealt{Andre2014} for a review]{Schneider1979,  Arzoumanian2011, Hacar2013, Schisano2014, Li2016b, kainulainen2017isf, Stutz2016}. Filaments are observationally defined as any elongated structures with an aspect ratio larger than $\sim 5$ and a clearly higher density than their surroundings \citep{Myers2009}. Given the link between filamentary structures and star formation, the processes driving the formation and evolution of filaments are linked with star formation rate and -efficiency. However, these processes are still not well understood.

Especially, the physics of filament fragmentation are not well known. This is mostly because determining the basic characteristics of filaments is observationally challenging, as the cold molecular hydrogen is invisible to observations. Therefore, different tracers and techniques are needed to determine its distribution and properties \citep[e.g.,][]{Lombardi2001, Goldsmith2008, Goodman2009, Andre2014}. Each of the techniques is sensitive to different density regimes and has different spatial resolution. For studies of the structures related to star formation, the resolution should clearly resolve the Jeans' length. This is about $0.1~\text{pc}$ for typical conditions of a molecular cloud (gas temperature $T=15~\rm K$, average density $\overline{n}(\text{H}) = \rm 10^5~cm^{-3}$). This currently limits the observations mostly on nearby ($<500~\text{pc}$) clouds. Interferometric observations can increase this resolution farther, but they have their own caveats (e.g. spatial filtering, slow mapping speed). 

However, the nearby clouds that can be systematically mapped in high-enough resolution are mainly low-mass clouds, containing mostly low line-mass filaments (mass per unit length of $(M/l) \lesssim$ a few $\times 10$ M$_\odot$) forming almost exclusively low-mass stars. An exception to this is the integral shaped filament in the Orion A cloud \citep[at distance $\rm 414~pc$,][]{Menten2007} whose fragmentation have been analyzed in high-resolution using interferometric data \citep[e.g.,][]{Takahashi2013,teixeira2016, kainulainen2017isf}. But in general, our current observational picture of filaments is mostly built by data on low-mass clouds. Filaments that have much higher line-masses $((M/l) \gg$ 100 M$_\odot$), which may also be able to form high-mass stars, have been identified in numbers, but they are typically located at further distances \citep[e.g.,][]{Jackson2010, Hernandez2012, Busquet2013, Kainulainen2013a, Ragan2014, Wang2014, Beuther2015, Abreu-Vicente2016, Henshaw2016filament, Li2016b, Wang2016}. Modern facilities are only approaching the ability to study them systematically in resolution that resolves the Jeans' scale.

Recently, \cite{Kainulainen2013b} developed a dust extinction based method that allows studying infrared dark molecular clouds at a resolution of $\sim$ $2\arcsec$ over a wide dynamic range of column densities, using a combination of near- and mid-infrared observations \citep[see also][]{Lombardi2001, Kainulainen2011irdcs, Butler2012}. This method allows us to analyze the internal structure of clouds up to several $\rm kpc$ distance at $\sim \rm 0.1$ pc resolution, enabling fragmentation studies of high line-mass filaments.  

With the high-resolution mapping technique in hand, we can address a basic question related to filament fragmentation: What are the fragmentation characteristics of massive filaments and are they in agreement with gravitational fragmentation models? 

In this paper, we take the advantage of the high-resolution provided by the \cite{Kainulainen2013b} extinction mapping technique and analyze the fragmentation characteristics of a $\sim 100~\text{pc}$ long, high line-mass filamentary cloud known as "Nessie" \citep{Jackson2010}. It is supposedly located within the Scutum-Centaurus Arm of the Milky Way \citep{Goodman2014, Ragan2014, Zucker2015, Abreu-Vicente2016}. The high resolution allows us to characterize the cloud structure and to gauge the fragmentation processes over a wide range of scales ($\sim 0.1~\text{pc}$ -- $100~\text{pc}$). 
We will use the dust extinction mapping technique in conjunction with the near-infrared (NIR) data from the ESO/VISTA telescope and mid-infrared (MIR) data from the \emph{Spitzer} satellite. We then analyze the derived column density map with a hierarchical structure-identification technique and examine the fragmentation of the cloud over multiple size-scales. The results will then be compared with theoretical models and other clouds in literature. Finally, we compare our identified small scale structures to clumps identified in low resolution ($\sim 20"$) dust emission maps by \cite{Csengeri2014}. This demonstrates how structures identified from data with ten times coarser resolution fragment when viewed in finer detail.

\section{Data}

\subsection{Infrared data and data reduction}
\begin{figure}[hbtp]
\centering
\includegraphics[width=0.5\textwidth, clip=true, trim= 0cm 0cm 0cm 0cm]{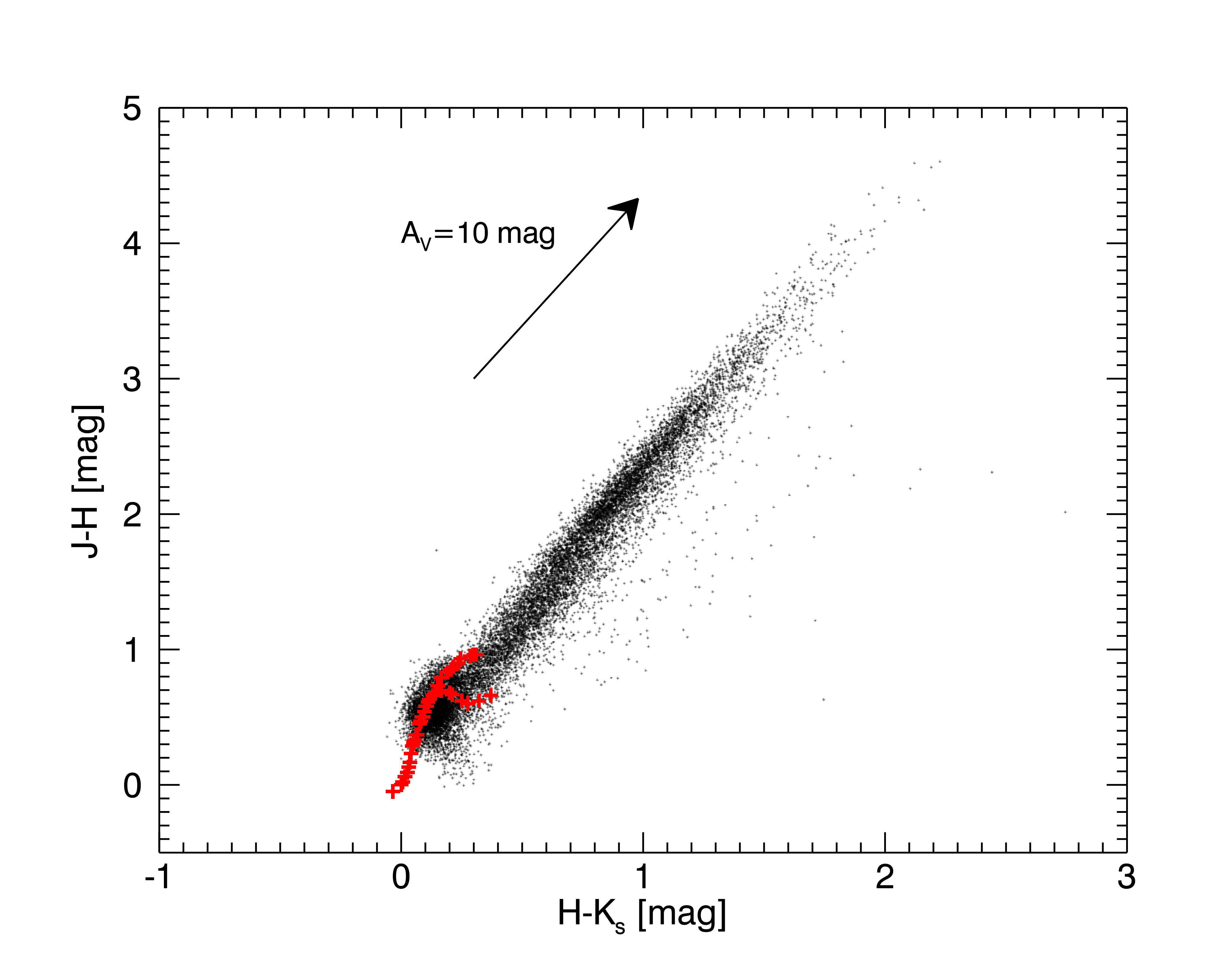}
\caption{Near-infrared color-color diagram of all sources in the mapped area extracted from the VVV survey with the photometric errors lower than $0.02~$mag. The blue crosses indicate non-redded intrinsic colors of stars \citep{Bessell1988}. The arrow shows the reddening for an extinction of $A_V=10\rm~mag$}.
\label{colplots}
\end{figure}

We employ NIR imaging data from the VVV (VISTA Variables in the Via Lactea) survey \citep{Saito2012} at the $4.1~m$ VISTA telescope of the Paranal Observatory. The calibrated and reduced data are publicly available in the ESO archive. Specifically, we used the J, H, K$_S$ spectral bands of the tiles d069 and d068. For each filter band there are two $t_\text{exp}=80~s$ exposures and additionally there are $8$ and $12$ $t_\text{exp}=16~s$ exposures of tiles d069 and d068 in the K$_S$ band, respectively. The pixel size of the images is $0.34"~\times~0.34"$. Detailed information about the observations can be found in Table \ref{Observation_tab} in the Appendix. We stacked the observations and performed PSF photometry with the daophot package \citep{Stetson1987} using the IRAF software. The PSF model was created from bright isolated stars with the model radius of $r_\text{PSF}=1.5"$. The different spatial resolutions of the single observation epochs has no significant effect on the photometry as we show in the Appendix \ref{PSF}. The daophot algorithm identifies and extracts extended sources and cosmic rays, and we expect only a very low contamination of the data by galaxies, because we are looking through the galactic mid-plane. The zero-point magnitudes were defined by comparing the resulting magnitudes of the stars with the corresponding stars of 2MASS, that are flagged as good photometric quality \citep{Skrutskie2006, Cutri2003}. This resulted in zero-points J$_\text{zpt}=21,21~\text{mag}$,  H$_\text{zpt}=21,22~\text{mag}$, K$_{S, \text{zpt}}=20,88~\text{mag}$. The resulting data shows the expected shape in the near-infrared color-color scatter plot (Fig. \ref{colplots}), with a bump for the main sequence stars and an elongated distribution for stars with varying reddening. We also tested the photometry measurements for completeness by adding artificial stars. We could identify all artificial stars up to a magnitude of about J$_\text{com}=16.5~\text{mag}$,  H$_\text{com}=15.5~\text{mag}$, nd  K$_{S, \text{com}}=15.0~\text{mag}$.

We also employ MIR 8 $\mu$m imaging data from the \emph{Spitzer}/GLIMPSE survey, data release 5 \citep{Benjamin2003, Churchwell2009}. The pipeline-reduced (S13.2.0 1v04) images were retrieved from the IRSA\footnote{\url{http://irsa.ipac.caltech.edu/data/SPITZER/GLIMPSE/}} database and used as such. The $\rm 8~\mu m$ image has a spatial resolution of $2.4\arcsec$ and a pixel size of $1.2\arcsec$ times $1.2\arcsec$. The used tile is centered around $RA=16:43:14.08$, $DEC=-16:00:15.92$. The effective integration time of the tile is $\rm 1.2~s$. 

\subsection{ATLASGAL data}
We also use data from the APEX telescope large area survey of the galaxy \citep[ATLASGAL, ][]{Schuller2009} for a comparison with our extinction data. The survey was obtained by the Millimeter and Submillimeter Group of the Max-Planck-Institut f\"ur Radioastronomie from 2007 to 2010 at the Atacama Pathfinder Experiment (APEX) located on Chajnantor in Chile. The survey instrument was the Large APEX Bolometer Camera (LABOCA) observing at $870~\mu m$, which traces the thermal dust emission. The resolution of the survey is $\Omega=19.2"$ with a sensitivity in the range of $40-70\rm~mJy/beam$. The maps covering the Nessie filament are centered at $l=-22.5\degree$, $b=0.0\degree$ and $l=-19.5\degree$, $b=0.0\degree$ and were observed on August 18th and 21st of 2007. The flux per beam, $F_\nu$, of the ATLASGAL map can be used to estimate the hydrogen column density $N(H_2)$ under the assumptions of a constant gas-to-dust ratio of $R=100$ and a dust opacity of $\rm \kappa_{\rm 345\,GHz}=1.85~cm^2\,g^{-1}$, which was extrapolated by \cite{Schuller2009} based on the work of \cite{Ossenkopf1994},
\begin{equation}
N(H_2)=\frac{F_\nu R}{B_\nu(T_d) \Omega \kappa_\nu \mu_{H_2} m_H}  $   .$
\label{atlascdequ}
\end{equation}
$B_\nu(T_d)$ is the Planck function at the dust temperature $T_d$, $m_H$ is the mass of a hydrogen atom, and $\mu_{H_2}$ the mean molecular weight of the interstellar medium with respect to hydrogen molecules, which is $2.8$ \citep{Kauffmann2008}.

\cite{Csengeri2014} have identified clump-like structures from the ATLASGAL data using 2D Gaussian fitting (Gauss Clump Source Catalog, GCSC). It provides the position, peak flux $F'_\nu$ and integrated flux $S_\nu$, the half maximum major and minor axes and the position angle of the clumps. We then calculated the masses of the clumps from \citep{Schuller2009}:
\begin{equation}
M=\frac{S_\nu R\,d^2}{B_\nu (T_d) \kappa_\nu}$   ,$
\label{atlasmassequ}
\end{equation}
where $R$ is the gas-to-dust ratio and $d$ the distance towards the clump.

\section{Extinction mapping technique}
We employ the technique from \cite{Kainulainen2013b}, which is based on combining extinction maps made at two wavelength regimes: in near-infrared using NICER \citep[Near-Infrared Color Excess Revisited, ][]{Lombardi2001} and in mid-infrared using the absorption against the Galactic background \citep[e.g., ][]{Peretto2009, Butler2012}. Below, the implementation of the two techniques is explained in detail.

\subsection{NICER-Method}
\label{NICER method}
We use the NICER method in conjunction with JHK$_S$ photometric data of the VVV survey. The method is based on near-infrared color measurements of stars shining through the molecular cloud and comparison of those with stars of a reference field that is (optimally) free from extinction. The observed reddening towards the cloud region is used to estimate the extinction by adopting a wavelength dependent reddening law. The extinction values towards each star are then used to derive a spatially smoothed dust extinction map.

This method is straightforward to apply for nearby clouds \citep[$d < 500~$pc, e.g.,][]{Lombardi2006, Froebrich2007, Juvela2008, Goodman2009, Kainulainen2009}, where the contamination due to stars between the cloud and the observer is small. The extinction towards more distant clouds might be underestimated because of these (mostly unreddend) foreground stars, especially in high extinction regions where the fraction of foreground sources is high \citep{Lombardi2005}. The foreground stars do not trace the dust reddening caused by the cloud, but only the reddening along the line of sight until the cloud. Therefore, foreground sources should be removed as accurately as possible, which is challenging in practice, because of the degeneracy between the intrinsic colors of stars and reddening caused by extinction. 

The subtraction of the foreground is also necessary for the reference field \citep[see, ][]{Kainulainen2011irdcs}. Due to diffuse dust in the Galactic plane stars in the reference field, located at the same distance as stars behind the cloud, are redder than the ones at closer distance. Therefore, foreground stars shift the mean color of the reference field towards blue, which leads to an overestimation of the extinction. For the implementation of the NICER method we have to find a reliable way to remove the effect of the foreground stars. This is described in the following.

\begin{figure*}[tbhp]
\centering
\includegraphics[width=\textwidth, clip=true, trim= 0cm 1cm 0cm 2cm]{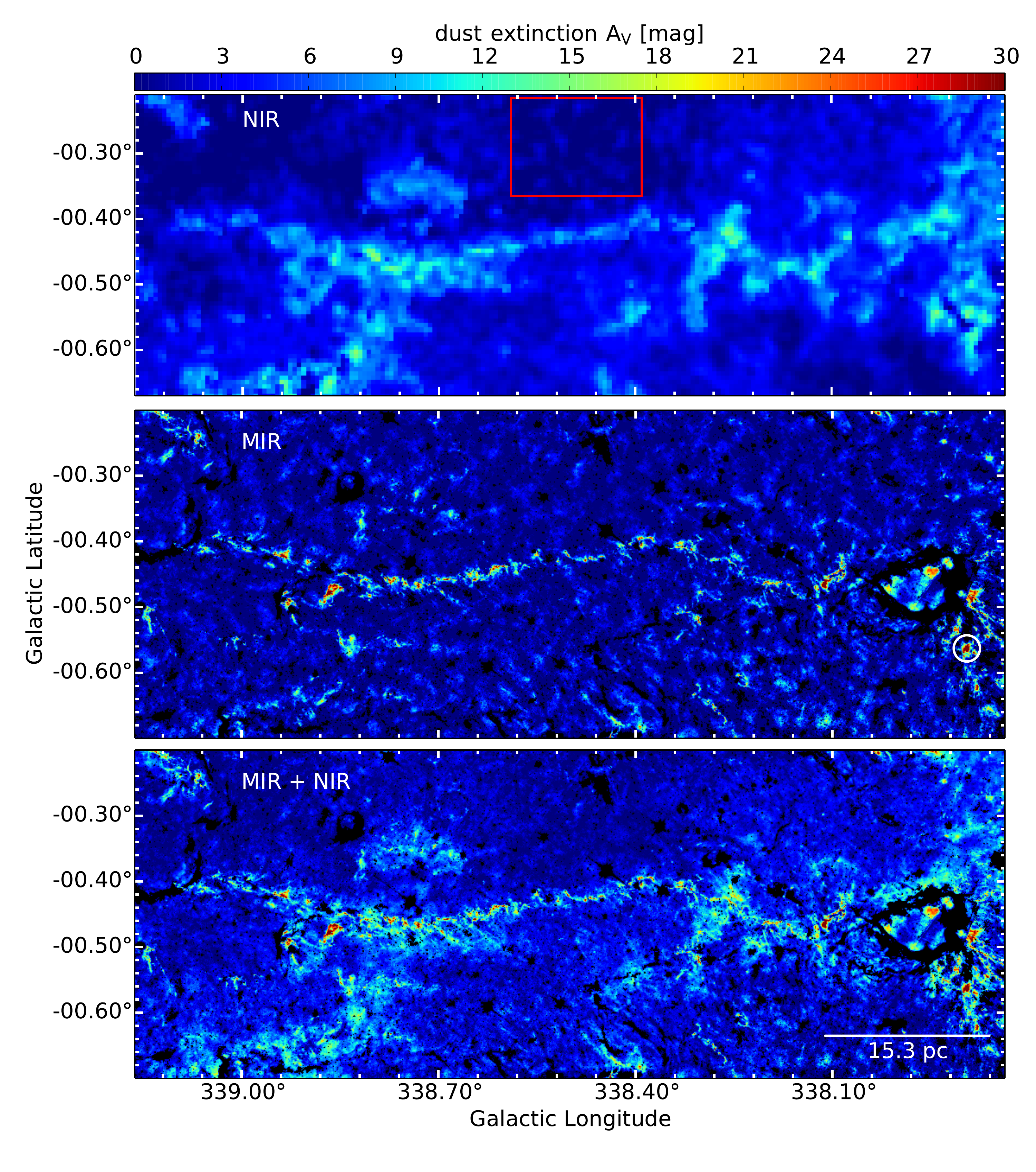}
\caption{Extinction maps of Nessie derived using the NIR data of the VVV survey (top), mid-infrared data of the \emph{Spitzer} Space Telescope (center) and their combination (bottom). The black areas indicate regions of bright mid-infrared emission that hampers extinction mapping. The red rectangle marks the area used for estimating the reference colors for the NICER method. The white circle marks the high extinction region used to estimate the mid-infrared foreground emission.}
\label{extinc maps}
\end{figure*}

\begin{figure}[tbhp]
\centering
\begin{minipage}{0.24\textwidth}
\includegraphics[width=\textwidth, clip=true, trim= 0cm 0cm 0.7cm 1.5cm]{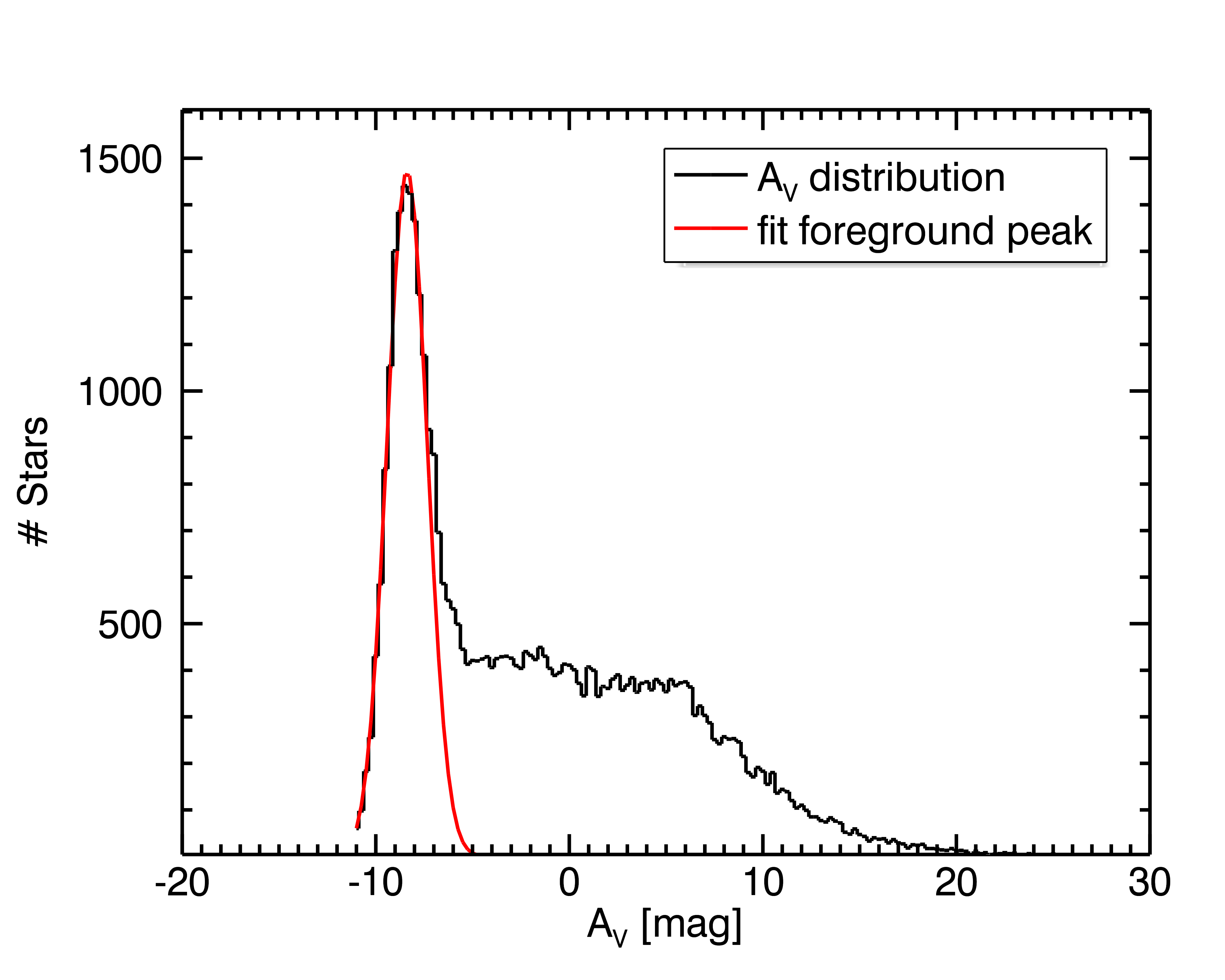}
\end{minipage}
\begin{minipage}{0.24\textwidth}
\includegraphics[width=\textwidth, clip=true, trim= 0cm 0cm 1.0cm 1.3cm]{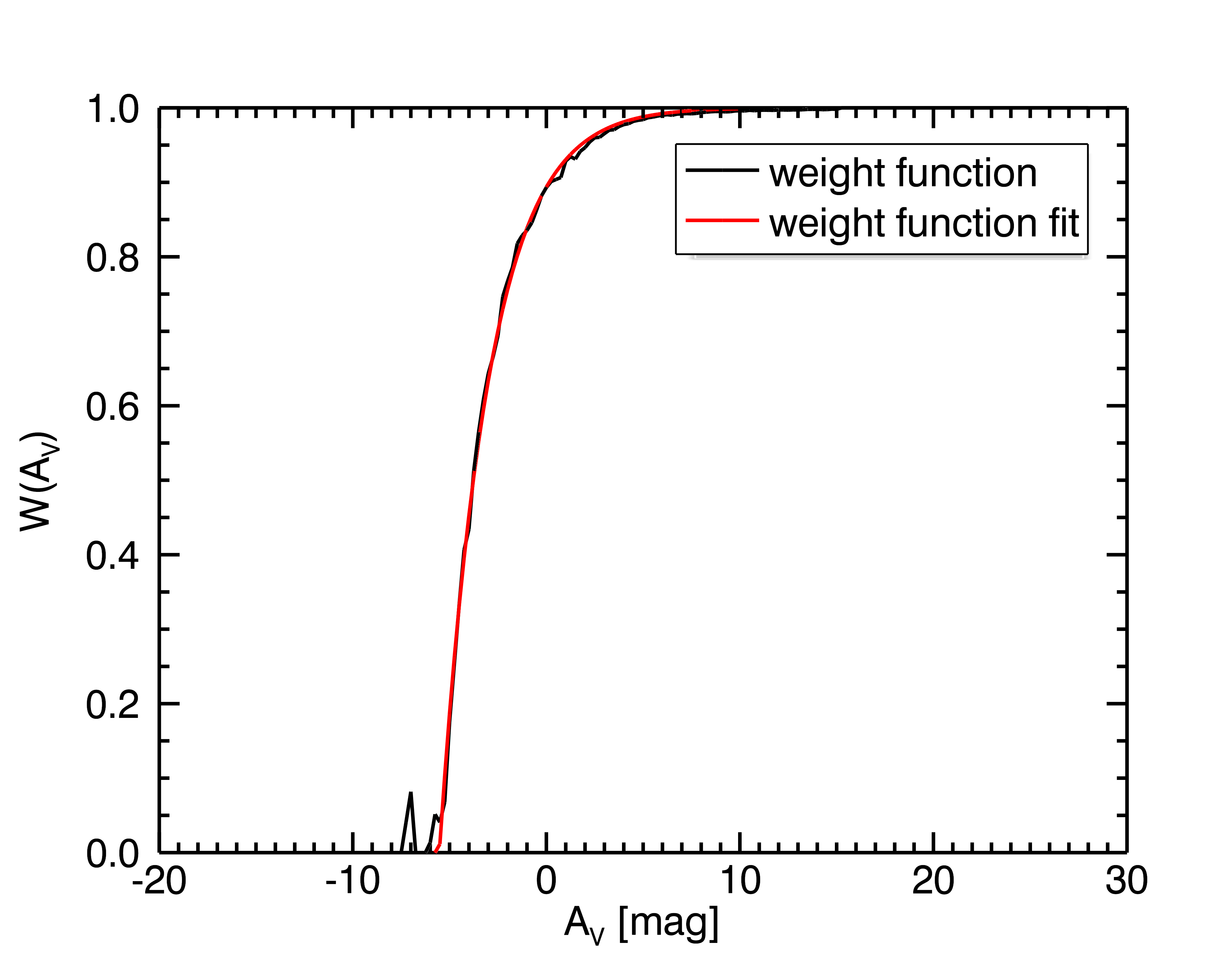}
\end{minipage}
\caption{{\bf Left: }The black line shows histogram of the calculated extinction from a high extinction region. The red line marks the Gaussian fitted to the peak of foreground stars. {\bf Right: }The black line shows the empirical weighting function, which is derived like shown in Eq. \ref{weight_equation1}. The red line shows the fitted function, which is then introduced into the weighting function of the NICER method (Eq. \ref{weight_equation2}).}
\label{Av_hist}
\end{figure}

First, we derive a "dirty" extinction map using arbitrary reference colors and use this map to identify low- and high-extinction regions. The low-extinction region (Fig. \ref{extinc maps}; $338.39\degree < l < 338.58\degree; -0.36\degree < b < -0.21\degree $) is then used as control field to estimate the reference colors, indicating the average star colors without dust reddening by the cloud. 
In the regions of high extinction, identifying foreground stars is simple: they appear as a distinct feature in the frequency distribution of individual extinction measurements \citep[cf.,][]{Kainulainen2011irdcs}. For regions of lower extinction the feature is less distinct, but under the assumption of uniformly distributed foreground stars the position and width of the frequency distribution remains the same; this fact can be used to statistically subtract the contribution of foreground stars to the reference field colors. To do this, we fit a Gaussian function $G_{\text{fg}}$, to the peak of the foreground stars in the extinction histogram $H(A_V)$ (Fig.\ref{Av_hist}) and subtract these stars in a statistical sense from the distribution. To achieve this, we add a weighting term ($W_{\text{fg}}(\widehat{A}_V^{(n)})$, see Fig. \ref{Av_hist}) into the original NICER method. This weighting term suppresses the contribution of stars that might be foreground stars, and it is calculated in the following way
\begin{equation}
W_\text{fg}(\widehat{A}_V^{(n)})=\frac{H(A_V)-G_\text{fg}}{H(A_V)} $   .$
\label{weight_equation1}
\end{equation}
The weighting term is introduced into equation (15) of \cite{Lombardi2001} as shown here:
\begin{equation}
W^{(n)}=\frac{W(\theta - \theta^{(n)}) \cdot W_\text{fg}(\widehat{A}_V^{(n)})}{\text{Var}(\widehat{A}_V^{(n)})} $   ,$
\label{weight_equation2}
\end{equation}
where $W^{(n)}$ is the weighting of the $n$th star, $W(\theta - \theta^{(n)})$ is the weight for the distance between the actual location $\theta$ and the location of the $n$th star $\theta^{(n)}$, $W_\text{fg}(\widehat{A}_V^{(n)})$ is the foreground weight based on the estimated extinction of the $n$th star, and $\text{Var}(\widehat{A}_V^{(n)})$ is variance of the estimated extinction of the $n$th star.

With this method the contribution of foreground stars was subtracted statistically from the mean color of the reference field to calculate an estimate of the mean color of the stars in the background of the cloud. The statistical subtraction is done in the JHK-color-color space, where the density of foreground stars was subtracted from the density of the reference field stars in each color-color bin. Then the foreground-corrected number of stars per bin was calculated from the resulting density in the reference field. The foreground-corrected mean color was calculated from this sample of stars, which is also the estimate of the background color. The JHK-color-color histograms of the reference field before and after correction are shown in Appendix \ref{color-correction}.

With the foreground-corrected reference color and the method for extracting foreground sources the "true"  near-infrared extinction map was calculated. The spatial resolution of the map is given by the width of the Gaussian smoothing function that is used to smooth the pencil-beam measurements towards the stars onto the map grid. The pixel size is chosen following the surface number density of background sources so that even in high extinction regions, where the density is lower, each pixel covers at least two stars. For the VVV data we concluded that a pixel size 24" is sufficient, which leads to a beam width of 48".

\subsection{Mid-infrared Extinction Measurement}
We use the MIR imaging data from the GLIMPSE survey to estimate extinction through the cloud at 8 $\mu$m. Generally, the technique is based on the extinction of the diffuse MIR emission from the Galactic plane by the dust of the cloud \citep[see, e.g.,][]{Johnstone2003, Peretto2009, Butler2012}. Consider a simplistic geometry in which the intensity of radiation behind the cloud is $I_\mathrm{0}$. Then, the intensity right in front of the cloud is $I_\mathrm{1} = I_\mathrm{0}e^{-\tau_8}$, in which $\tau_\mathrm{8}$ refers to the optical depth at the \emph{Spitzer} 8 $\mu$m band. An observer detects the intensity $I_\mathrm{obs, 1}$, which in addition to $I_\mathrm{1}$ contains the intensity $I_\mathrm{fg}$ that is emitted from between the cloud and the observer, i.e., $I_\mathrm{obs, 1} = I_\mathrm{1} + I_\mathrm{fg}$. A line-of-sight off the cloud does not exhibit extinction and the observed intensity is $I_\mathrm{obs, 0} = I_\mathrm{0} + I_\mathrm{fg}$. Combining these relations, one can solve the optical depth
\begin{equation}
\tau_8 = \ln{\frac{I_\mathrm{obs, 0} - I_\mathrm{fg}}{I_\mathrm{obs, 1} - I_\mathrm{fg}}}.
\label{eq:tau8}
\end{equation}
Thus, the optical depth along the line of sight can be estimated through measurements of the off-cloud and foreground intensities.

Various approaches have been used in the past to estimate the off-cloud and foreground intensities \citep[see, e.g.,][]{Johnstone2003, Peretto2009, Ragan2009, Butler2012}. We follow an approach similar to \citet{Butler2012} to which we refer to for a thorough description and discussion; we describe here only the implementation of the technique in our case. The off-cloud intensity is estimated using a median-filtered 8 $\mu$m map. Prior to the filtering, the most prominent dark features are masked from the map by using a threshold intensity of 46 MJy sr$^{-1}$. The filter size defines the upper limit of the structures the map is sensitive to. However in our case, we will later combine the MIR-derived map with the NIR-derived map that probes spatial scales larger than $24\arcsec$. Therefore, the filter function width is not a crucial choice for us, as long as there is some overlap of scales probed by the MIR and NIR maps. Following the discussion in \citep[][]{Ragan2009}, we chose the filter width of $3\arcmin$. 

The foreground intensity is estimated with the help of the pixels with lowest intensities (i.e., highest extinctions) in the 8 $\mu$m data. If several independent high-extinction regions show similar intensities, one can assume that such locations are opaque and the intensity towards them is a reasonable estimate of the foreground intensity. The smallest intensities detected in the cloud area are $I_\mathrm{obs, 1} = 24.6$ MJy sr$^{-1}$. There are three independent locations in the cloud where the intensity is within $2\sigma_\mathrm{rms}$ of this value \citep[the rms noise, $\sigma_\mathrm{rms}$, of the GLIMPSE data is $\sim$0.6 MJy sr$^{-1}$,][]{Reach2006}. One of them ($l$, $b = 337.895\degree$, $-0.563\degree$) is extended, containing tens of pixels, which indicates that the region indeed is saturated. The number of the saturated regions is relatively low given the large extent of the cloud on the sky; it would be preferable to have numerous saturated regions along the cloud. Regardless, we adopt the value of 24 MJy sr$^{-1}$ for the foreground intensity. We note that the resulting fraction of foreground emission, i.e., $I_\mathrm{fg} / I_\mathrm{obs, 0} \approx 45 \%$, well in the range of the foreground intensities typically determined for IRDCs \citep[e.g.,][]{Butler2012}.
 
Following the estimation of the off-cloud and foreground intensities, Eq. \ref{eq:tau8} is used to compute an optical depth map for Nessie. Finally, the map is converted into units of visual extinction by adopting the ratio between 8 $\mu$m and $V$ band optical depths \citep[based on][see \citealt{Kainulainen2013b}]{Cardelli1989, Ossenkopf1994}
 \begin{equation}
A_\mathrm{V} = 33.6 \tau_8.
 \end{equation}
The resulting extinction map is shown in Fig. \ref{extinc maps}.

\subsection{Combined Near- and Mid-infrared Extinction Measurement}
We have now derived the near-infrared and mid-infrared extinction maps; both show some advantages and disadvantages. The near-infrared data is sensitive to low column densities, but have a low resolution. The mid-infrared data have a good resolution, but are much less sensitive. Therefore, we now want to combine them and use the near-infrared data to recalibrate the mid-infrared data, thus prevailing high spatial resolution of the mid-infrared data while imposing the good calibration of the near-infrared data on them. The combination of near- and mid-infrared extinction maps follows the scheme described in \cite{Kainulainen2013b}. The combined maps deliver a higher dynamic range of extinction compared to maps computed from near- or mid-infrared data alone (Fig. \ref{extinc maps}). The correlation between the two maps is shown in the Appendix \ref{MIR-NIR-correlation}.

The combined map is then converted to molecular hydrogen column density by applying the conversion of \cite{Savage1977,Bohlin1978,Rachford2002}: 
\begin{equation}
N(\text{H}_2)=A_V \cdot 0.94 \cdot 10^{21} \text{cm}^{-2}\,\text{mag}^{-1}  $   ,$
\end{equation}
using a typical reddening constant of $R_V = 3.1$ \citep{Schultz1975} and assuming all hydrogen atoms are in molecular form.

\section{Results}

\subsection{Distance determination}

\begin{figure}[thbp]
\centering
\includegraphics[width=0.5\textwidth, clip=true, trim= 0cm 6cm 0cm 7cm]{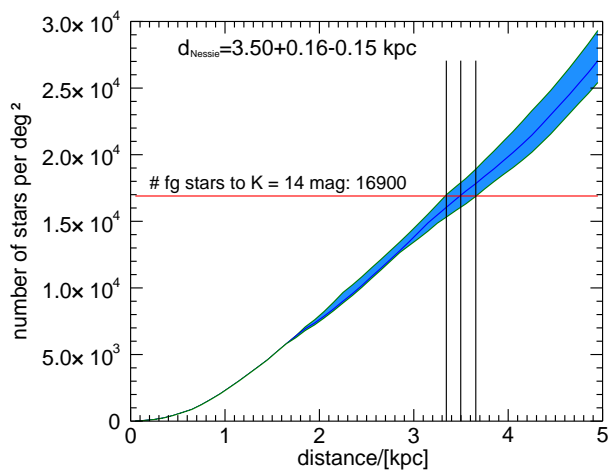}
\caption{Predicted stellar surface density based on the Besan\c{c}on stellar distribution model \citep{Robin2003}. The blue area indicates the uncertainty arising from the scatter in the diffuse extinction measurements. The horizontal line represents the measured foreground star surface density and the vertical lines the resulting estimates of the distance and its uncertainty.}
\label{distfit}
\end{figure}

The foreground star density measurements (see Section \ref{NICER method}) allow us to estimate the distance of Nessie independently of previous, kinematic distance estimates. We can compare the measured surface density of foreground stars with a distance-dependent stellar surface density model of the Galaxy. We used the Besan\c{c}on Galactic stellar distribution model \citep{Robin2003} to estimate the distance, see Fig. \ref{distfit}. For a more detailed description of the method see \cite{Kainulainen2011irdcs,Ioannidis2012}. The most important input parameter of the stellar distribution model is the extinction caused by diffuse interstellar dust. We used the measurements by \cite{Marshall2006} to estimate the mean extinction along the line of sight towards Nessie. For an estimate of the uncertainty we also estimated the minimum and maximum extinction, which indicate the upper and lower limit of the surface density (Fig. \ref{distfit}). We neglected other, potentially significant uncertainties in our distance calculations such as the uncertainty of the measured number surface density of the foreground stars or of the stellar distribution model. Therefore, the uncertainty of the distance is underestimated and it is more likely in the order of $15~\%$ corresponding to $\Delta d \approx 0.5\rm~kpc$ \citep{Kainulainen2011irdcs}.

The result of our distance estimate is $d_\text{extinction}=3.5 \pm 0.5~\text{kpc}$, which is in agreement with the kinematic distance estimations of \cite{Jackson2010}, $d_\text{HCN}=3.1~\text{kpc}$. We find also dynamical distance measurements from \cite{Wienen2015} for $14$ ATLASGAL sources likely embedded in the Nessie cloud. Their distances range between $3.0~\text{kpc}$ and $3.5~\text{kpc}$, which is also in agreement with our estimate. The distance of $\sim 3.5~\text{kpc}$ suggests that Nessie is associated with the Scutum-Centaurus spiral-arm of the Milky Way as suggested by \cite{Goodman2014} and \cite{Ragan2014}. 

\subsection{The large-scale structure}
\label{mappingresult}
\begin{figure*}[tpbh]
\centering
\includegraphics[width=0.95\textheight, clip=true,trim=1cm 1cm 1.7cm 1cm, angle=90]{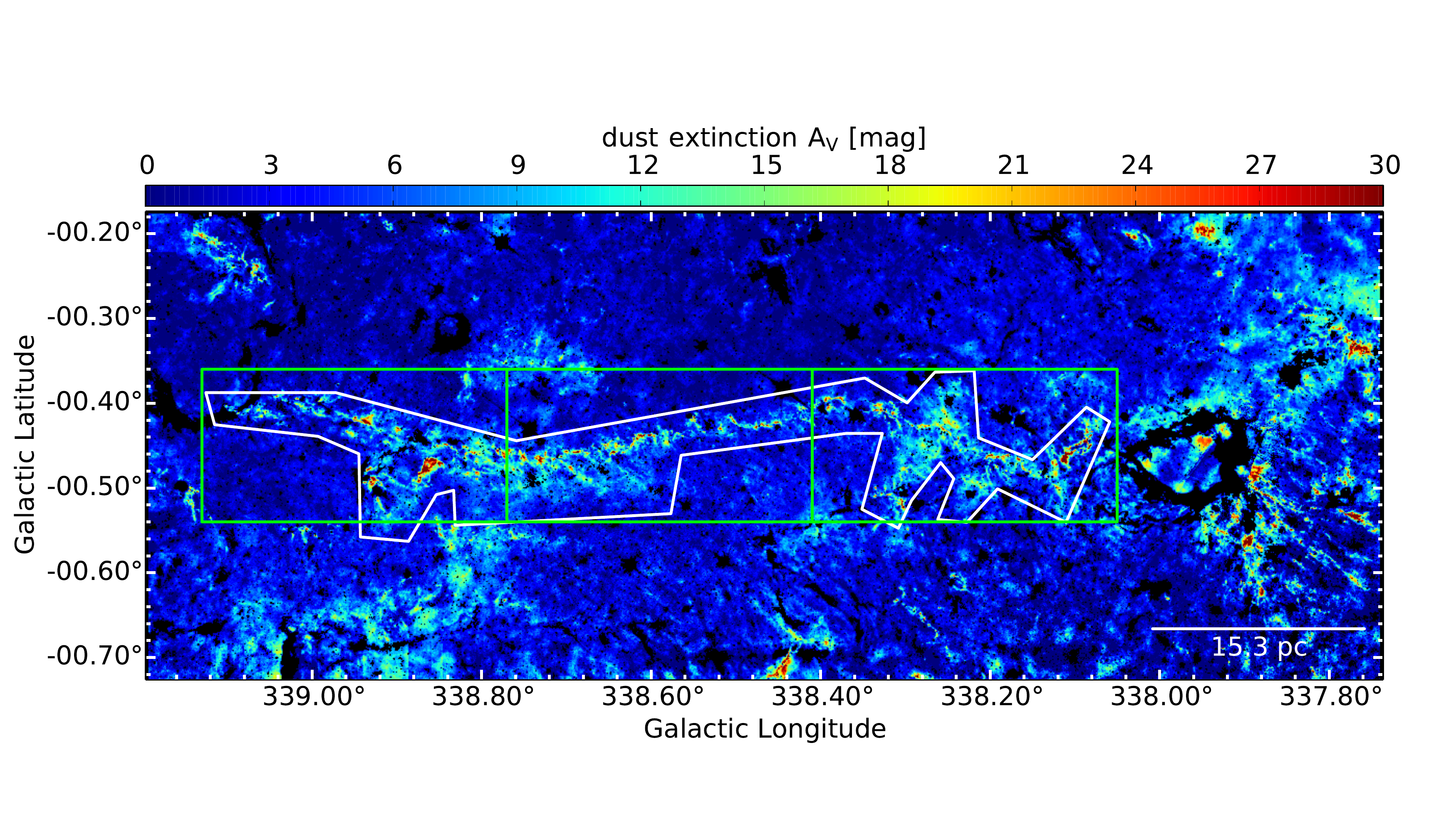}
\caption{Column density map of the Nessie filament. The white polygon marks the area chosen for the mass estimate of the cloud. The green rectangles show the positions of the zoom-ins shown in Figs. \ref{zoomed}, \ref{zoomed2}, and \ref{zoomed3}. }
\label{Nessie complete}
\end{figure*}

\begin{figure*}[hptb]
\centering
\includegraphics[trim=0cm 0cm 1cm 0cm, angle=90, clip=true, width=0.77\textwidth]{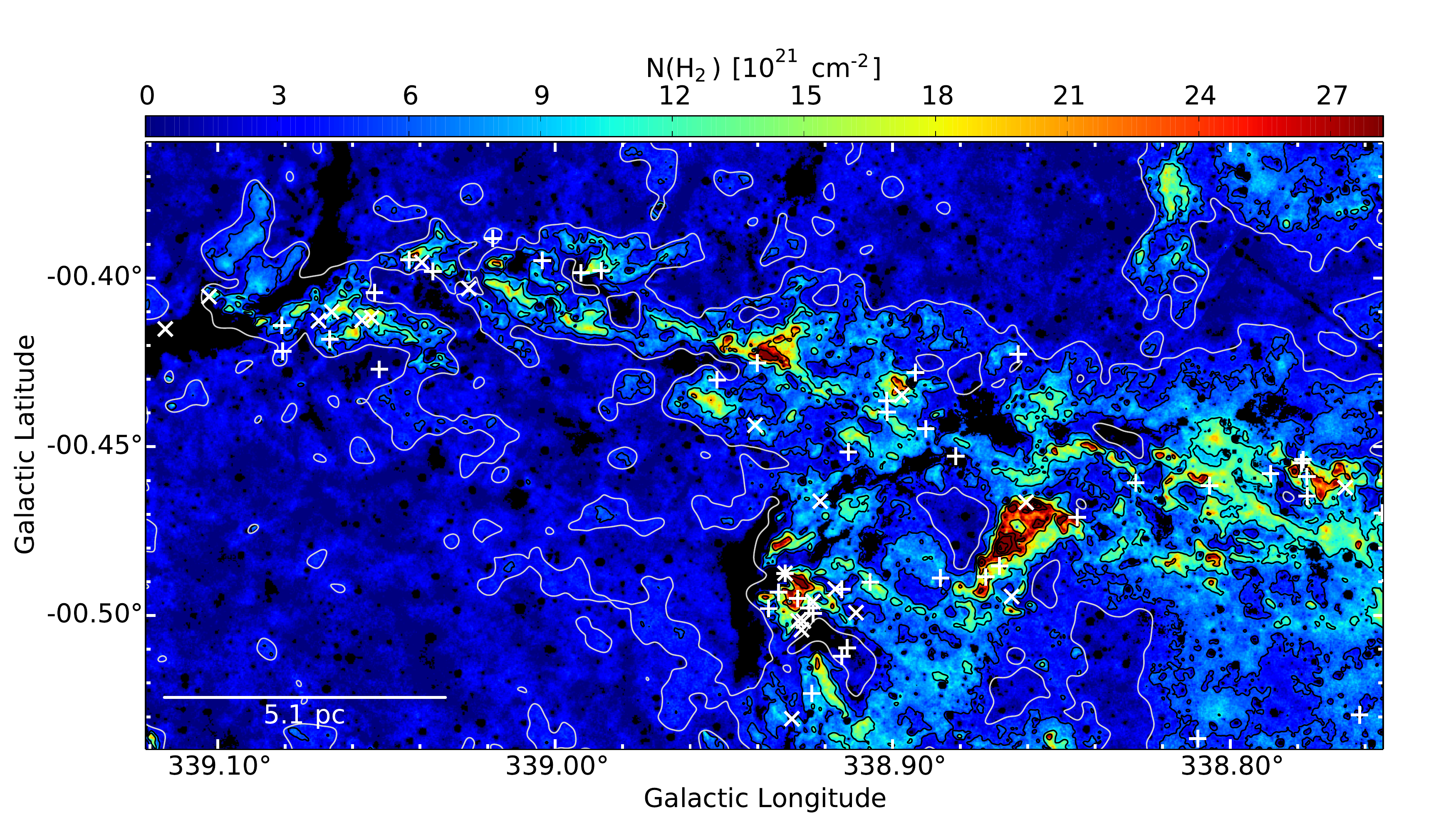}
\caption{Zoom-in number one of the column density map (Fig. \ref{Nessie complete}). The black contours indicate the levels of $\rm {5,10,20,30,40,50,60}\cdot 10^{21}~cm^{-2}$. The white contour indicates the smoothed $A_V = \rm 3~mag$ level. Additionally, the Class1 ('x') and Class2 ('+') YSOs are marked in white. }
\label{zoomed}
\end{figure*}

\begin{figure*}[hptb]
\centering
\includegraphics[trim=0cm 0cm 1cm 0cm, angle=90, clip=true, width=0.77\textwidth]{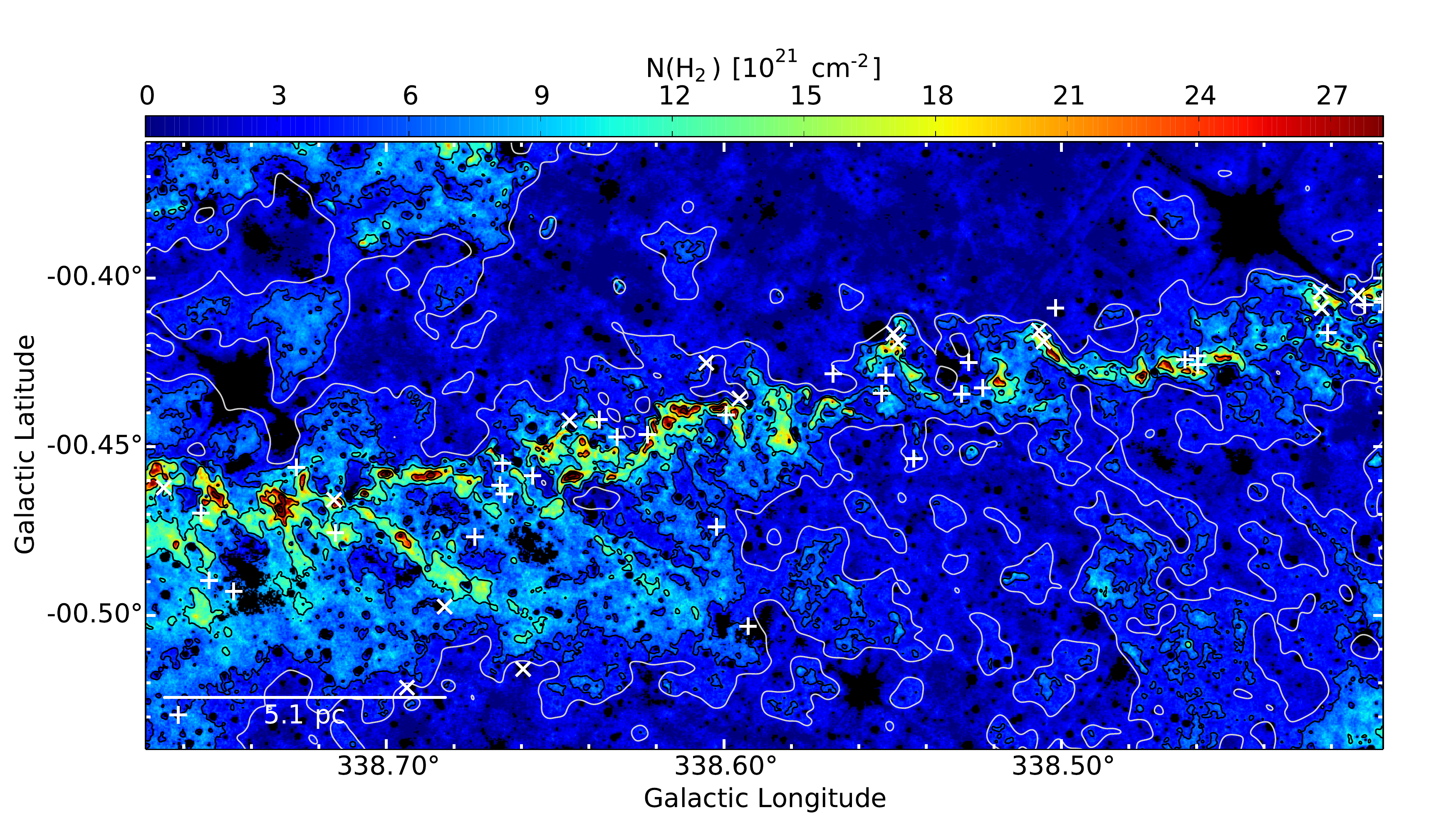}
\caption{Zoom-in number two of the column density map (Fig. \ref{Nessie complete}).The black contours indicate the levels of $\rm {5,10,20,30,40,50,60}\cdot 10^{21}~cm^{-2}$. The white contour indicates the smoothed $A_V = \rm 3~mag$ level. Additionally, the Class1 ('x') and Class2 ('+') YSOs are marked in white.}
\label{zoomed2}
\end{figure*}

\begin{figure*}[hptb]
\centering
\includegraphics[trim=0cm 0cm 1cm 0cm, angle=90, clip=true, width=0.77\textwidth]{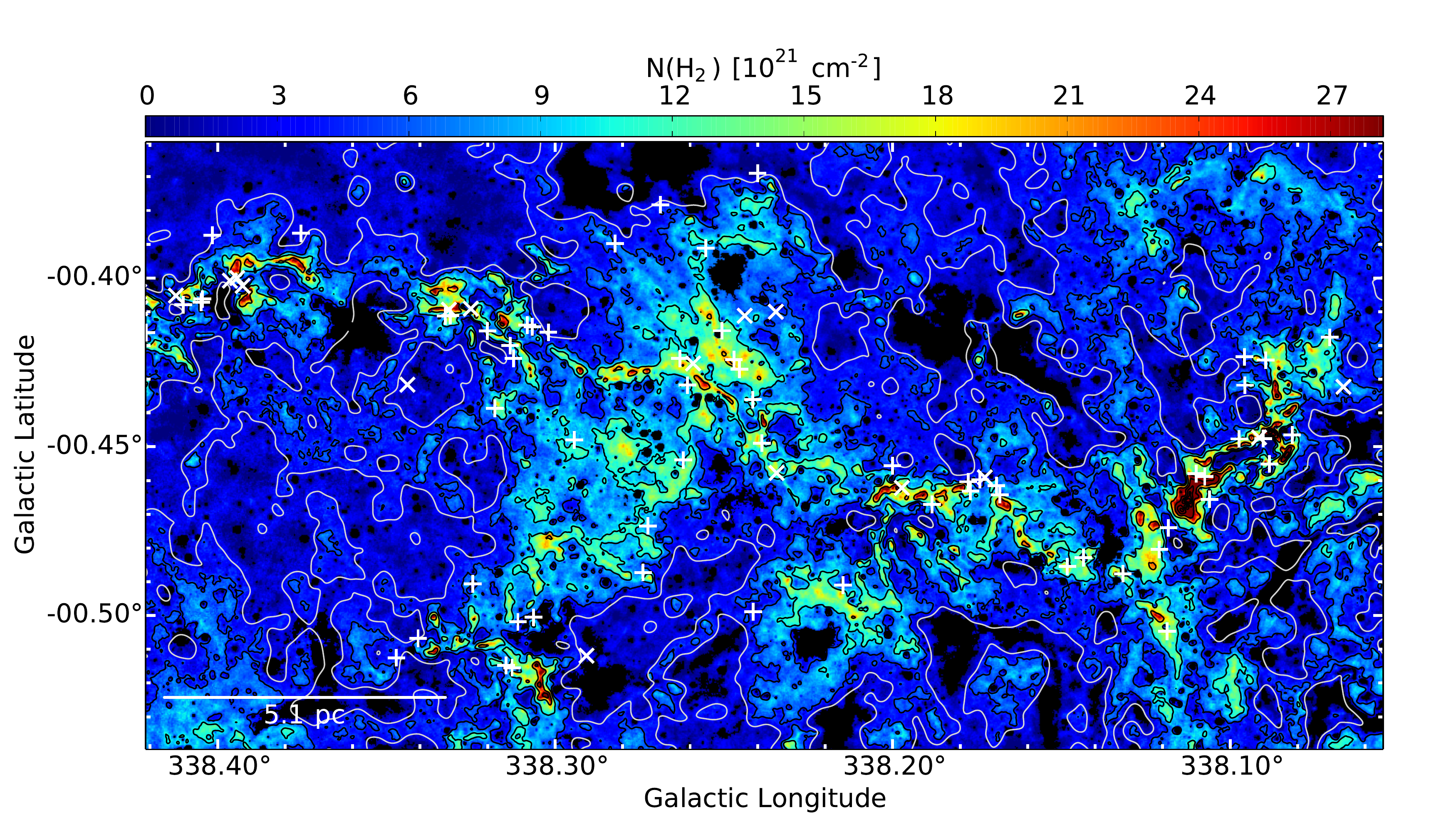}
\caption{Zoom-in number three of the column density map (Fig. \ref{Nessie complete}). The black contours indicate the levels of $\rm {5,10,20,30,40,50,60}\cdot 10^{21}~cm^{-2}$. The white contour indicates the smoothed $A_V = \rm 3~mag$ level. Additionally, the Class1 ('x') and Class2 ('+') YSOs are marked in white.}
\label{zoomed3}
\end{figure*}

The combined near- and mid-infrared extinction map of the Nessie cloud is shown in Fig. \ref{Nessie complete} and zoom-ins in Figs. \ref{zoomed}, \ref{zoomed2}, and \ref{zoomed3}. For comparison, Fig. \ref{extinc maps} shows the near-infrared based map, mid-infrared based map, and their combination.

The filament has a length of $\sim 1.1^\circ$ following the central, dense main axis (neglecting inclination) and perpendicular a width of $\sim 0.05^\circ$. This corresponds to a physical size of $67~\text{pc} \times 3~\text{pc}$ at the distance of $d = 3.5~\text{kpc}$. The width of the extinction structures, defined at the column density contours of about $A_V=3\rm~mag$, varies along the filament. This can be seen in the zoomed in map of Nessie (Fig. \ref{zoomed}). In the region between $338.57^\circ< \textit{l} < 338.95^\circ$ the low column density material is located only towards the south of dense main axis, between $338.23^\circ< \textit{l} < 338.30^\circ$ towards north and south and the rest of the filament shows almost no surrounding low column density material. These two low column density regions show also some less dense structures, which are mainly orientated almost perpendicular to the main filament. 

We need to identify which structures that we see in the map are actually part of Nessie. This is difficult because we miss information about the line-of-sight velocities of the structures. However, the Nessie filament was confirmed as a velocity coherent structure by \cite{Jackson2010}. Additionally, some areas lack the mid-infrared extinction data and cannot be used in the further analysis, such as the H\small{II}-bulb at $(\textit{l}; \textit{b}) = (337.95^\circ; -0.46^\circ)$ (Fig. \ref{Nessie complete}), which is part of Nessie in \cite{Jackson2010}. Therefore, the map needs to be cropped to the Nessie filament. To do this, we introduce a polygon around the cloud (see Fig. \ref{Nessie complete}). The area selection is mainly based on by eye inspection of the derived column density map with orientation on the $A_V=3\rm~mag$ contour and the observations published by \cite{Jackson2010}.

We derive an estimate of the total cloud mass from the column density map, given by:
\begin{equation}
M_{\text{Nessie}}=\sum_{i,j}(N(\text{H}_2)_{i,j})\cdot p^2 \cdot m_{\text{H}} \cdot \mu_{\text{H}_2}     $	,$ 
\label{massequ}
\end{equation}
where $N(\text{H}_{2})_{i,j}$ is the column density of the ($i,j$) pixel of the map, $p=\tan{(1.2")}\cdot d_{\text{Nessie}}$ is the physical size of a pixel, $m_{\text{H}}$ is the mass of the hydrogen atom, and $\mu_{\text{H}_2}=2.8$ the mean molecular weight of the interstellar medium \citep{Kauffmann2008}. The total mass of the Nessie cloud within the polygon (Fig. \ref{Nessie complete}) is $M_{\text{Nessie}}=4.2 \cdot 10^4~\text{M}_{\odot}$. 

From the length and mass we calculate the mean line-mass of the filament (mass per unit length along the main axis of the filament). The mean line-mass of Nessie is $(M/l)=\rm 627~M_\odot\, pc^{-1}$. As we neglected an inclination of the filament, which would increase its length, the derived line-mass is an upper limit. We note that there are variations in the line mass along the filament, both at large scales due to the varying amount of diffuse extinction and at small scales due to the substructure of the cloud.

\subsection{Fragmentation Analysis}

We analyzed fragmentation of Nessie simultaneously over a wide range of spatial scales using an algorithm explained in \cite{Kainulainen2014}, which employs wavelet filtering to identify structures at various spatial scales. In short, the algorithm uses a spatial filtering algorithm based on the à Trous wavelet transform \citep{2002aida} to decompose the column density map into scale-maps that describe structure at different scales. The different scales are defined as $2^i$ pixels, with $2 \leq i \leq 8$, where the limits are given by the pixel size for small scales and the cloud size for large scales. Individual structures are then identified from each scale-map using the clumpfind-2D algorithm \citep{Williams1994}. This provides the position, the size in $x$ and $y$ direction, and the total amount of column density of the structures $N(\text{H})_\text{tot}$. 

For reliable detection of structures, it is necessary to estimate the noise level of each scale-map. The noise level is estimated as the standard deviation $\sigma$ of an (almost) extinction free area. The size of the area corresponds to the size-scale of the largest scale map. To test the robustness of the structure identification, we tested the clumpfind-2D algorithm for contour level separations of $\rm 1.5\sigma, 3\sigma, 4\sigma \, and \, 5\sigma$ with the lowest level at $3\sigma$. The results do not show a significant difference and we chose the level separation of $3\sigma$.

The numbers of structures identified at each scale using the chosen technique are listed in Table \ref{frag-tab}. The number of structures increases towards smaller scales, but drops significantly for the smallest scales ($i=2$, see Table \ref{frag-tab}). This behavior was seen for all tested algorithm parameters and therefore, it is not likely to be an artifact. In the data these smallest structures trace only the densest clumps, which are predominantly located along the dense spine of the filament, but not in the surrounding low column density gas. This suggests, that only in the densest parts the filament is able to fragment into the smallest scales.

Table \ref{frag-tab} shows the properties of structures at each scale $i$: the total number of identified structures $N_\text{strc}$, the total mass of these structures $\sum(M_\text{strc})$, the median hydrogen number density $\widetilde{n}(\text{H})$, and the median separation $\widetilde{s}$. The sum of the masses over all scales, including scale $i > 8$, results in a total cloud mass of about $M_\text{Nessie}^\text{scales}=4.9 \cdot 10^4 \text{M}_\odot$. This is slightly higher than the mass derived from the combined column density map (see Section \ref{mappingresult}). The difference is a consequence of the used spatial filtering algorithm, which may not reproduce the true shapes of the structures accurately. 

\begin{figure}[tbh]
\centering
\begin{minipage}{0.24\textwidth}
\includegraphics[width=\textwidth, clip=true, trim= 0cm 0.0cm 0.8cm 0.8cm]{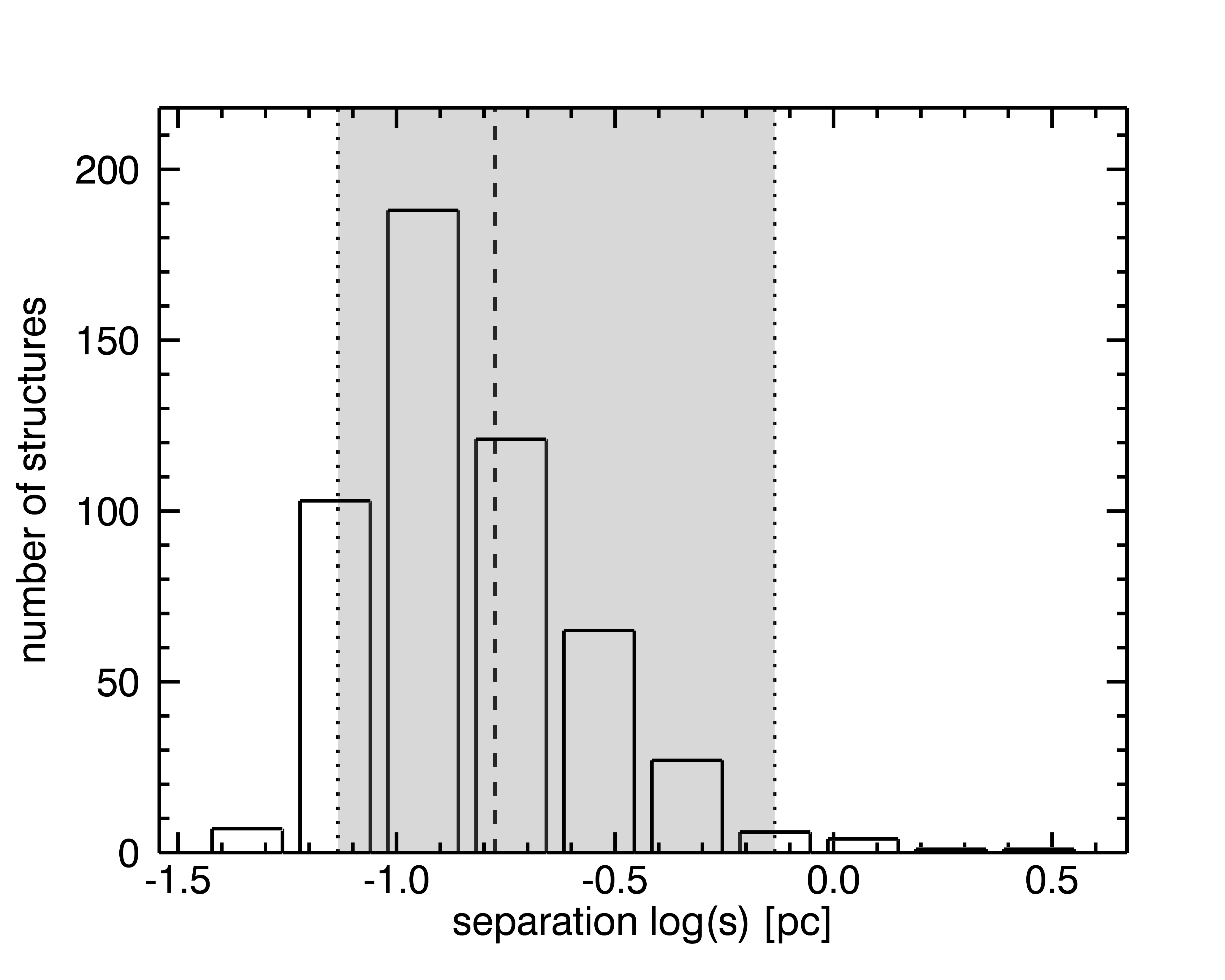}
\end{minipage}
\begin{minipage}{0.24\textwidth}
\includegraphics[width=\textwidth, clip=true, trim= 0cm 0.0cm 0.8cm 0.8cm]{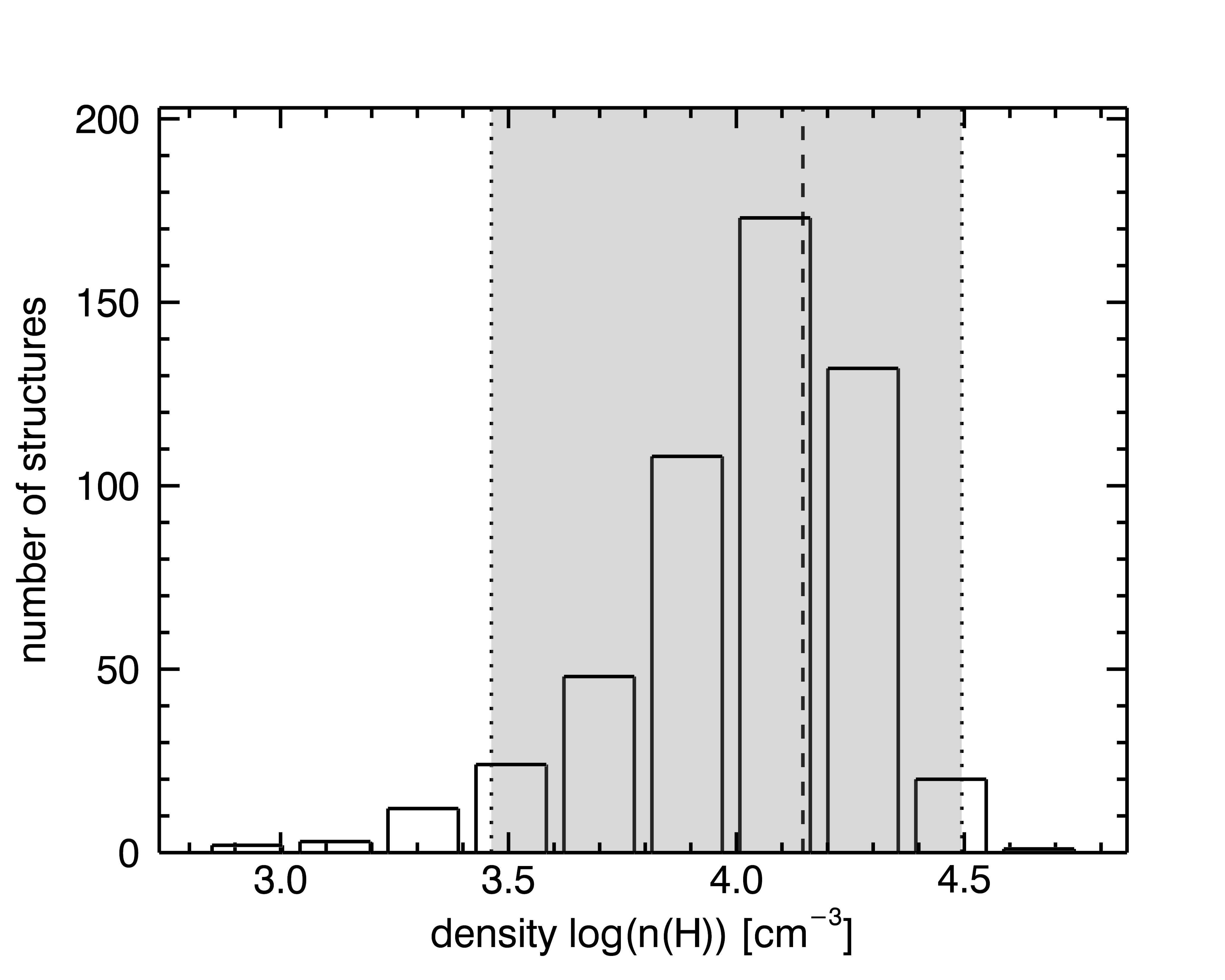}
\end{minipage}

\begin{minipage}{0.24\textwidth}
\includegraphics[width=\textwidth, clip=true, trim= 0cm 0cm 0.8cm 0.8cm]{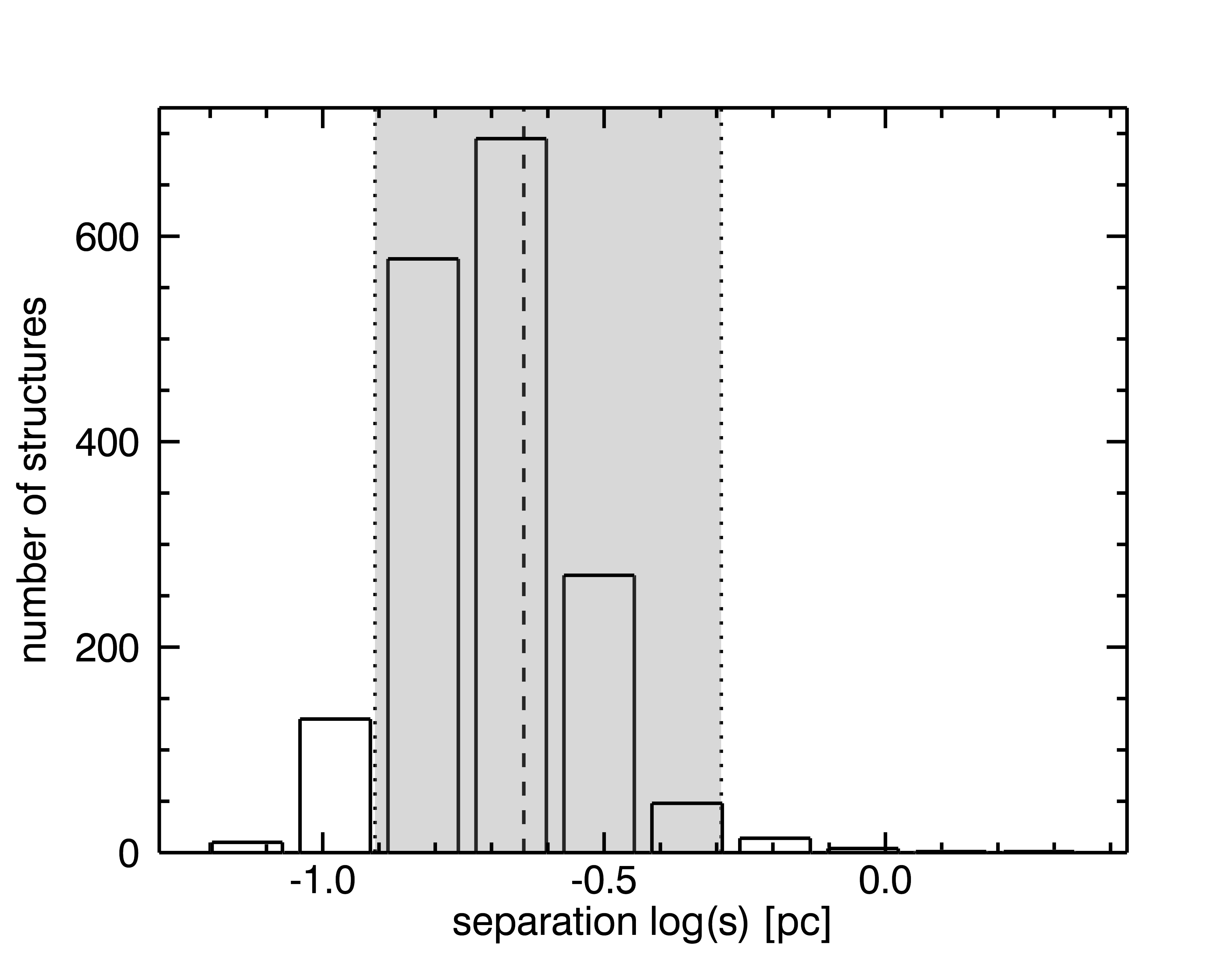}
\end{minipage}
\begin{minipage}{0.24\textwidth}
\includegraphics[width=\textwidth, clip=true, trim= 0cm 0cm 0.8cm 0.8cm]{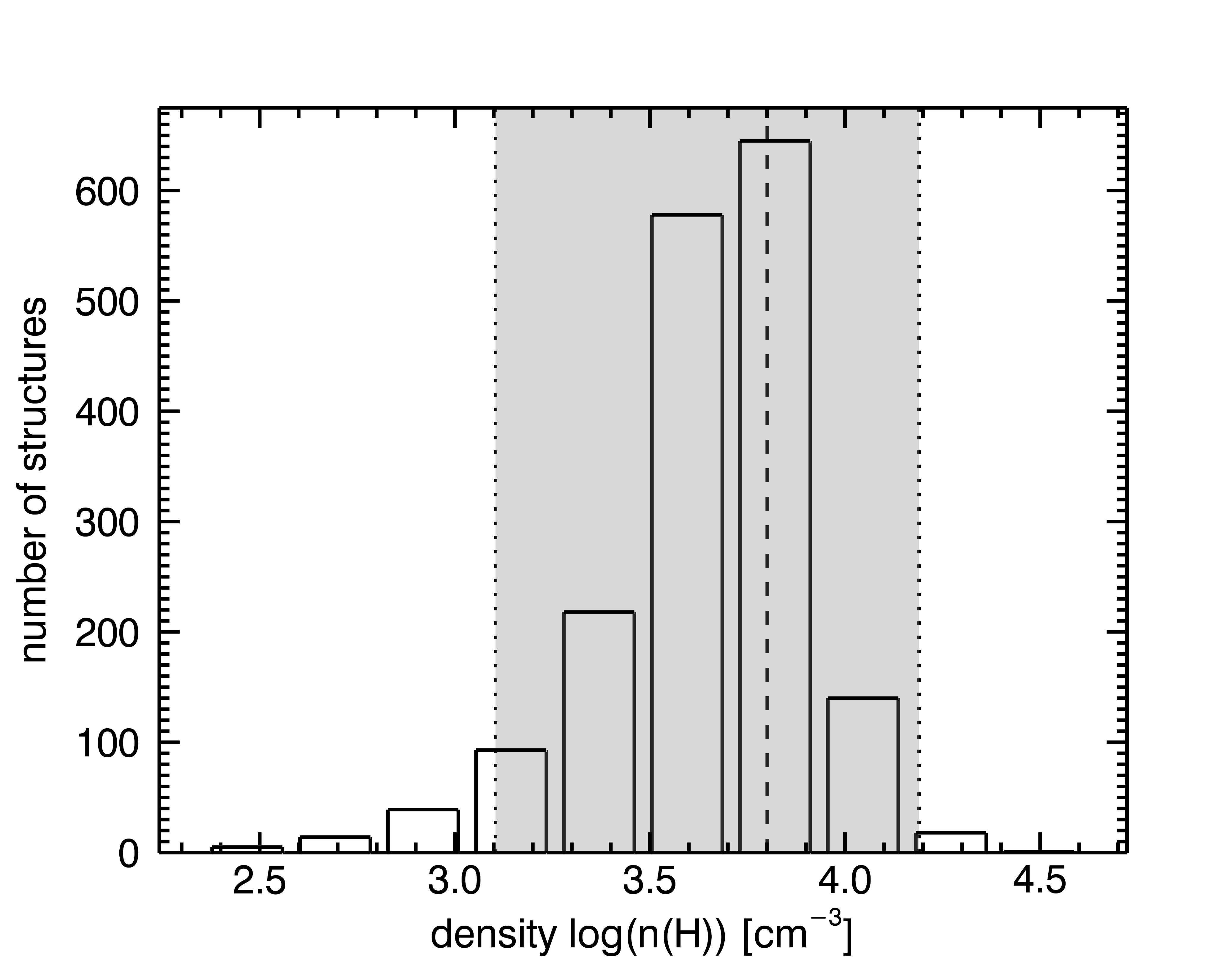}
\end{minipage}
\caption{Distributions of the separations (left) and densities (right) of the structures identified from the scale maps $i=2$ (top) and $i=3$ (bottom). The dashed line indicates the median and the dotted lines the 95\% quantiles of each distribution. }
\label{histograms}
\end{figure}

We include in the fragmentation analysis all structures identified at scales $i=2\text{ -- }8$. We only include structures within the Nessie filament area (see the polygon in Fig. \ref{Nessie complete}). We computed the projected nearest neighbor distances of the structures. The separation distributions of the scales $i=2,3$ are shown in Fig. \ref{histograms}. They are non-Gaussian in shape and we adopt the median separation as a diagnostic of the separations (given in Table \ref{frag-tab}). 

\begin{figure}[tbhp]
\centering
\includegraphics[width=0.5\textwidth, clip=true, trim= 0cm 0cm 0cm 0cm]{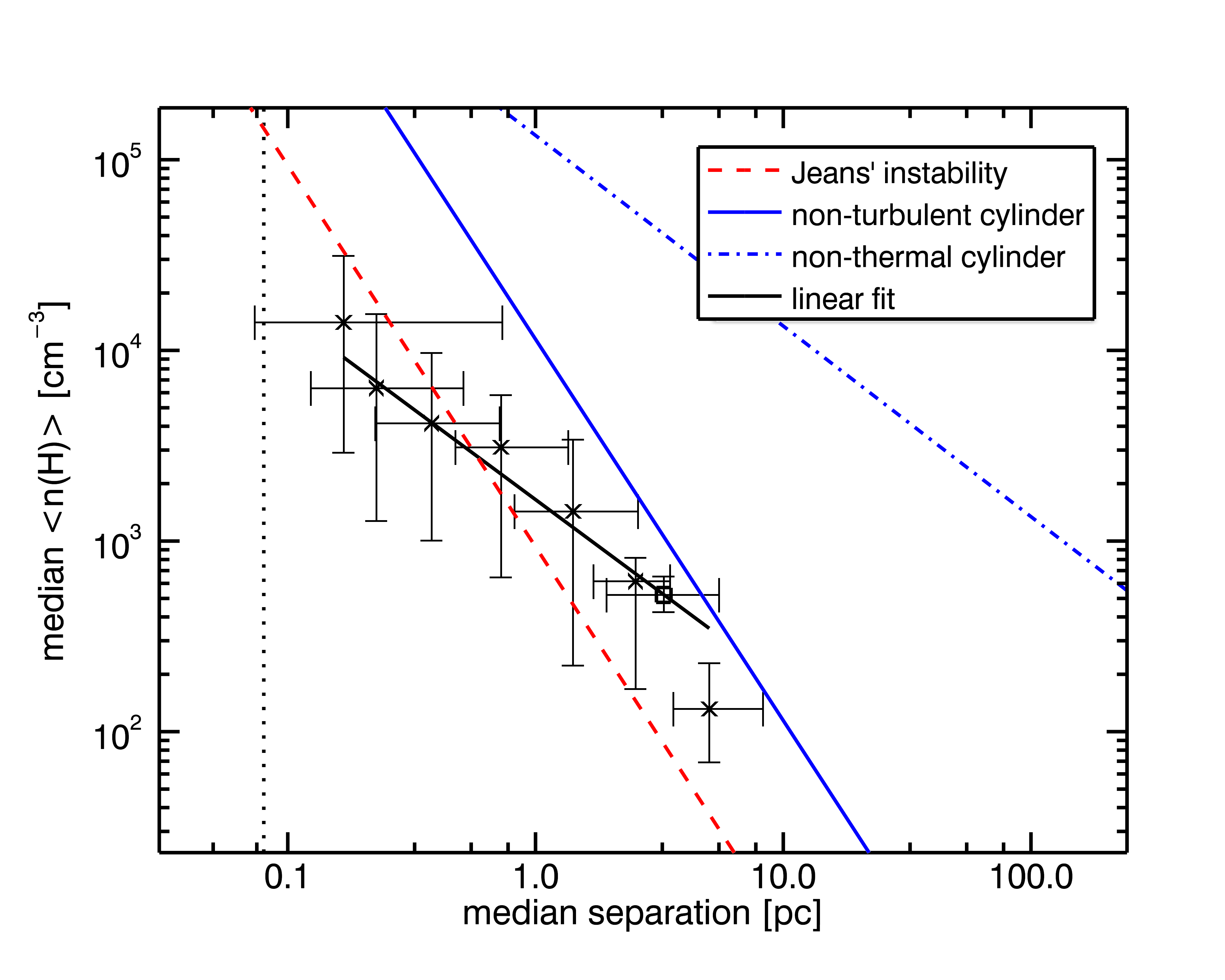}
\caption{Median number density of structures at different spatial scales as a function of their median separation. Measurements of this study are marked with crosses. The square marks the data point derived from HNC observations of \cite{Jackson2010}. The error bars show the $95\%$ quantiles of both measurements. The blue lines indicate the scale dependency of a infinite long cylinder in the non-turbulent case (solid), and non-thermal case (dash-dotted), and the dashed, red line indicates the scale dependency of Jeans' fragmentation. The black line shows a power law fit to the data.}
\label{densepplot}
\end{figure}

For the fragmentation analysis an estimate of the structure density is interesting; we estimate this from the outputs of the clumpfind-2D algorithm. The size of a structure was given by clumpfind-2D as the number of pixels, $N_\text{pix}$, in the $\rm FWHM$ area. For the calculation of the structure volume we assume the shape of a prolate spheroid, that has been found to be among the shapes that best quantify the structures at the scales we are looking at \citep[e.g.,][]{Kainulainen2014}. The depth of the prolate spheroids is estimated as the shorter of the projected $x$ and $y$ dimensions. Therefore, the volume of a fragment is 
\begin{equation}
V=4/3~\pi\cdot x\cdot y\cdot \text{min}\lbrace x,y\rbrace $	.$
\end{equation}
The average column density, $\overline{N}(H)$, is given by: $\overline{N}(\text{H})=N(\text{H})_{\text{tot}}/N_\text{pix}$, and therefore, the hydrogen number density of one structure is: $n(H)=\overline{N}(\text{H})\cdot \pi \cdot x \cdot y /V$. The median number density and the $95 \%$ interval for structures at each scale are shown as a function of their median separation in Fig. \ref{densepplot}.

Additionally, we estimated the median separation and density from the HNC molecular line observations of \cite{Jackson2010}. We used the shown positions to estimate their separation at the distance of $d= \rm 3.5~kpc$. The density was calculated assuming a spherical geometry with a radius of $r=\sqrt{\Omega/\pi}$, using the angular size $\Omega$ of the identified clumps, and their mass $M$. The hydrogen number density is given by: 
\begin{equation}
n(H)=\frac{M}{\mu_\text{H} m_\text{H} (4/3\,\pi r^3)}$   ,$
\end{equation}
where $\mu_\text{H}=1.4$ is the mean molecular weight of the interstellar medium with respect to atomic hydrogen and $m_\text{H}$ is the mass of a hydrogen atom.
 
We estimated the uncertainty of the median separations and median mean densities using bootstrapping, because their probability distributions are not Gaussian (see Fig. \ref{histograms}).
For the separation and mean density on every scale, we drew a new sample of values from among the observed values of separations and mean densities. This new sample had the same amount of data points as originally detected at that scale. We then calculated the median of these new, simulated samples. The resulting distribution of the median values then estimates the sampling function of the observed median and was used to estimate the uncertainty using the standard deviation. The uncertainties vary between $1 \%$ and $14 \%$ for the separation and between $1 \%$ and $25 \%$ for the density on scales of $i=3$ and $i=8$. The uncertainty values of all scales are given in Table \ref{frag-tab}. 

The scatter shown in the separation density plot represent the $95\%$ quantiles of the measured parameters. Large uncertainties, which are neglected here, are the opacity at different wavelength (J, H, K, $8~\mu m$) and their ratios contributing in the extinction measurement and the conversion factor from extinction to column density. For measuring masses also the uncertainty of the distance, as discussed before, introduces an significant contribution. For more detail see \cite{Kainulainen2011irdcs,Kainulainen2013b}.

The density-separation relation (Fig. \ref{densepplot}) shows a clear decrease of the mean densities for larger separations. We perform a linear least-square-fit in the log-log space to the data, which represents a power law of the form $\widetilde{n}(\text{H})=A \cdot \widetilde{s}\,^p$ as $log(\widetilde{n}(\text{H}))=p \cdot log(\widetilde{s})+log(A)$. 
The resulting parameters are $p=-0.96 \pm 0.05$ and $log(A)=3.22 \pm 0.02$, which is $A=1669^{+91}_{-86}\, \rm cm^{-3}$.
The fitted model is shown as black line in figure \ref{densepplot}.

A commonly used fragmentation model is the spherical Jeans' instability model \citep{Jeans1902}, where the separation is linked to the mean density $\overline{\rho}$ via the Jeans' length 
\begin{equation}
l_J=c_s(\pi /(G \overline{\rho}))^{1/2}$   ,$
\label{Jeans}
\end{equation}
where $c_s$ is the sound speed within the medium, and $G$ the gravitational constant. We compute the prediction from this assuming a gas temperature of $T=\rm 15~K$. At all scales, the observed mean separations are in agreement with the Jeans' scale within a factor of $\sim 3$. However, for the smallest scales $i=2\text{ -- }4$ the measurements are systematically below the predicted relationship and for the largest scales systematically above (see the discussion about the slope of the relationship later in this section).

A more shallow slope of the Jeans' fragmentation can be achieved by assuming a non-isothermal medium \citep[e.g.][]{Takahashi2013}. The innermost dense ($\rm \sim 10^4~cm^{-3}$) regions of the cloud are shielded from the interstellar radiation field and therefore, can reach temperatures down to $\rm 10~K$. As the surrounding low density gas ($\rm \sim 10^2~cm^{-3}$) is exposed to the radiation, we assume a higher temperature of $\rm 20~K$. This leads to a slope of about -1.7, which still does not solve the systematic deviations from the observation.

Another commonly used model describes the fragmentation an infinitely long, self-gravitating cylinder \citep{Chandrasekhar1953, Inutsuka1992}. This model predicts the separation, $\lambda$, depending on the scale-height $H=\sigma_\text{v}(4\pi G \rho_c)^{-1/2}$, where $\rho_c$ is the central density of a filament in virial equilibrium, $\sigma_\text{v}$ the velocity dispersion of the medium, and $G$ the gravitational constant. In the case of a non-turbulent medium, the velocity dispersion $\sigma_\text{v}$ is given by the sound speed $c_s$ within the medium (we assume $T=15~\text{K}$ to calculate the sound speed). In the regime of the filament radius $R \gg H$ the separation is given by $\lambda = 22~H$. If we assume a central density at the largest scale of $n_c(H) \approx \rm 10^{3}~cm^{-3}$, then we derive a scale-hight of $H \approx \rm 0.15~pc$. This is smaller than the typical radius of Nessie, $R \approx \rm 1.5~pc$ (see Section \ref{mappingresult}). Therefore, the separation is predicted to be \begin{equation}
\lambda = 22 \cdot c_s(4\pi G \rho_c)^{-1/2}$  ,$
\label{non-turbulent}
\end{equation}
which is shown in Fig. \ref{densepplot} and it is in agreement with the measurements within a factor of $\sim 3$ for scales larger than $i=5$, but systematically above the measured densities. However, the model predicts central densities while we derived mean densities, and therefore, the model predicts an upper limit of the mean densities.

The above models describe fragmentation in non-turbulent medium. However, observations show that high line mass filaments have a non-thermal linewidth \citep{Jackson2010,Kainulainen2013a}, which is higher than the sound speed $c_s$ in the non-turbulent case. \cite{Larson1981} found an relation between the size of a molecular cloud and its observed linewidth. Such a linewidth-size relation might also apply to the here observed structures and therefore, we adopted a typical relation of $\rm \sigma_\text{v}=0.72~km\,s^{-1} \cdot (\lambda / 1~pc)^{0.5}$ \citep{Solomon1987, Heyer2004, Pillai2006a, Shetty2012, Colombo2015}, where the linewidth $\sigma_\text{v}$ depends on the observed size scale $\lambda$. The non-thermal linewidth exceeds non-turbulent motion, given by the sound speed $c_s$, at large scales. But the linewidth-size relation can also be partially explained by the non-isothermal behaviour of the gas.

\begin{equation}
\lambda^{0.5} = 22 \cdot {\rm 0.72~km\,s^{-1}}(4\pi G \rho_c)^{-1/2}$  ,$
\label{turbulent}
\end{equation}
where $\rho_c$ is the central density of a filament in virial equilibrium, and $G$ the gravitational constant (Fig. \ref{densepplot}). 

Therefore, the relation between the central density and the separation is $\rho_c \propto \lambda^{-1}$, which is in agreement with the observed slope of $p=-0.96 \pm 0.05$. However, again we have to mention that the model predicts central densities while we derived mean densities. Additionally, without informations about the kinematics of the cloud, we cannot constrain the scaling velocity of the linewidth-size relation.

\begin{table*}[tbhp]
\begin{center}
\caption{Results of the fragmentation analysis}
\begin{tabular}{cccccccccc}
\hline 
\noalign{\vspace{0.5mm}}
Scale $i$ & Scale & $N_\text{strc}$ & $\sum(M_\text{strc})$ & $\overline{M}_\text{strc}$ & $\widetilde{n}(\text{H})\tablefootmark{a}$ & $\sigma (\widetilde{n}(\text{H}))$ & $\widetilde{s}\, \tablefootmark{b}$ & $\sigma (\widetilde{s})$ & $\lambda_\text{J}$ \\ 
 & [pc] &  & [$10^3~\text{M}_\odot$] & [$\text{M}_\odot$] & [$10^3~\text{cm}^{-3}$] & [$10^3~\text{cm}^{-3}$] & [pc] & [pc] & [pc] \\ 
\hline 
\noalign{\vspace{0.5mm}}
>8 & >5.2 & 1 & 34.5 &  &  &  &  &  & \\ 
8 & 5.2 & 11 & 4.0 & 373 & 0.13 & 0.04 & 5.0 & 0.8 & 2.7 \\ 
7 & 2.6 & 31 & 3.2 & 108 & 0.61 & 0.05 & 2.5 & 0.2 & 1.2 \\ 
6 & 1.3 & 72 & 2.4 & 33.2 & 1.4 & 0.12 & 1.4 & 0.1 & 0.81 \\ 
5 & 0.65 & 242 & 2.0 & 8.28 & 3.1 & 0.12 & 0.73 & 0.02 & 0.55 \\ 
4 & 0.33 & 903 & 1.9 & 2.06 & 4.1 & 0.07 & 0.38 & 0.004 & 0.48 \\ 
3 & 0.16 & 1751 & 1.2 & 0.66 & 6.3 & 0.08 & 0.23 & 0.002 & 0.38 \\ 
2 & 0.08 & 523 & 0.20 & 0.40 & 14.2 & 0.47 & 0.17 & 0.004 & 0.26 \\ 
\hline 
\end{tabular} 
\label{frag-tab}
\end{center}
\tablefoot{\tablefoottext{a}{median of the mean density of the identified structures}
\tablefoottext{b}{median of the separation between identified structures}}
\end{table*}

\subsection{Comparison with ATLASGAL}
\label{ATLASGAL sus}

\begin{figure}[tbhp]
\centering
\includegraphics[width=0.5\textwidth, clip=true, trim= 4.5cm 1cm 5cm 0cm]{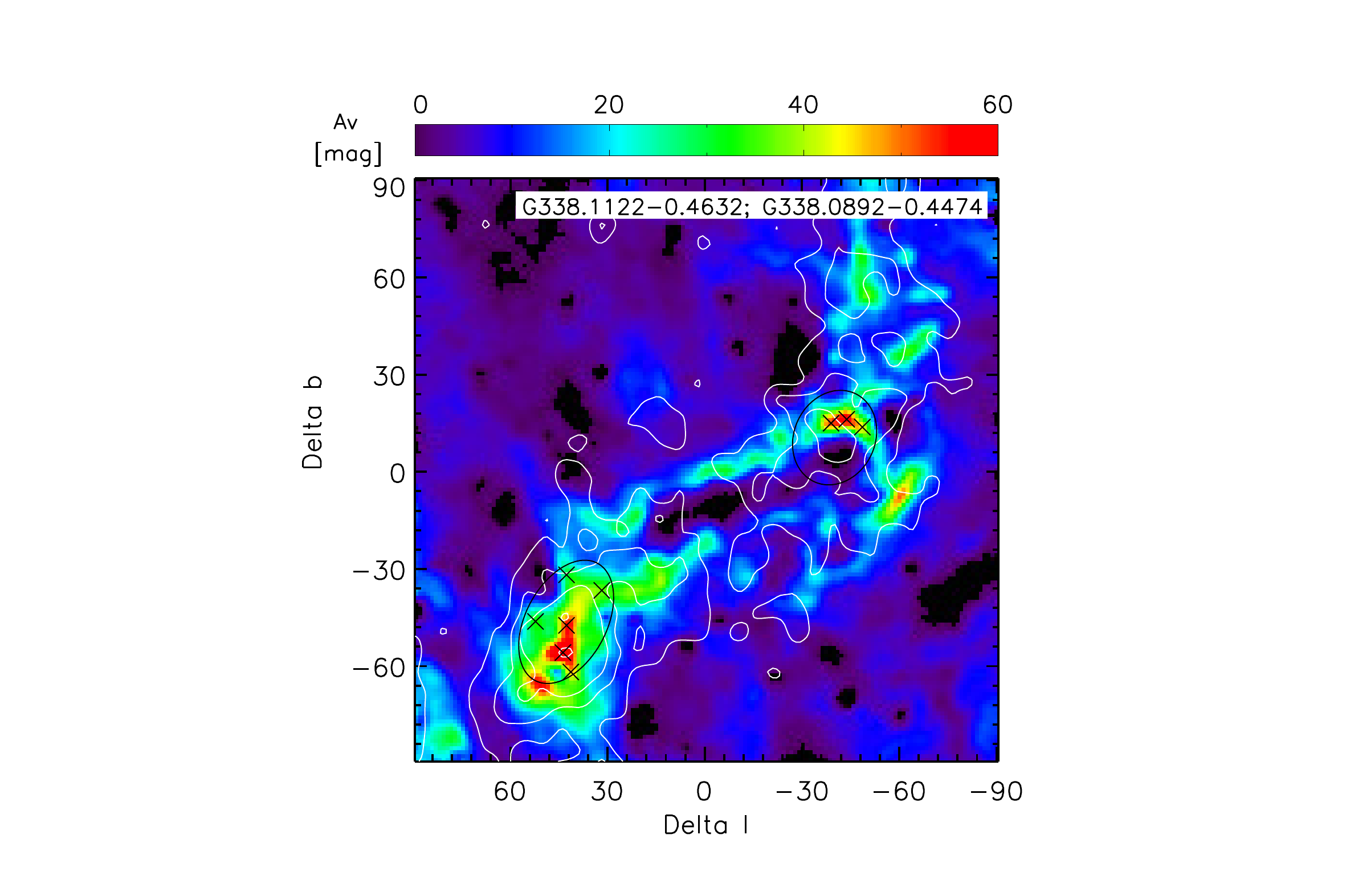}
\caption{Combined NIR and MIR extinction map ($l=338.10^\circ, b=-0.45^\circ$) overlaid with the half power contour of two ATLASGAL GCSC sources (black ellipses) and their covered sources identified with clumpfind-2D from the scale 2 map. The white lines show the contours of the ATLASGAL emission.}
\label{compare sources}
\end{figure}

We describe shortly how the parsec-scale structures identified in Nessie from ATLASGAL data \citep[resolution of $18\arcsec$, ][]{Schuller2009} break down into substructures when viewed at about $10$ times finer resolution of the extinction data. For this, we considered the $16$ sources from the ATLASGAL GCSC catalog \citep{Csengeri2014} that are likely embedded in the cloud. We calculated the number of structures within the FWHM ellipse of the ATLASGAL sources at the two smallest scales ($i=2, 3$) of the extinction map (see Fig. \ref{compare sources}). We also estimated the mass of the ATLASGAL clumps by adopting Equation (\ref{atlasmassequ}) and assuming a dust temperature of $T_\text{d} \approx \rm 15~K$. These masses are then compared to the total mass of the small scale structures. The resulting ratios are shown in Tab. \ref{ATLAS-result}.

In particular, we found that, on average, the number of small scale structures within the half power ellipse of the clump is $\overline{N}_{\text{strc}, 2}=2.9$ and $\overline{N}_{\text{strc}, 3}=2.8$. These contain $2\%$ and $6\%$ of the mass of the ATLASGAL clump. The half power ellipses of the clumps and the $i=2$ structures identified within the clumps are shown in Fig. \ref{ATLAS clumps} overlaid on the extinction map. While half of the ATLASGAL clumps are clearly associated with high extinction peaks, especially the four most massive ones ($\rm >500~M_\odot$) contain  no or only low extinction peaks. This is dominantly because of the caveats of the extinction mapping technique. The massive clumps commonly exhibit MIR emission of Polycyclic Aromatic Hydrocarbons (PAHs) in the $8\rm~\mu m$ band \citep{Benjamin2003}; this interferes with the extinction mapping procedure. Also bright foreground stars cause a lack of mid-infrared extinction and influence our results. In total this likely leads to an underestimated number of substructures per clump and to underestimating some of their masses. It also shows that our method is excellent for identifying the youngest and densest regions, but it starts to fail as soon as star formation progresses and the regions show strong MIR emission.\\

\begin{table*}[htbp]
\begin{center}
\caption{ATLASGAL GCSC clumps \citep{Csengeri2014} likely embedded in the Nessie cloud}
\begin{tabular}{cccccccccc}
\hline 
\noalign{\vspace{0.5mm}}
Name & Size & PA & $M_\text{clump}$ & $N_\text{strc, 2}$ & $M_\text{strc, 2}$ & $\frac{M_\text{strc, 2}}{M_\text{clump}}$ & $N_\text{strc, 3}$ & $M_\text{strc, 3}$ & $\frac{M_\text{strc, 3}}{M_\text{clump}}$\\ 
 & ["] & [$^\circ$] & [M$_\odot$] &  & [M$_\odot$] &  &  & [M$_\odot$] &  \\
\hline
\noalign{\vspace{0.5mm}}
G338.9380-0.4231: & 46 x 20 & -12 & 221 & 4 & 3.15 & 0.014 & 4 & 12.33 & 0.056 \\
G338.9362-0.4808: & 28 x 22 & 52 & 197 & 2 & 1.36 & 0.007 & 1 & 4.69 & 0.024 \\
G338.9371-0.4919: & 41 x 34 & 134 & 1094 & 3 & 3.27 & 0.003 & 2 & 10.77 & 0.010 \\
G338.9275-0.5018: & 39 x 26 & 102 & 523 & 0 & 0.00 & 0.000 & 3 & 7.77 & 0.015 \\
G338.8688-0.4796: & 32 x 23 & 71 & 248 & 5 & 4.34 & 0.018 & 1 & 15.63 & 0.063 \\
G338.7790-0.4591: & 39 x 23 & -24 & 176 & 4 & 3.97 & 0.022 & 4 & 15.95 & 0.090 \\
G338.7314-0.4691: & 32 x 19 & 90 & 116 & 3 & 4.76 & 0.041 & 3 & 13.14 & 0.114 \\
G338.5519-0.4190: & 27 x 24 & 71 & 134 & 2 & 2.57 & 0.019 & 4 & 7.65 & 0.057 \\
G338.4236-0.4101: & 28 x 26 & 111 & 292 & 0 & 0.00 & 0.000 & 2 & 2.20 & 0.008 \\
G338.3937-0.4053: & 42 x 31 & 72 & 632 & 2 & 2.22 & 0.004 & 3 & 10.43 & 0.016 \\
G338.3923-0.3972: & 34 x 19 & 16 & 124 & 2 & 1.75 & 0.014 & 3 & 9.91 & 0.080 \\
G338.3271-0.4096: & 36 x 27 & -20 & 534 & 4 & 3.72 & 0.007 & 3 & 11.02 & 0.021 \\
G338.1991-0.4642: & 27 x 25 & 36 & 181 & 2 & 2.61 & 0.014 & 3 & 10.09 & 0.056 \\
G338.1122-0.4632: & 41 x 25 & 62 & 202 & 6 & 8.27 & 0.041 & 5 & 20.46 & 0.101 \\
G338.0892-0.4474: & 30 x 25 & 65 & 147 & 3 & 6.57 & 0.045 & 1 & 16.08 & 0.109 \\
G338.3048-0.5223: & 47 x 22 & 95 & 216 & 4 & 4.21 & 0.019 & 3 & 15.57 & 0.072 \\
mean:             &        &  & 315 & 2.88 & 3.30 & 0.017 & 2.81 & 11.48 & 0.056 \\
stddev:           &        &  & 261 & 1.63 & 2.16 & 0.014 & 1.17 &  4.66 & 0.037 \\

\hline
\end{tabular}
\label{ATLAS-result}
\end{center}
\end{table*}

\section{Discussion}

\subsection{Scale-dependent fragmentation of Nessie}

In the following, we discuss the scale-dependent fragmentation of Nessie (Fig. \ref{densepplot}) in the context of the analytic gravitational fragmentation models. We showed that the upper limit of the average line-mass of Nessie is $(M/l)=\rm 627~M_\odot\, pc^{-1}$. For a thermally supported filament at a temperature of $T=15~\text{K}$ the critical line-mass is $(M/l)_\text{crit}=\rm 20~M_\odot\, pc^{-1}$. Thus, the filament is clearly thermally supercritical. There are no analytic theories that would self-consistently explore the evolution of such highly thermally super-critical filaments.

In the absence of directly applicable models, a common approach in the recent literature is to assume that the non-thermal motions provide a straightforward, idealized supporting force for the filament, increasing its critical line-mass \citep[e.g.,][]{Jackson2010, Hernandez2012, Busquet2013, Beuther2015}. This commonly leads to a conclusion that the line-masses of high line-mass filaments are close to their critical line-masses. This is true for Nessie, too. \citet{Jackson2010} showed that the non-thermal motions in Nessie increase the critical line-mass to $(M/l)_\text{vir}=\rm 525~M_\odot\, pc^{-1}$, which is similar to our observed value. 

Building on the above agreement, observations are commonly compared to the predictions of gravitational fragmentation models developed for near-equilibrium  cylinders. These models typically proceed from a static initial configuration with a linear perturbation analysis. In short, such models predict a periodic fragmentation pattern with a specific wavelength, i.e., the fragmentation pattern predicted by the models is not scale-dependent. However, the fragmentation wavelength depends on the density of the filaments as described by Eqs. \ref{turbulent}, \ref{non-turbulent}, and \ref{Jeans}; filaments with different densities have different fragmentation wavelengths. This should be kept in mind when interpreting the relationship between the data and models presented in Fig. \ref{densepplot}. 

In this context, the observed slope of the mean density -- separation relationship in Nessie is in agreement with that of a non-thermal, self-gravitating cylinder that has a Larson-like linewidth-size relation \citep[$\sigma_v \propto \lambda^{0.5}$, ][]{Larson1981, Solomon1987, Heyer2004, Shetty2012, Colombo2015}. 
As the cloud shows non-thermal velocity dispersions \citep{Jackson2010}, this relation could be a result of turbulent motions within the cloud, but also systematic motions, such as collapse, could affect the linewidth. The observed median nearest-neighbor separations of the fragments are within a factor of two of the predictions of the isothermal and non-isothermal Jeans' fragmentation \citep{Jeans1902}. However, the slope is significantly steeper than the observed one. Additionally, on the large scales the separations also are in agreement with the fragmentation model of a non-turbulent, self gravitating, infinite long cylinder \cite{Chandrasekhar1953,Inutsuka1992}, but again the slope of the model is significantly steeper than observed. Note, that the cylindrical models predict central densities, which can only be seen as upper limits for the derived mean densities.

Previously, a change of fragmentation mode between large and small scales has been seen at the size-scale of $\rm \sim 0.5 pc$, e.g., in the studies of the young high-mass cloud G11.11-0.12 \citep{Kainulainen2013a}, the Taurus cloud \citep{Hacar2013}, and the integral-shaped filament in Orion \citep{teixeira2016, kainulainen2017isf}. While we do not detect such feature in Nessie, the data are in agreement with the presence of such a feature, i.e., cannot rule it out (c.f., Fig. \ref{densepplot}). One possible explanation for the change of fragmentation modes can be changing influence of the environment \citep{Pon2011}. While on large scale fragmentation is driven by the characteristics of the cylindrical, filamentary structure, the smaller scales approach a more spherical shape, which is independent of larger scales. Also, recent numerical simulations have explored possibilities to explain scale-dependent fragmentation through dynamical processes \citep[e.g.,][]{Clarke2017filaments, Gritschneder2017}.

\subsection{Star formation potential}

Ultimately, one would like to link the fragmentation in Nessie to star formation. To take the first step towards this, we estimated the young stellar object (YSO) content of Nessie using publicly available multi-band photometric catalogs. The detailed methods used to identify the YSOs and estimate the SFR are explained in Zhang et al. 2017, submitted. Here we give a short description of the method. 

For the YSO selection we used near-infrared data \citep[we did the PSF photometry on VVV images, VISTA Variables in the Via Lactea, ][]{Saito2012}, {\emph Spitzer} GLIMPSE \citep[Galactic Legacy Mid-Plane Survey Extraordinaire, ][]{Benjamin2003, Churchwell2009} and MIPSGAL \citep[ Multiband Imaging Photometer Galactic Plane Survey, ][]{Carey2009, Gutermuth2015} archival catalogues, AllWISE catalogue \citep[Wide-field Infrared Survey Explorer, ][]{Wright2010}, Herschel Hi-GAL catalogue \citep[Herschel infrared Galactic Plane Survey, ][]{Molinari2010HiGAL, Molinari2016}, and Red MSX source catalogue \citep[Midcourse Space Experiment, ][used to include massive protostars]{Lumsden2013} and the methods from \cite{Gutermuth2009, Koenig2014, Saral2015, Robitaille2008, Veneziani2013}. Our YSO selection scheme uses the SEDs of sources from 1 to 500\,$\mu$m and can efficiently mitigate the effects of contamination. In Nessie, we finally obtain 298 sources with the excessive infrared emission, of which 35 are classified as AGB candidates using the multi-color criteria. 

\begin{figure}
\centering
\includegraphics[width=0.9\linewidth]{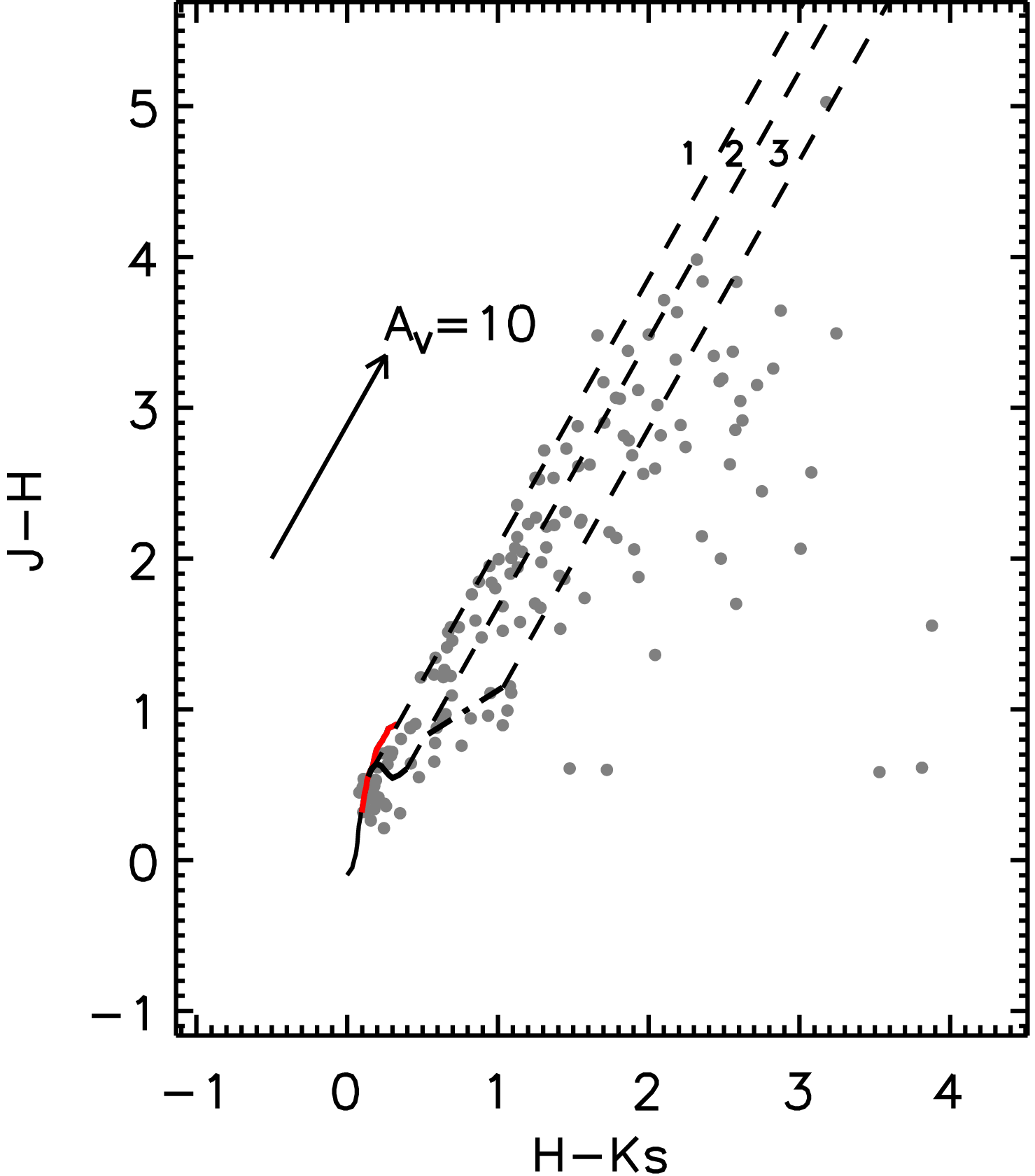}
\caption{The H-K$_{S}$ versus J-H color–color diagram for the YSO candidates in Nessie. The solid curves show the intrinsic colors for the main-sequence stars (black) and giants \citep[red;][]{Bessell1988}, and the dash–dotted line is the locus of T Tauri stars from \citet{tt}. The dashed lines show the reddening direction, and the arrow shows the reddening vector. The extinction law we adopted is from \citet{xue16}. Note that the dashed lines separate the diagram into three regions marked with numbers 1, 2, and 3 in the figure. We use different methods to estimate the extinction of YSO candidates in different regions (see the text for details).}
\label{fig:fig1mz}
\end{figure}

Considering the distance of Nessie, it is necessary to correct the flux densities of the YSO candidates for extinction. We use the method suggested by \cite{Fang2013, ZhangM2015} to estimate the foreground extinction towards each YSO candidate and de-redden their photometry. Here we also give a short description about this method. 
\begin{itemize}
\item[1]{For the sources with J, H, K$_S$ detections, the extinction is obtained by employing the JHK$_S$ color-color diagram. Figure~\ref{fig:fig1mz} shows the J-H versus H-K$_S$ color-color diagram of the YSO candidates in Nessie. Given the different origins of intrinsic colors of YSO candidates, the color-color diagram is divided into three subregions. In region 1, the intrinsic color of [J-H]$_0$ is simply assumed to be 0.6; in region 2, the intrinsic color of a YSO is obtained from the intersection between the reddening vector and the locus of main sequence stars \citep{Bessell1988}; in region 3, the intrinsic color is derived from where the reddening vector and the classical T Tauri star (CTTS) locus \citep{tt} intersects. Then the extinction values of YSO candidates are estimated from observed and intrinsic colors with the extinction law of \citet{xue16}.}
\item[2]{For other sources (outside these three regions or without detections in JHK$_S$ bands), their extinction is estimated with the median extinction values of surrounding Class II sources that have extinction measurements in step 1.}
\end{itemize}

\begin{figure*}
\centering
\includegraphics[width=0.95\linewidth]{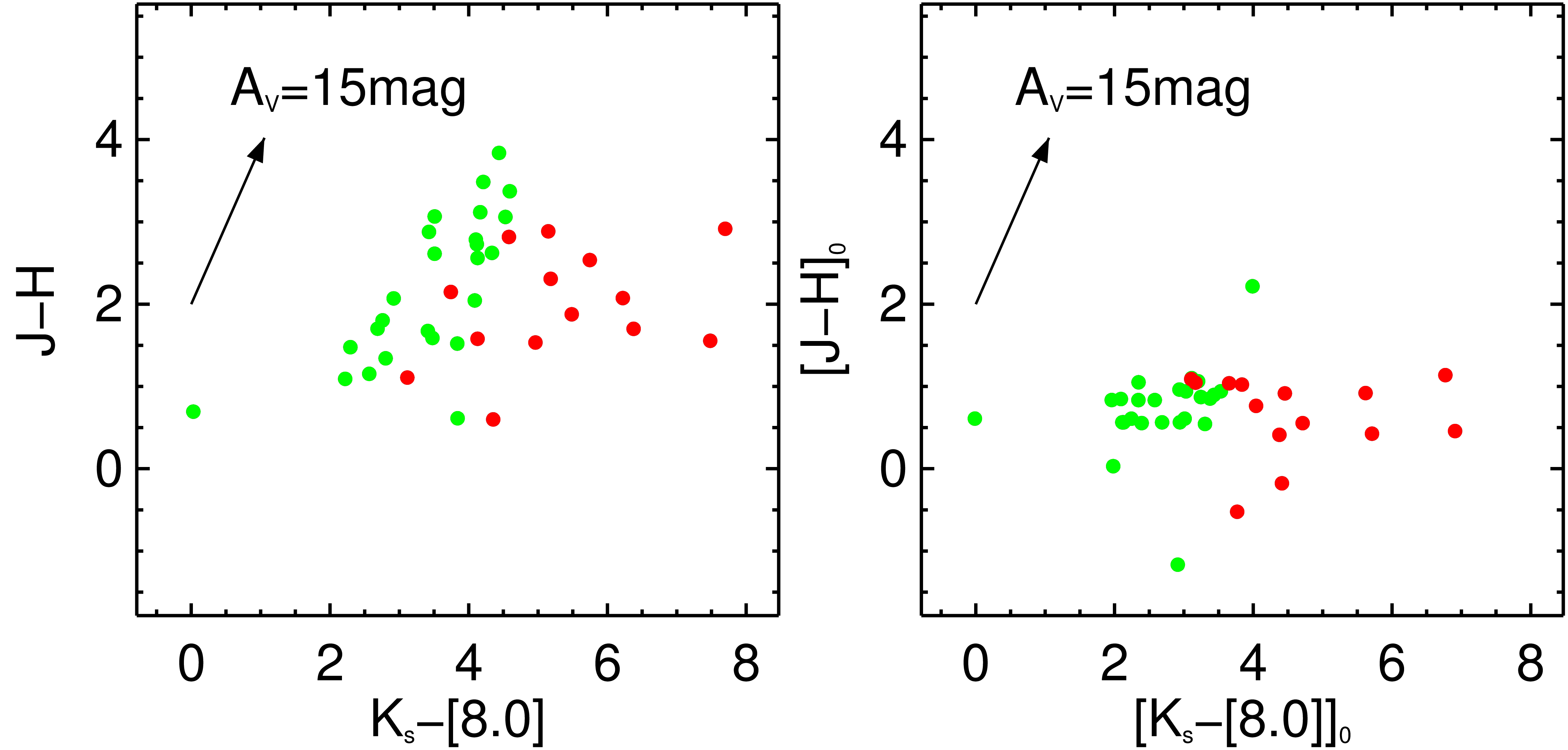}
\caption{The observed (\textit{left}) and de-reddened (\textit{right}) K$_S-[8.0]$ versus J-H color-color diagrams for Class I+Flat (red) and Class II (green) sources in Nessie. The black arrows show the extinction vectors.}
\label{fig:fig2mz}
\end{figure*}

Using the de-reddened SEDs, we re-classify the YSO candidates into Class I, Flat, and Class II sources based on their spectral indices and bolometric temperatures \citep{Greene1994, Chen1995}. Figure~\ref{fig:fig2mz} shows the K$_S-[8.0]$ versus J-H color-color diagrams before and after de-reddening for Class I+Flat and Class II sources in Nessie.

Although we have removed some contamination during the YSO selection process, our YSO candidates in Nessie are still contaminated by the foreground and background sources.

The foreground contamination mainly includes the foreground AGBs and the foreground YSOs which are associated with the molecular clouds that are located between us and Nessie. We use the $A_V$ values of YSOs obtained previously and the 3D extinction map \citep{Marshall2006} to isolate the foreground contamination. Based on the distance of Nessie, we can estimate the foreground extinction in different lines of sight towards Nessie with the 3D extinction map. If the extinction value of a YSO is lower than the corresponding foreground extinction of Nessie, this YSO would have high probability to be a foreground contamination. We checked the YSOs in Nessie and marked the possible foreground contamination using this method. The fraction of foreground contamination in Nessie is $10\%$ in Class I+Flat sources and $9\%$ in Class II sources.

Our YSOs are also contaminated by background sources, including  extragalactic objects, background AGBs, and background YSOs which are associated with the molecular clouds that are located behind Nessie. We think that the extragalactic contamination is not important in our YSOs because we are observing through the Galactic plane. Many background AGBs have been removed using the multi-color criteria during the YSO identification process. The residual contamination of background AGBs is estimated with the control fields. We select five nearby fields with weak CO emission as the control fields and apply the YSO selection scheme to all the control fields to select YSOs. Assuming that there is no YSOs in each control field, all selected `YSOs' in the control fields are actually contamination of AGBs (if neglecting the extragalactic contamination). With an assumption of a uniform distribution for AGB stars, we can estimate the number of residual background AGBs in the Nessie using the mean value of the surface density of background AGBs in five control fields. Combining the numbers of background AGBs identified by color criteria and estimated using control fields, we found that the fraction of background contamination is $22\%$ in Class I+Flat sources and $11\%$ in Class II sources. Note that we did not try to eliminate the contamination from background YSOs because they are difficult to remove without the information of radial velocities of YSOs.

\begin{figure}
\centering
\includegraphics[width=1.0\linewidth]{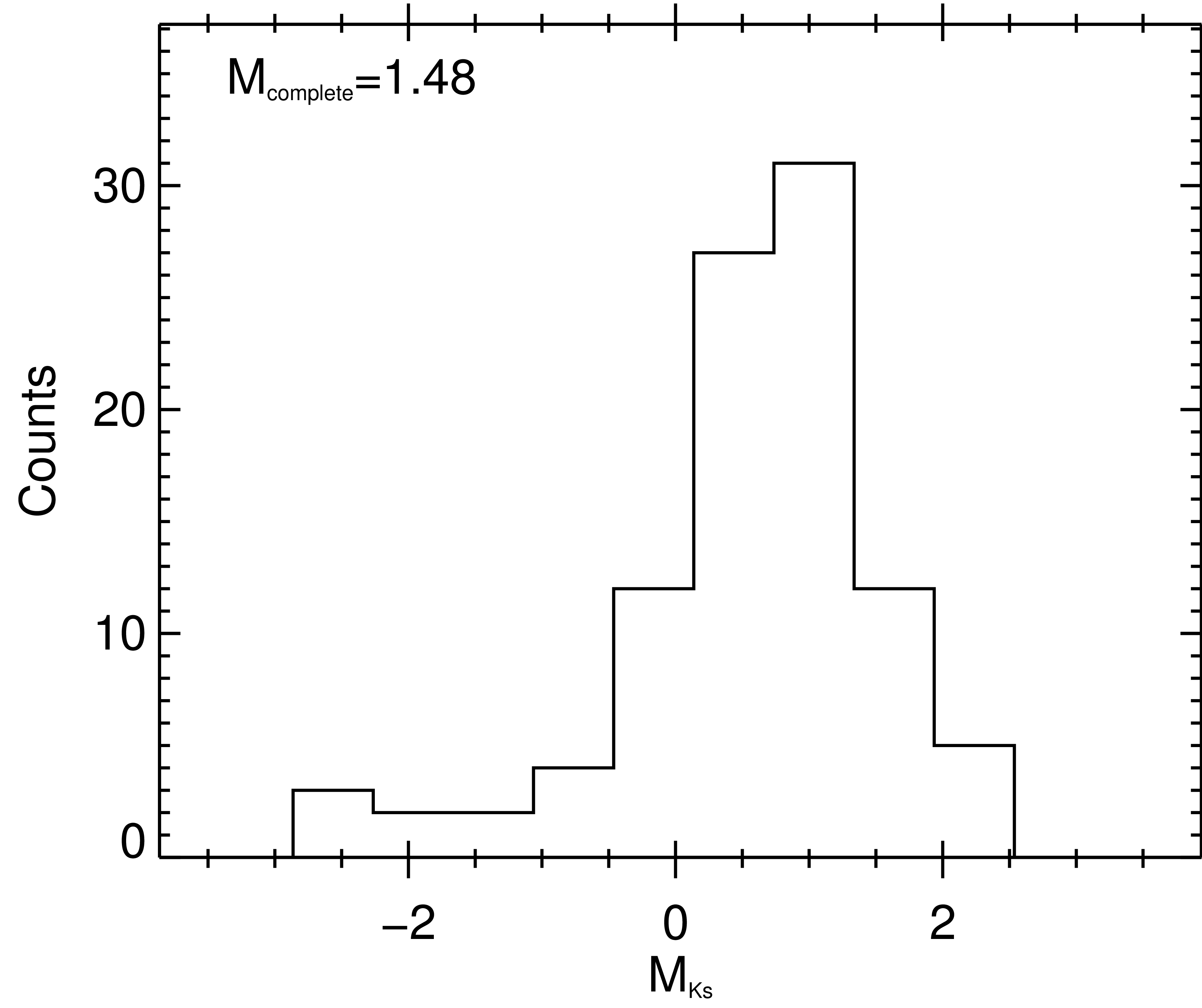}
\caption{K$_S$ absolute magnitude ($M_{\text{K}_S}$) histogram of Class II sources in Nessie.}
\label{fig:fig4mz}
\end{figure}

\begin{figure}
\centering
\includegraphics[width=1.0\linewidth]{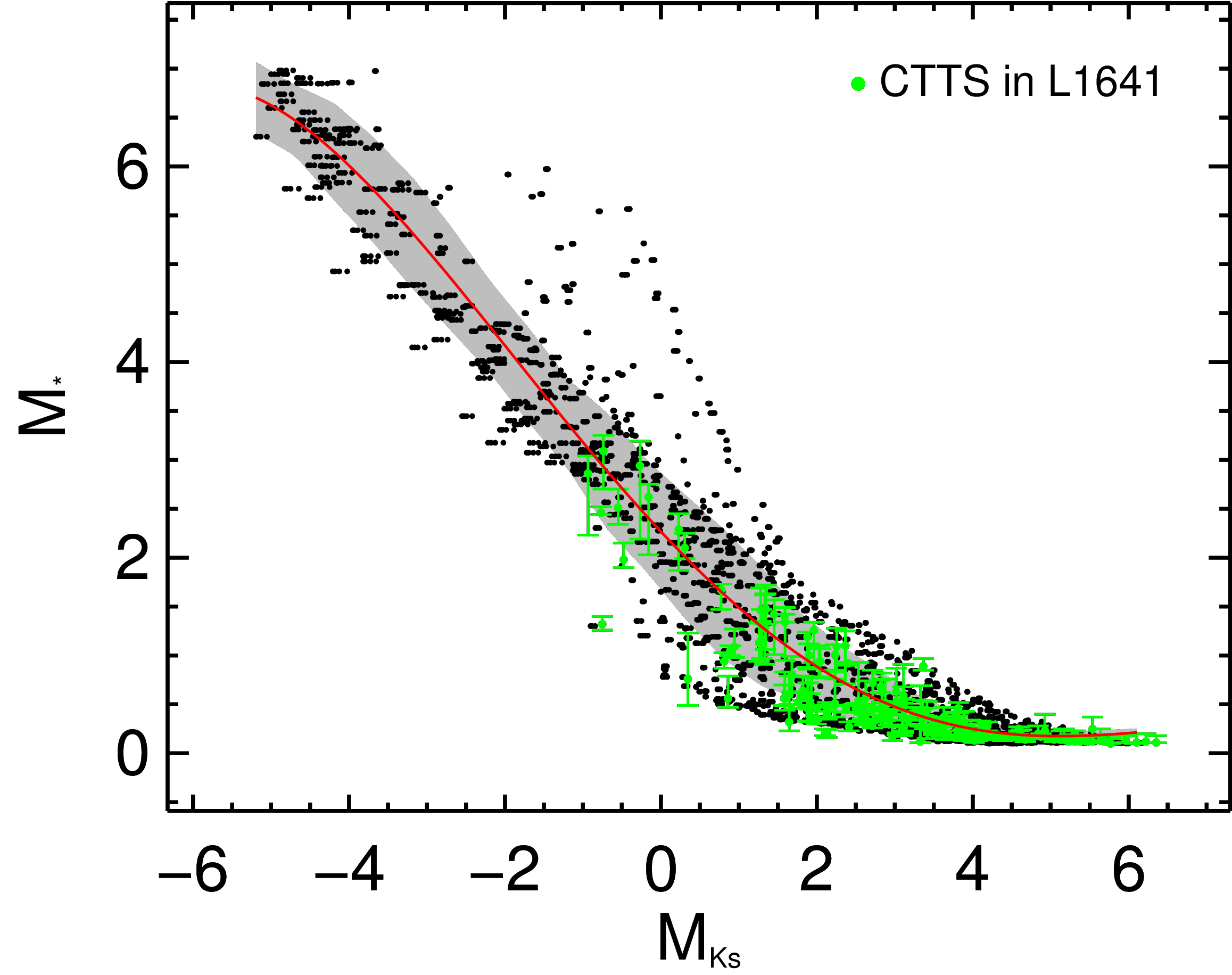}
\caption{The relation between stellar mass and K$_S$ absolute magnitude of Class II source. The black dots represent the \citet{Robitaille2006} Stage 2 models with $0.001<M_{\textrm{disk}}/M_{*}<0.01$, $0.08<M_{*}<7\,M_{\sun}$, and $30\degree <$ inclination angle $< 60\degree$. The red curve shows the robust polynomial fitting while the gray region shows the $1\sigma$ uncertainty of the fitting. The CTTS in L1641 from \citet{Fang2013} are marked with green filled circles. Most of CTTS are located in the gray region, which confirms that this $M_{\text{K}_S}-M_{*}$ relation for Class II sources is consistent with the observational results.}
\label{fig:fig5mz}
\end{figure}

\begin{figure}
\centering
\includegraphics[width=1.0\linewidth]{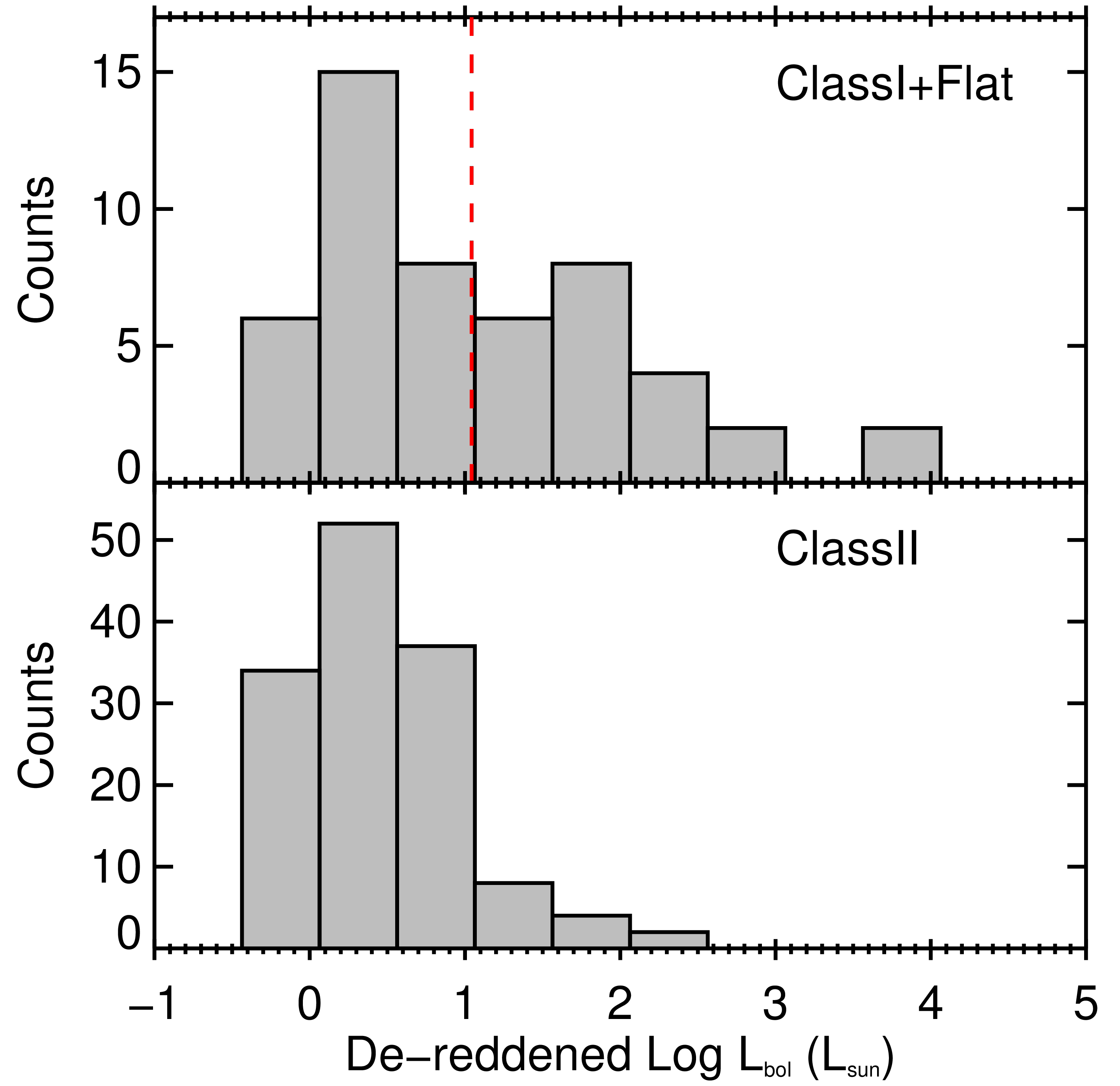}
\caption{De-reddened luminosity functions of Class I+Flat (top panel) and Class II (bottom panel) sources in Nessie. The red vertical line shows the de-reddened luminosity completeness.}
\label{fig:fig3mz}
\end{figure}

After removing the contamination, we obtain 51 Class I and flat spectrum objects and 137 Class II sources in the Nessie. In order to calculate the star formation rate (SFR), we must estimate the total mass of YSOs in Nessie. In this work, we use different methods to estimate the total mass of Class I+Flat and Class II populations:

\begin{itemize}
\item{We use the de-reddened photometry of Class II sources  in Nessie to estimate the flux completeness. Figure~\ref{fig:fig4mz} shows the K$_S$ absolute magnitude histogram of Class II sources in Nessie. We simply adopt the peak position of histogram as the completeness of K$_S$ band ($\rm \sim 1~mag$). Figure~\ref{fig:fig5mz} shows the $M_{\text{K}_S}-M_{*}$ relation for Class II sources constructed from YSO models presented by \cite{Robitaille2006}. Using this relation, we transfer the K$_S$ band completeness to the mass completeness of 1.48$\pm$0.65 $M_{\odot}$. Assuming a universal IMF \citep{Kroupa2001}, we estimated the number of Class II sources to be $1282_{-614}^{+1228}$ and the total mass of Class II sources to be  $698.4_{-355.9}^{+711.8}\rm \, M_\odot$.}
\item{For Class I+Flat sources, we used the observed luminosity functions constructed by \cite{Kryukova2012} as the template to estimate the total number of Class I+Flat sources. We calculate the bolometric luminosities of Class I+Flat sources using the trapezoid rule to integrate over the finitely sampled de-reddened SEDs \citep{dunham08,dunham15}. Figure~\ref{fig:fig3mz} shows the the de-reddened luminosity function of Class I+Flat sources in Nessie and the corresponding luminosity completeness that is calculated with the method suggested by \citet{Kryukova2012} is also marked with the red line. As a comparison, we also plot the luminosity function of Class II sources in the Nessie. Assuming an universal luminosity function, we estimate the total number of Class I+Flat sources in Nessie to be $185_{-51}^{+52}$. Assuming the average mass of 0.5 solar mass for each Class I/Flat source, we estimated the total mass of Class I+Flat sources to be $92.7_{-25.7}^{+25.8}\rm \, M_\odot$.}
\end{itemize}

Adopting the lifetime of Class II sources, $\rm 2\,Myr$ \citep{Evans2009}, as the star formation time-scale, we obtain a star formation rate of $\text{SFR} = 389_{-182}^{+364}\rm \, M_\odot \, Myr^{-1}$ for Nessie. The star formation efficiency within the star-forming time-scale is estimated by the total mass of YSOs, $M_\text{YSOs}$, and the gas mass of Nessie, $M_\text{Nessie}$,  $\text{SFE} = M_\text{YSOs} / (M_\text{Nessie}+M_\text{YSOs}) = $ $0.018_{-0.008}^{+0.017}$. The uncertainty is mainly from the uncertainty of transferring $\rm K_s$ magnitudes to stellar masses and the small number of observed Class I and Class II sources. To place these values in context, the SFR of Nessie is comparable to those of the most active nearby star forming regions like Perseus ($150~\text{M}_\odot \, \text{Myr}^{-1}$), Orion A ($715~\text{M}_\odot \, \text{Myr}^{-1}$) and Orion B \citep[$159~\text{M}_\odot \, \text{Myr}^{-1}$; all values from][]{Lada2010}.

It is immediately interesting to compare this direct SFR estimate to other measures commonly linked with the star formation potential of molecular clouds. One such measure is the mass of dense gas in the cloud \citep[e.g.,][]{Kainulainen2009, Lada2010}. Specifically, \cite{Lada2010} found that in the Solar Neighborhood clouds (distance $\lesssim$ 500 pc) SFRs correlate best with the mass above a column density threshold of $A_V\approx 7.3~\text{mag}$. Adopting this threshold results in the dense gas mass of $M_{\text{dg}}=8.7~\cdot~10^3~\text{M}_\odot$ in Nessie. Following the prescription of \cite{Lada2010} for the Solar Neighborhood clouds, the SFR of $\text{SFR}=4.6\cdot 10^{-8}~\text{yr}^{-1} \cdot M_\text{dg}=400~\text{M}_\odot\,\text{Myr}^{-1}$ follows. This is in agreement with the SFR derived from the YSOs; in Nessie the mass of dense gas above $A_V\approx 7.3~\text{mag}$ is a reasonable predictor of the SFR.

Yet another measure commonly connected with SFR is the dense core population of the molecular clouds \citep[e.g.,][]{motte1998, Alves2007, marsh2016}. To analyze this population in Nessie, we can take the advantage of the high spatial resolution of our column density map: we can directly count the cores that might form stars or multiple stellar systems and estimate their mass. The mass enclosed in the dense structures smaller than $\rm \sim 0.1~pc$ is likely to take part in star formation processes. Therefore, the number of structures at the smallest scale of the wavelet-filtered map ($i=2,\rm \sim 0.08 pc$) provides a first-order estimate for the number of stars forming in the cloud in the near future. To account for possible accretion processes during the collapse of a core, we assume the gas at the scales $i=2$ and $i=3$ (size $\rm <~0.16~pc$) can participate in the collapse. This will then give an upper limit for the mass available for star formation. The mass of stars formed by these cores is then estimated by assuming a star formation efficiency of $\rm SFE=30\%$ \citep[e.g.,][]{Alves2007, Rathborne2009, Andre2010}. This results in the stellar mass of $M_{i=2,3}=409~\text{M}_\odot$. Adopting again the star formation time of $t_\text{SF} \approx 2~\text{Myr}$ leads to a star formation rate of $\text{SFR}= M_\ast / t_\text{SF}=205~\text{M}_\odot \, \text{Myr}^{-1}$ for the Nessie cloud. This estimate is within a factor of $2$ of the values derived previously. We can also simply use the number of detected cores to gain a crude estimate of the star formation potential. If we assume that each structure at scale $i=2$ will form at least one star, Nessie will form 523 stars. This is within a factor of two of the actual number of (completeness corrected) Class I and II sources. If we further divide the total mass in the cores in Nessie by 523, the predicted average mass of a star of 0.78 M$_\odot$ follows; this is relatively close to the mean stellar mass of 0.5 M$_\odot$ of the initial mass function \citep[e.g.,][]{Kroupa2002}. Altogether, the above considerations suggest that the dense core population identified from Nessie using the approach of this paper is a reasonable proxy of Nessie's star formation potential.

\section{Conclusions}

We analyzed the column density structure of the (projected) $67~\text{pc}$ long filamentary Nessie cloud using a combined near- and mid-infrared extinction mapping method on data of the VVV survey and $8~\mu$m \emph{Spitzer}/GLIMPSE images. Our results are as follows:

\begin{enumerate}
\item We derived a high-resolution ($ \sim 0.03~\text{pc} $), high dynamic range ($ N(\text{H}_2)= 3 \text{ -- } 100 \cdot 10^{21}~\text{cm}^{-2}$) column density map for Nessie and estimated the distance towards it to be $ d = 3.5~\text{kpc}$ based on near-infrared source-counts. The mass of Nessie is $4.2 \cdot 10^4 ~ \text{M}_\odot$, considering regions above $ N(\text{H}_2) \gtrsim \rm 3 \cdot 10^{21}~cm^{-2}$. This leads to the mean line-mass of about $ 627~\text{M}_\odot \, \text{pc}^{-1}$. 

\item We analyzed the fragmentation of the cloud across a wide range of scales between $\rm 0.1\text{ -- }10~pc$ and detected fragmentation at all scales. We characterize the fragments and find that their masses decrease and densities increase as a function of size-scale. At the smallest scale, the typical masses of the fragments are $\rm 0.4~M_\odot$ and mean densities are $\sim 10^4$. The mean densities of the fragments decrease with their nearest-neighbor separations, following approximately a power-law with an exponent of $-0.96 \pm 0.05$. The previous determination of the $\rm 4~pc$ fragmentation length by \citet{Jackson2010} is in agreement with this relationship, however, our data shows that determining the fragmentation length at any one particular scale does not capture the full, scale-dependent picture of fragmentation in Nessie. 

\item In the context of analytic gravitational fragmentation models, the observed nearest-neighbor separations are within a factor of two of the Jeans' length at all size-scales. However, the slope of the observed mean density -- separation relationship is significantly shallower than the scale-dependency of the Jeans' length. The observed relationship is in agreement with a gravitationally fragmenting near-equilibrium cylinder that is supported by non-thermal motions that exhibits a Larson-like velocity-size scaling, i.e., a power-law with an exponent of $0.5$. This scaling could result, e.g., from turbulent motions in the cloud, because the cloud shows clearly non-thermal velocity dispersions \citep{Jackson2010}. 

\item We estimated the SFR of Nessie to be $\rm 389~M_\odot\,Myr^{-1}$ based on the number of identified YSOs in the cloud. An estimate based on the number of $\rm \sim~0.1~pc$ scale column density "cores" yields $\rm 205~M_\odot\,Myr^{-1}$. We also estimate the SFR based on the total amount of dense gas ($A_\mathrm{V} > 7.3$ mag; \citealt{Lada2012}) in the cloud, resulting in $\rm 400~M_\odot\,Myr^{-1}$. These results suggest that both the number of dense cores and the amount of dense gas above $A_\mathrm{V} > 7.3$ mag are relatively good proxies of the star-forming content of Nessie. We further derive the SFE of $0.018$ for Nessie. These numbers indicate that the star-forming content of Nessie is similar to the Solar neighborhood giant molecular clouds like Orion A.

\item The ATLASGAL clumps identified in Nessie typically harbor $2\text{ -- }3$ small-scale structures ($< 0.16~\text{pc}$). These structures contain about $7\,\%$ of the mass of the parental clump. However, this is a lower limit as the extinction mapping is susceptible for incompleteness arising from mid-infrared bright objects, like foreground stars, and warm/hot gas. 
  
\end{enumerate}

We showed that the filamentary Nessie cloud has scale-dependent fragmentation characteristics. These characteristics are in agreement with some of the predictions of gravitational fragmentation models. However, self-consistent scale-dependent fragmentation models are needed to gain understanding of the structure and evolution of filamentary clouds.

\section*{Acknowledgments}
We thank the referee for constructive comments.
M. M. is supported for this research through a stipend from the International Max Planck Research  School (IMPRS) for Astronomy and Astrophysics at the Universities of Bonn and Cologne. The work of J. K. was supported by the Deutsche Forschungsgemeinschaft priority program 1573 ("Physics of the Interstellar Medium"). This project has received funding from the European Union's Horizon 2020 research and innovation program under grant agreement No 639459 (PROMISE). H. B. acknowledges support from the European Research Council under the Horizon 2020 Framework Program via the ERC Consolidator Grant CSF-648505. M. Z. acknowledges support from the National Natural
Science Foundation of China (grants No. 11503086). This research has made use of the NASA/ IPAC Infrared Science Archive, which is operated by the Jet Propulsion Laboratory, California Institute of Technology, under contract with the National Aeronautics and Space Administration.\\
This work is based on observations made with ESO Telescopes at the La Silla Paranal Observatory under programme ID 179.B-2002. The ATLASGAL project is a collaboration between the Max-Planck-Gesellschaft, the European Southern Observatory (ESO) and the Universidad de Chile. It includes projects E-181.C-0885, E-078.F-9040(A), M-079.C-9501(A), M-081.C-9501(A) plus Chilean data.

\bibliography{library}{}

\bibliographystyle{aa}

\begin{appendix}

\section{Used observations}
\label{observations}
For the calculation of the near-infrared extinction map of the Nessie filament we use observations conducted by the VVV (VISTA Variables in the Via Lactea) survey \citep{Saito2012} in the JHK$_S$ photometric bands. This calibrated and reduced (science ready) data is publicly available in the ESO archive. The exact observations used in this study are listed in Table \ref{Observation_tab}.

\begin{table*}
\caption{List of observations}
\label{Observation_tab}
\centering
\begin{tabular}{c c c c c c}
\hline\hline
Filter & RA & DEC & Exposure time & beamsize  & date \\
       & hh:mm:ss.ss & dd:mm:ss.ss &       s       & $\arcsec$ &     \\
\hline
\multicolumn{6}{l}{near-infrared VVV tile d068}\\
J  & 16:40:50.52 & -47:19:13.08 & 80 & 0.82 & 2010-03-27 \\
J  & 16:40:50.52 & -47:19:13.08 & 80 & 0.93 & 2010-05-09 \\
H  & 16:40:50.52 & -47:19:13.08 & 80 & 0.84 & 2010-03-27 \\
H  & 16:40:50.52 & -47:19:13.08 & 80 & 0.94 & 2010-05-09 \\
K$_S$ & 16:40:50.52 & -47:19:13.08 & 80 & 0.82 & 2010-03-27 \\
K$_S$ & 16:40:50.52 & -47:19:13.08 & 80 & 0.96 & 2010-05-09 \\
K$_S$ & 16:40:50.52 & -47:19:13.08 & 16 & 0.93 & 2010-03-06 \\
K$_S$ & 16:40:50.52 & -47:19:13.08 & 16 & 0.93 & 2010-06-26 \\
K$_s$ & 16:40:50.52 & -47:19:13.08 & 16 & 0.74 & 2011-05-14 \\
K$_S$ & 16:40:50.52 & -47:19:13.08 & 16 & 0.82 & 2011-05-15 \\
K$_S$ & 16:40:50.52 & -47:19:13.08 & 16 & 0.74 & 2011-05-16 \\
K$_S$ & 16:40:50.52 & -47:19:13.08 & 16 & 0.71 & 2011-05-16 \\
K$_S$ & 16:40:50.52 & -47:19:13.08 & 16 & 0.90 & 2011-05-18 \\
K$_S$ & 16:40:50.52 & -47:19:13.08 & 16 & 1.09 & 2011-08-31 \\
K$_S$ & 16:40:50.52 & -47:19:13.08 & 16 & 0.93 & 2011-09-01 \\
K$_S$ & 16:40:50.52 & -47:19:13.08 & 16 & 0.88 & 2011-09-05 \\
K$_S$ & 16:40:50.52 & -47:19:13.08 & 16 & 0.93 & 2011-09-17 \\
K$_S$ & 16:40:50.52 & -47:19:13.08 & 16 & 0.81 & 2011-09-21 \\
\\
\multicolumn{6}{l}{near-infrared VVV tile d069}\\
J  & 16:46:25.56 & -46:13:07.32 & 80 & 0.79 & 2010-03-27 \\
J  & 16:46:25.56 & -46:13:07.32 & 80 & 0.96 & 2010-05-09 \\
H  & 16:46:25.56 & -46:13:07.32 & 80 & 0.81 & 2010-03-27 \\
H  & 16:46:25.56 & -46:13:07.32 & 80 & 0.89 & 2010-05-09 \\
K$_S$ & 16:46:25.56 & -46:13:07.32 & 80 & 0.83 & 2010-03-27 \\
K$_S$ & 16:46:25.56 & -46:13:07.32 & 80 & 0.87 & 2010-05-09 \\
K$_S$ & 16:46:25.56 & -46:13:07.32 & 16 & 1.02 & 2010-03-06 \\
K$_S$ & 16:46:25.56 & -46:13:07.32 & 16 & 0.79 & 2010-08-18 \\
K$_S$ & 16:46:25.56 & -46:13:07.32 & 16 & 0.97 & 2011-06-15 \\
K$_S$ & 16:46:25.56 & -46:13:07.32 & 16 & 0.71 & 2011-05-14 \\
K$_S$ & 16:46:25.56 & -46:13:07.32 & 16 & 0.83 & 2011-05-15 \\
K$_S$ & 16:46:25.56 & -46:13:07.32 & 16 & 0.72 & 2011-05-16 \\
K$_S$ & 16:46:25.56 & -46:13:07.32 & 16 & 1.04 & 2011-08-09 \\
K$_S$ & 16:46:25.56 & -46:13:07.32 & 16 & 1.01 & 2011-09-06 \\

\hline
\end{tabular}
\end{table*}

\section{Photometry of different observations}
\label{PSF}

For the photometry of the near-infrared data we use a set of different observations (see Appendix \ref{observations}), which show different spatial resolutions due to different conditions. Therefore, the point-spread-function (PSF) for point sources will be different in the single observations, and also effect the stacked data. This might be especially relevant in the K$_s$ filter where we use a larger set of observations. To test the significance of this quality difference we compare the results of photometry in the K$_S$ filter performed on tile 068 of the stacked data, one 80 s exposure (from 2010-05-09), and the lowest resolution 16 s exposure (from 2011-08-31). For all data-sets we used the same parameters as described before and also calibrated the found magnitudes with the 2MASS data \citep{Skrutskie2006, Cutri2003}. We then identified stars seen in the stacked and 16 s data, and stacked and 80 s data, and plotted the derived magnitudes against each other.

\begin{figure}
\centering
\includegraphics[width=0.5\textwidth]{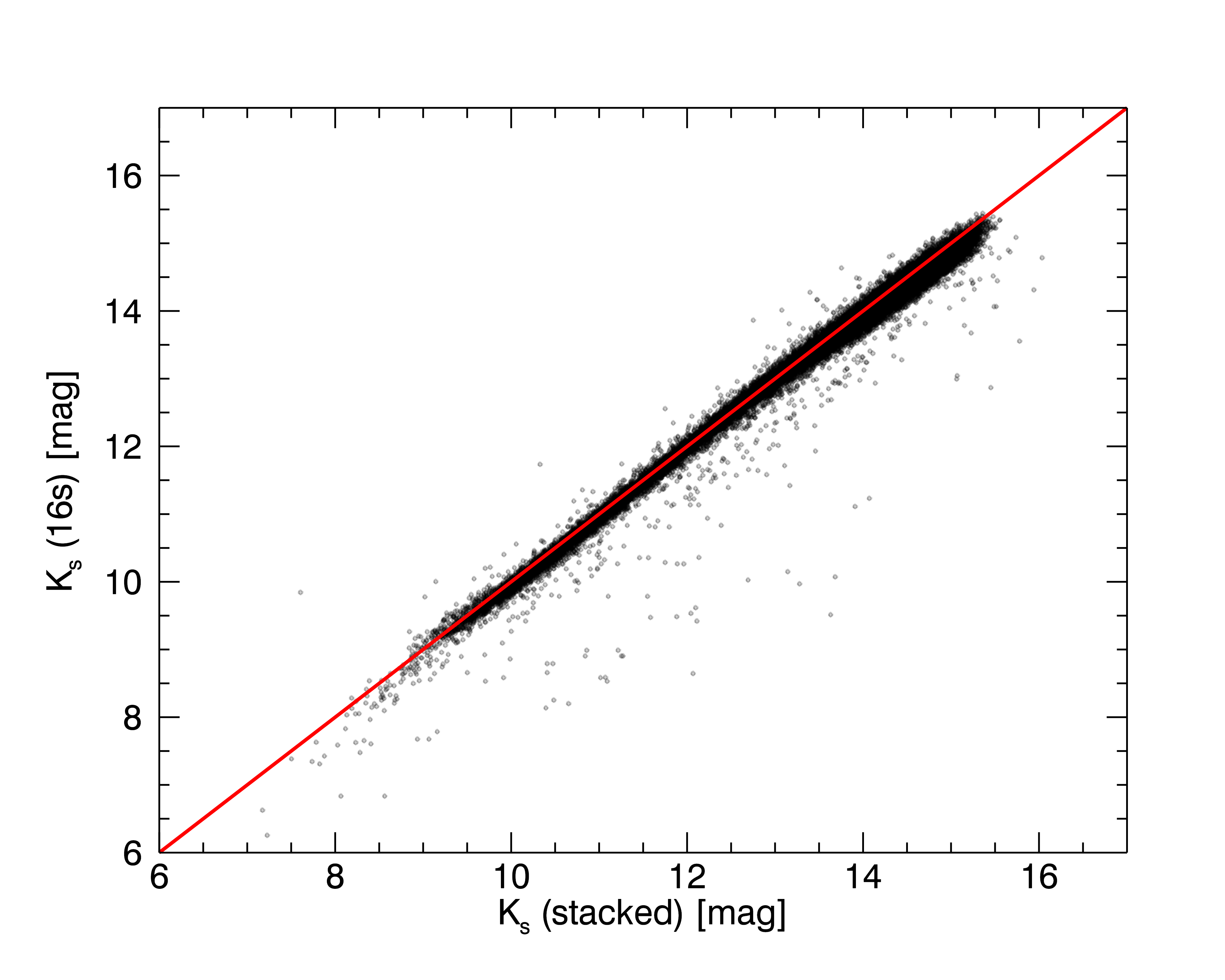}
\caption{Comparison of K$_S$ magnitudes between the stacked and 16 s data. The red line indicates the one-to-one correlation. The shown stars have an photometric uncertainty lower than 0.05 mag.}
\label{PSF_16s}
\end{figure}

\begin{figure}
\centering
\includegraphics[width=0.5\textwidth]{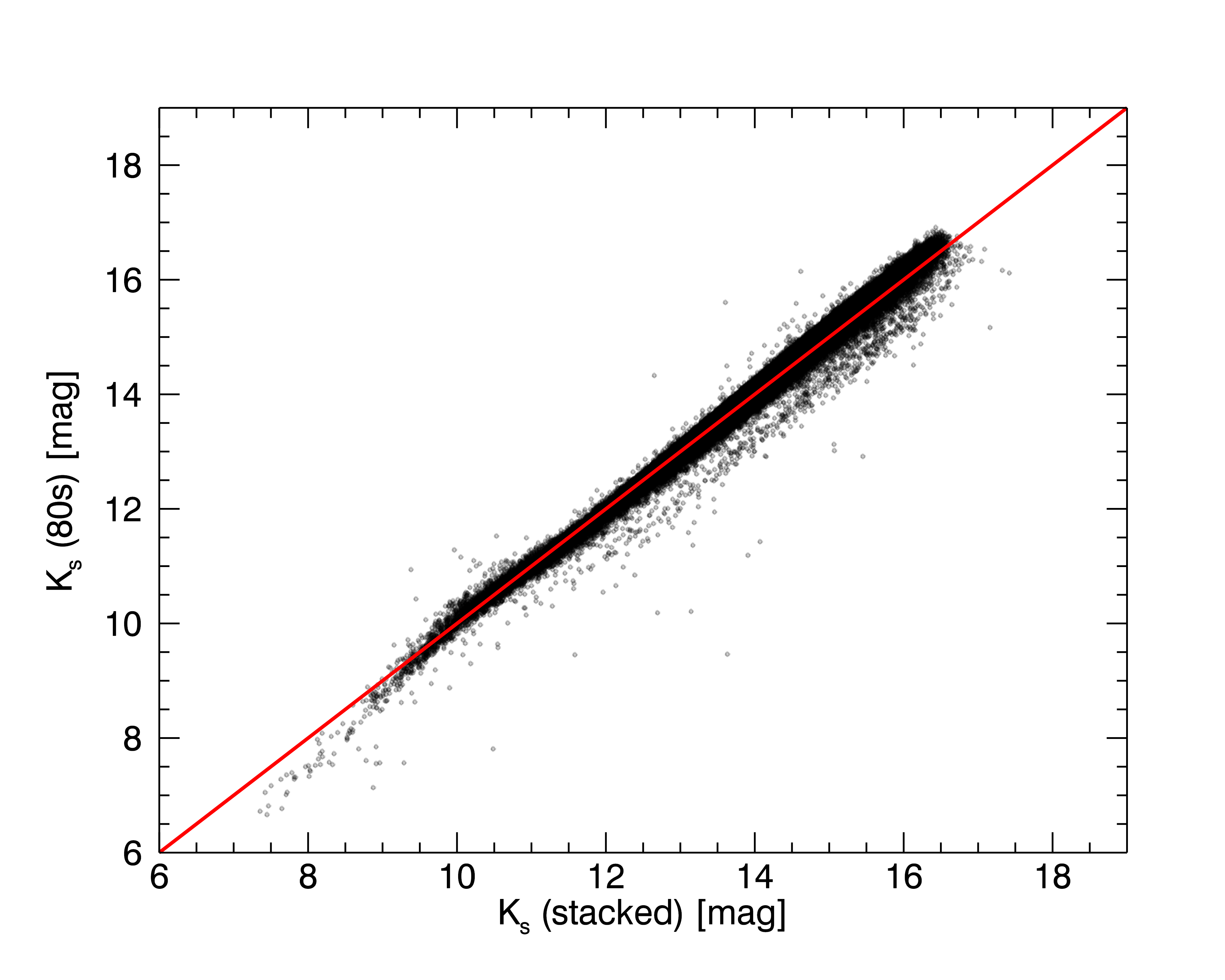}
\caption{Comparison of K$_S$ magnitudes between the stacked and 80 s data. The red line indicates the one-to-one correlation. The shown stars have an photometric uncertainty lower than 0.05 mag.}
\label{PSF_80s}
\end{figure}

We find a good correlation for the three data-sets. However, there is a larger scatter for the 16 s data because of the lower sensitivity of the data. In general, we find a significantly increased number of stars for the longer exposure or stacked data. Especially, more faint stars are detected because of the higher sensitivity of the data. As the number of sources is important for our applied method of near-infrared extinction measurement we except small uncertainties introduced by the PSF fitting on the stacked data as they are not significant, which we could show here.

\section{Reference color correction}
\label{color-correction}
\begin{figure}
\centering
\includegraphics[width=0.5\textwidth]{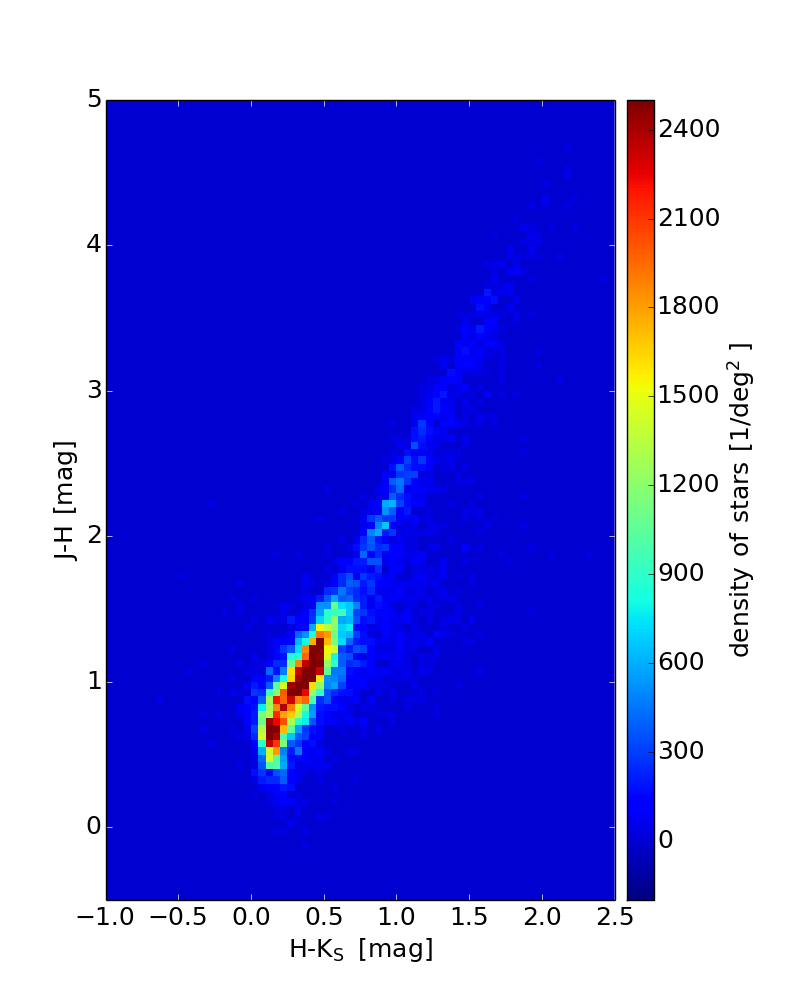}
\caption{JHK$_s$ color-color histogram of the reference field before correction.}
\label{col-col-org}
\end{figure}

\begin{figure}
\centering
\includegraphics[width=0.5\textwidth]{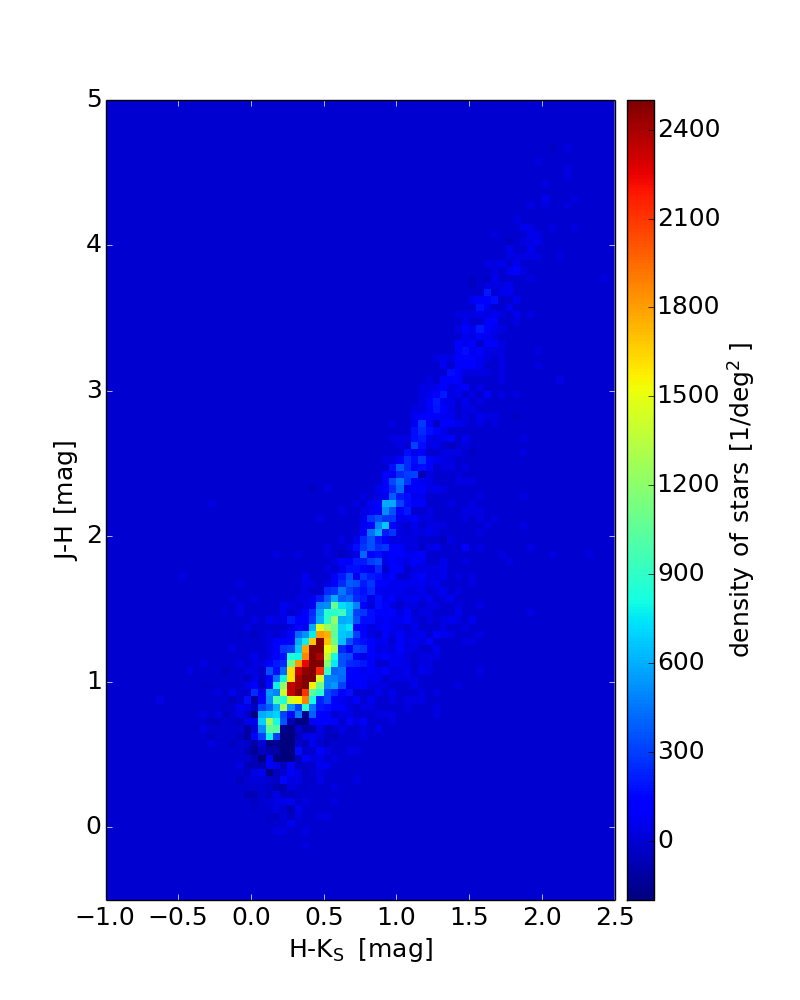}
\caption{JHK$_s$ color-color histogram of the reference field after correction.}
\label{col-col-ref}
\end{figure}

For the estimation of the dust extinction of a molecular cloud we need to calculate the average color of the stars behind the cloud. For good measurements we need to address two problems. First the cloud itself causes a strong shift of the star colors. This is exactly the effect we want to measure, but a direct measure of the color on the farther side of the cloud is impossible. Therefore, we assume the colors of stars in a nearby cloud-free region are the same as behind the cloud. Second, diffuse dust in the Milky Way causes a steady dust reddening with distance from the observer. Therefore, stars located in between the cloud and the observer will confuse the measurement of the background color and need to be removed. We address this problem by statistical subtraction of foreground stars in the JHK$_S$ color-color-space. We first bin the stars in the J-H and H-K$_S$ colors and scale the numbers with the size of the reference field, which leads to a 2-dimensional histogram shown in Fig. \ref{col-col-org}. Then, we do the same for stars located towards the highest extinction regions of the cloud. These stars are either in front of the cloud and show almost now color excess or they are behind the cloud, in which case the show a strong color excess and can be ignored. Again, we scale the number of stars per bin with the area in which they where observed. We subtract the number of stars per bin of foreground histogram from the number of stars in the corresponding bin of the reference field histogram. The resulting histogram is shown in Fig. \ref{col-col-ref} and represents the distribution of star colors behind the cloud. Some bins show negative number of stars, but neighbouring bins show still 'unreddend' stars, so they cancel in deriving the average J-H, and H-K$_S$ colors.  

\section{mid-infrared near-infrared correlation}
\label{MIR-NIR-correlation}
\begin{figure}
\centering
\includegraphics[width=0.5\textwidth]{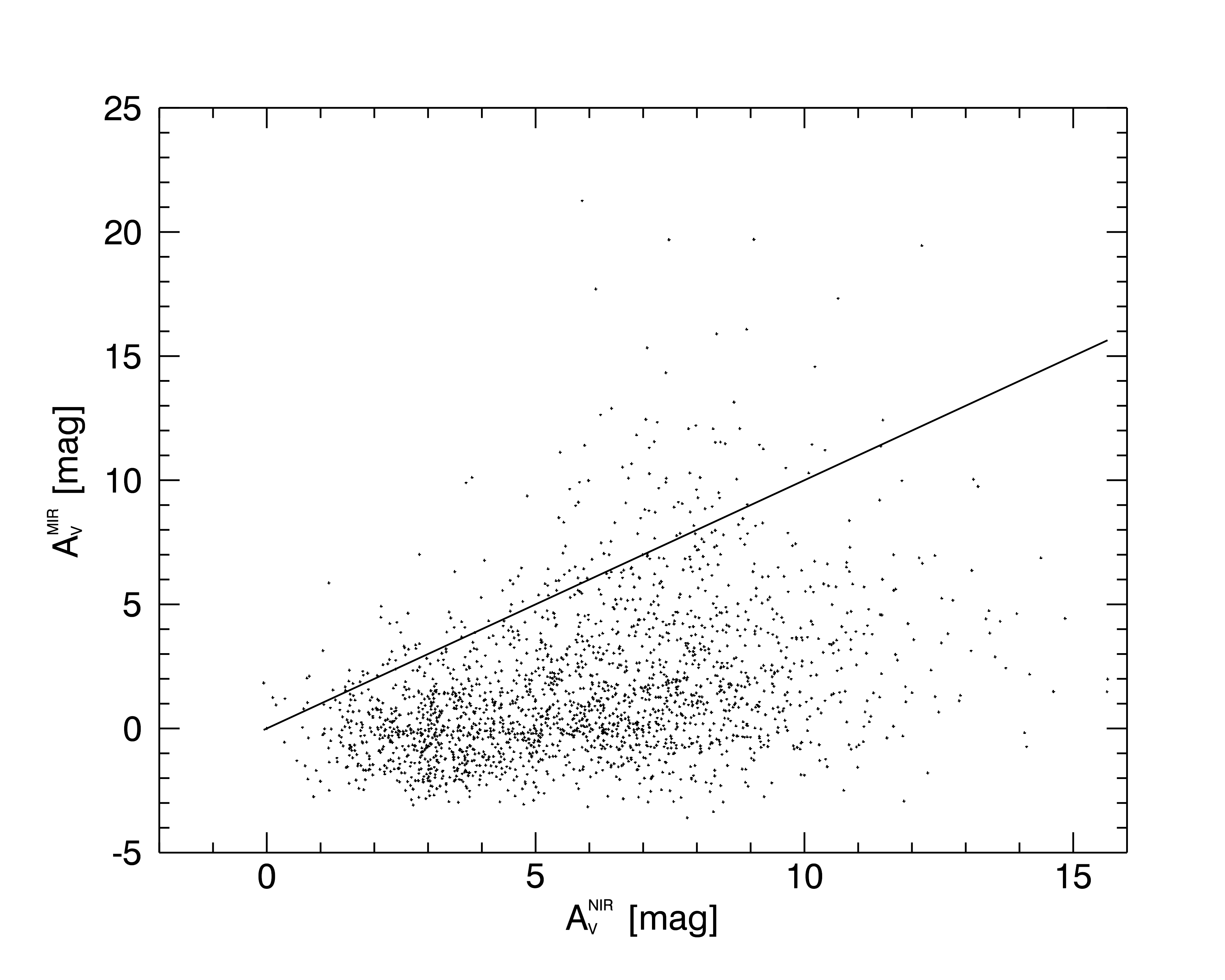}
\caption{Pixel-to-pixel comparison of the mid- and near-infrared extinction values restricted to the filament area (polygon in Fig. \ref{Nessie complete}). The black line indicates the one-to-one correlation. }
\label{correlation_plot}
\end{figure}

For the combination of the near- and mid-infrared extinction maps we convolve the mid-infrared data ($\rm FWHM = 2.4"$ ) to the significantly lower resolution of the near-infrared data ($\rm FWHM = 48"$ ). Then, we perform a pixel-to-pixel comparison between the two maps to investigate their correlation. Fig. \ref{correlation_plot} shows only a poor correlation of the data and a large scatter. For $A_\mathrm{V}^{NIR} \rm \lesssim 10\,mag$ the mid-infrared extinction at most positions is underestimated by a factor of $\sim 5$, but at some positions the data is correlated. This can be explained by the spatial filtering of the mid-infrared mapping, which is not able to trace the diffuse cloud component. Therefore, the correlation arises only from the very inner parts of the filament. Additionally, at extinctions higher than $A_\mathrm{V}^{NIR} \rm \lesssim 5\text{ -- }10\,mag$ the near-infrared data begins to underestimate the extinction, because of a lower number of background stars in the line-of-sight. A similar behavior of the correlation can be seen in the study of \cite{Kainulainen2013b}.

\section{ATLASGAL clumps}
Here we show cut-outs from the combined near- and mid-infrared extinction map of the $16$ ATLASGAL GCSC sources contained in Nessie. In section \ref{ATLASGAL sus} we describe how these parsec-scale structures identified from ATLASGAL (white contours) break down into possibly star-forming substructures. Therefore, we show the positions of identified scale $i=2$ structures with black crosses.

\begin{figure*}[h]
\begin{minipage}{0.28\textwidth}
\hfill\vspace{0.28\textwidth}
\end{minipage}
\begin{minipage}{0.46\textwidth}
\includegraphics[width=\textwidth, clip=true, trim= 0cm 0cm 0cm 0cm]{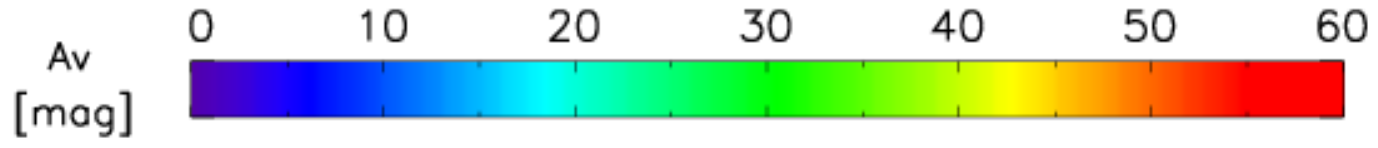}
\end{minipage}
\begin{minipage}{0.23\textwidth}
\hfill\vspace{0.23\textwidth}
\end{minipage}

\begin{minipage}{0.28\textwidth}
\includegraphics[width=\textwidth, clip=true, trim= 5.0cm 2.3cm 5.8cm 3cm]{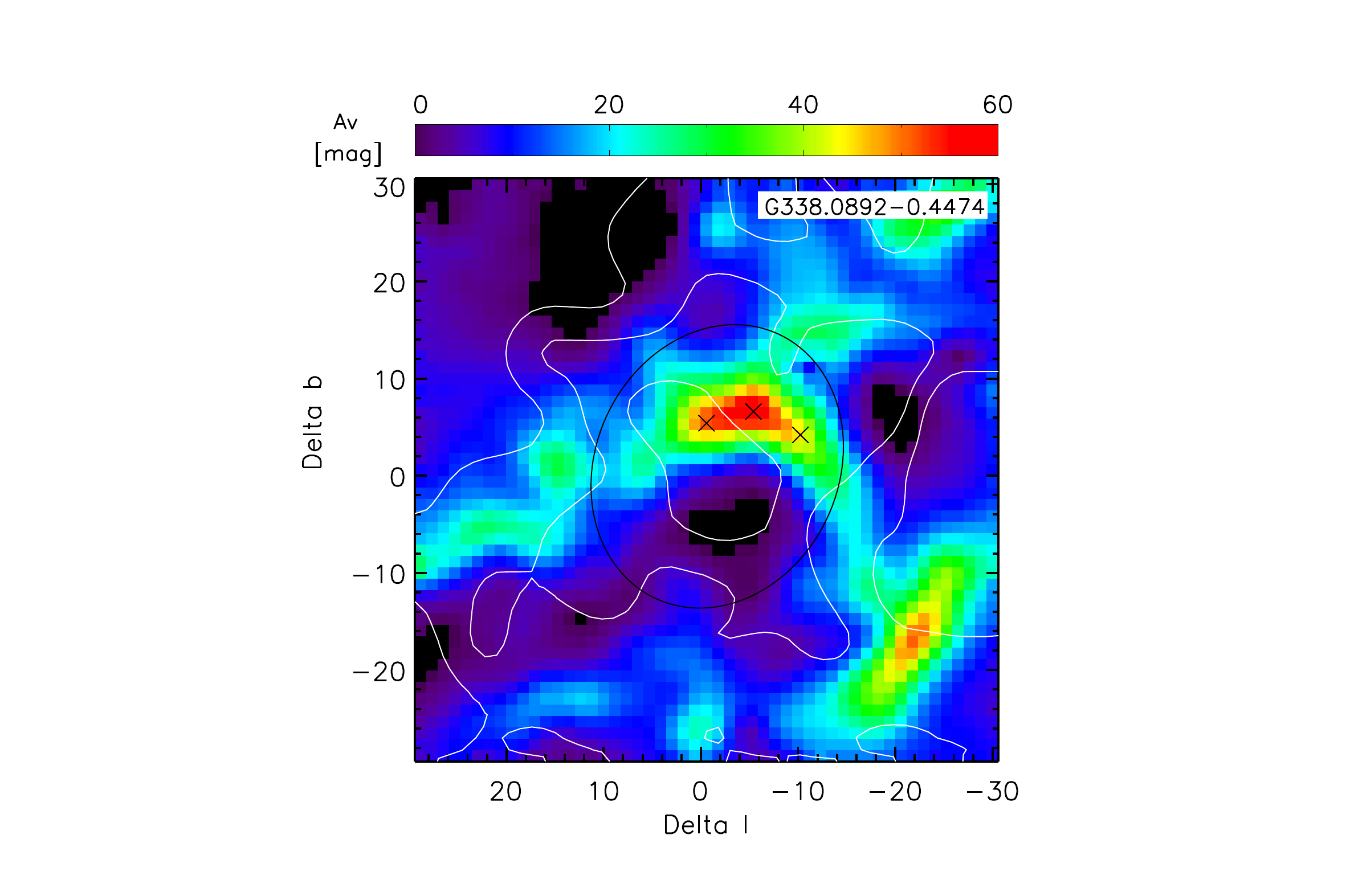}
\end{minipage}
\begin{minipage}{0.23\textwidth}
\includegraphics[width=\textwidth, clip=true, trim= 7.4cm 2.3cm 5.8cm 3cm]{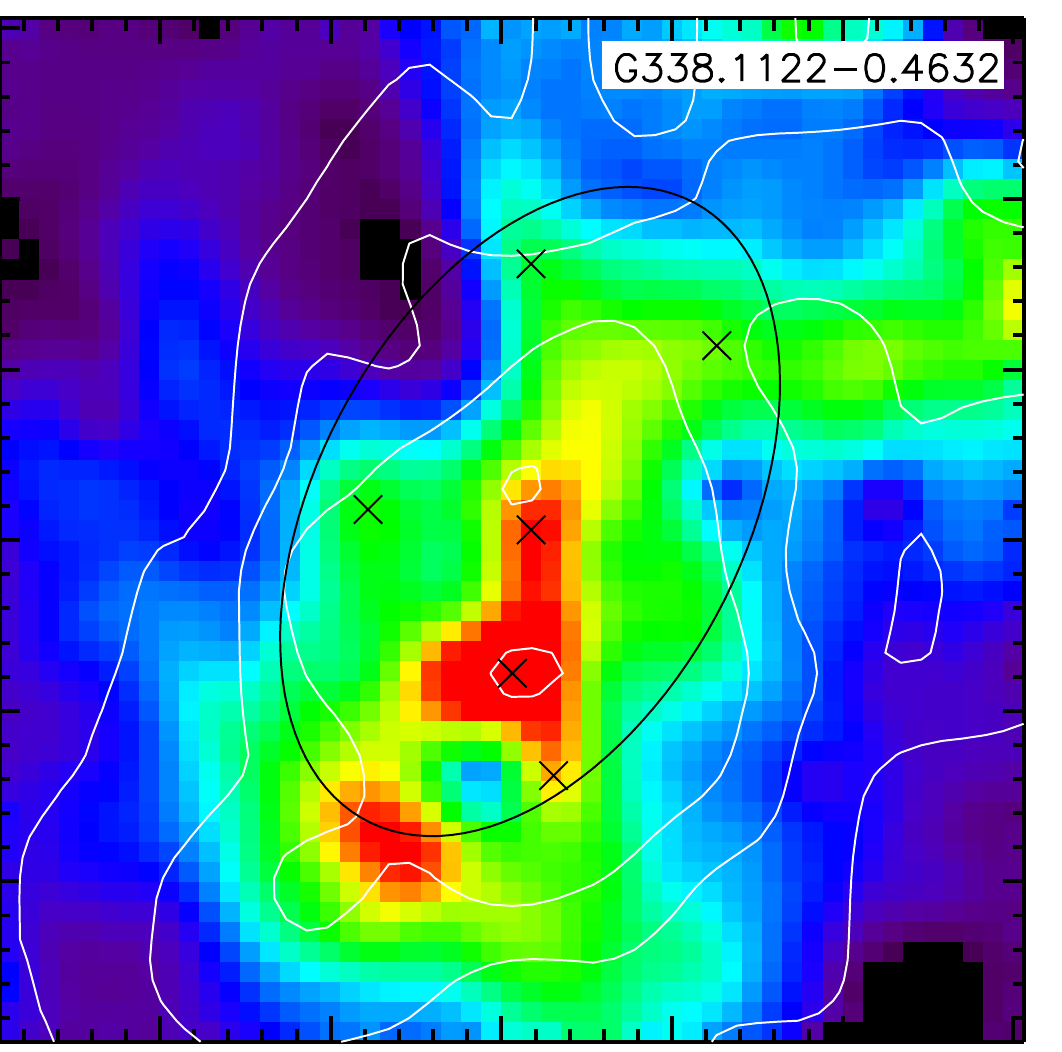}
\end{minipage}
\begin{minipage}{0.23\textwidth}
\includegraphics[width=\textwidth, clip=true, trim= 7.4cm 2.3cm 5.8cm 3cm]{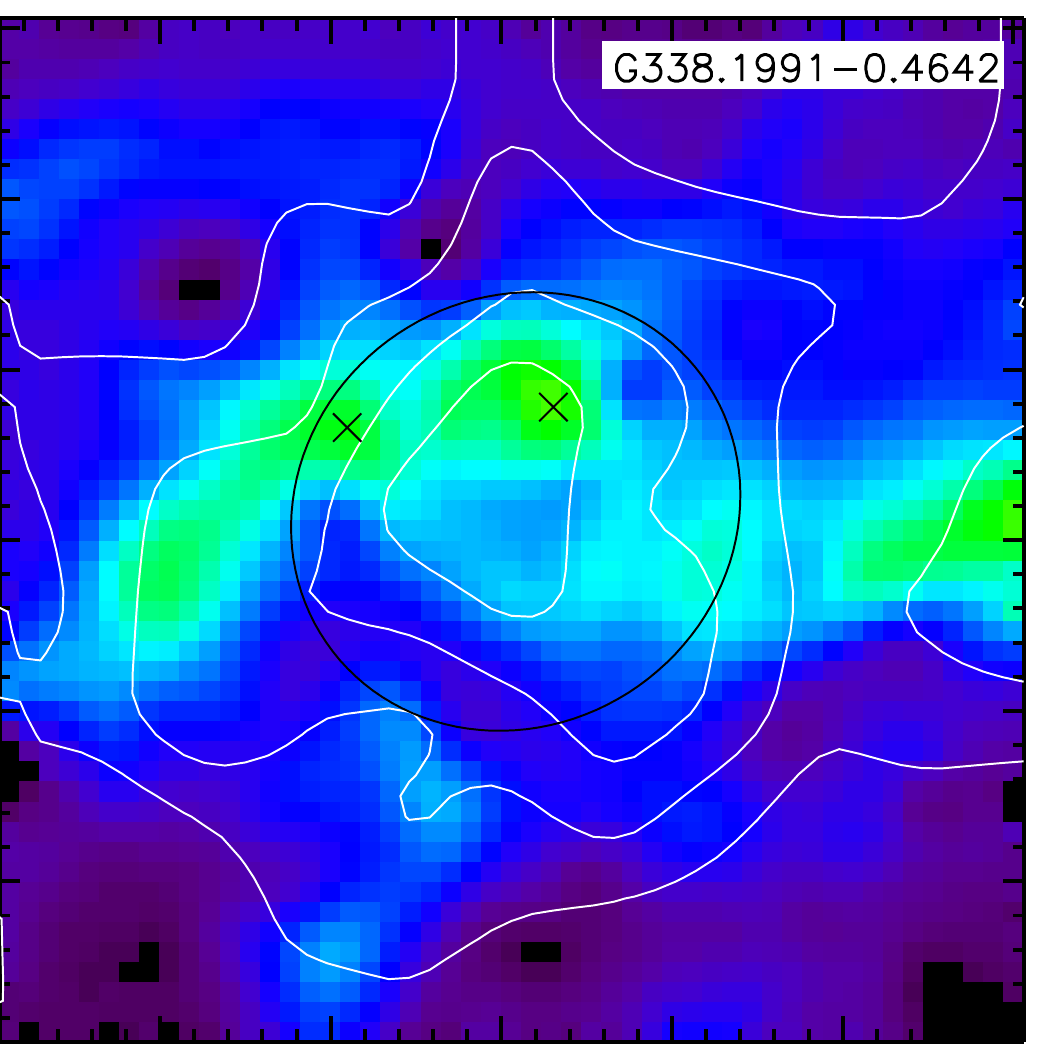}
\end{minipage}
\begin{minipage}{0.23\textwidth}
\includegraphics[width=\textwidth, clip=true, trim= 7.4cm 2.3cm 5.8cm 3cm]{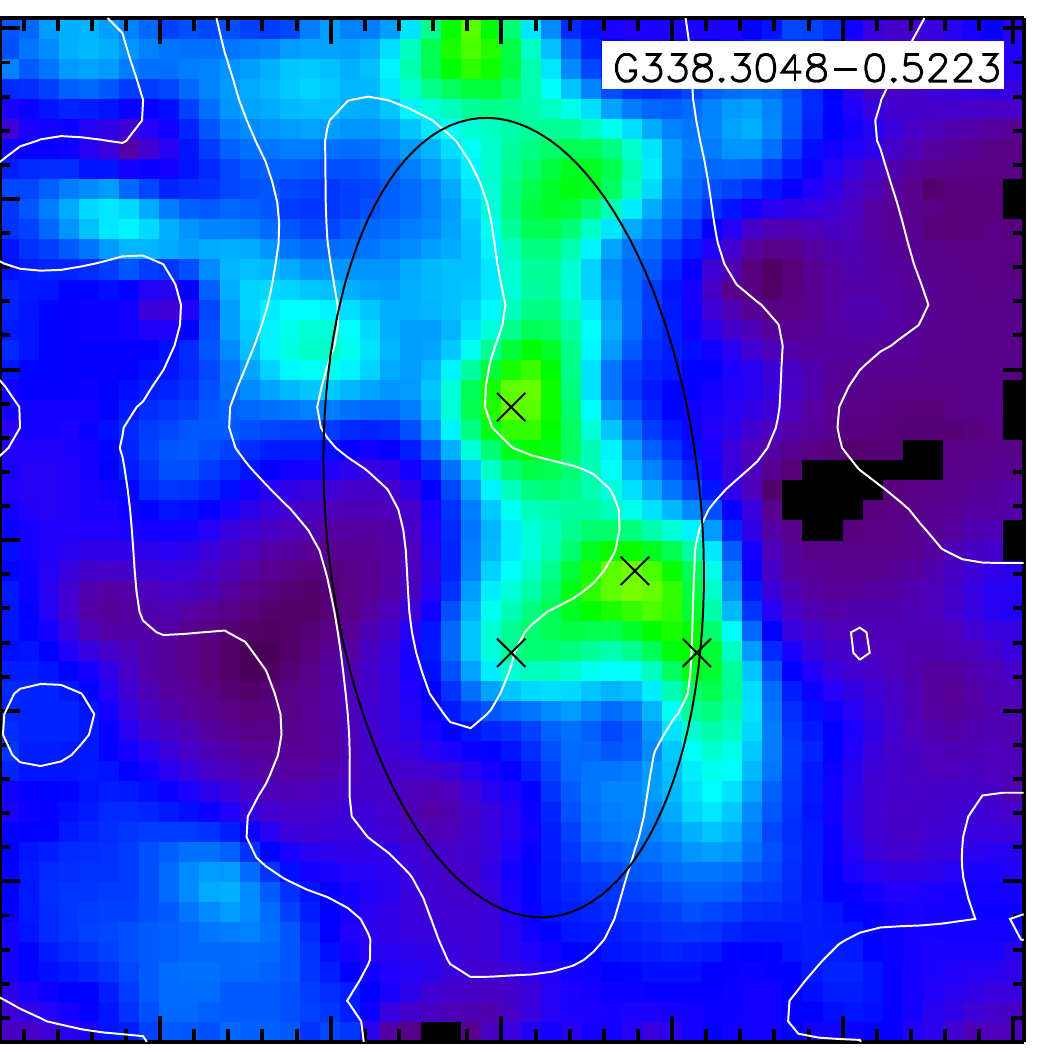}
\end{minipage}

\begin{minipage}{0.28\textwidth}
\includegraphics[width=\textwidth, clip=true, trim= 5.0cm 2.3cm 5.8cm 3cm]{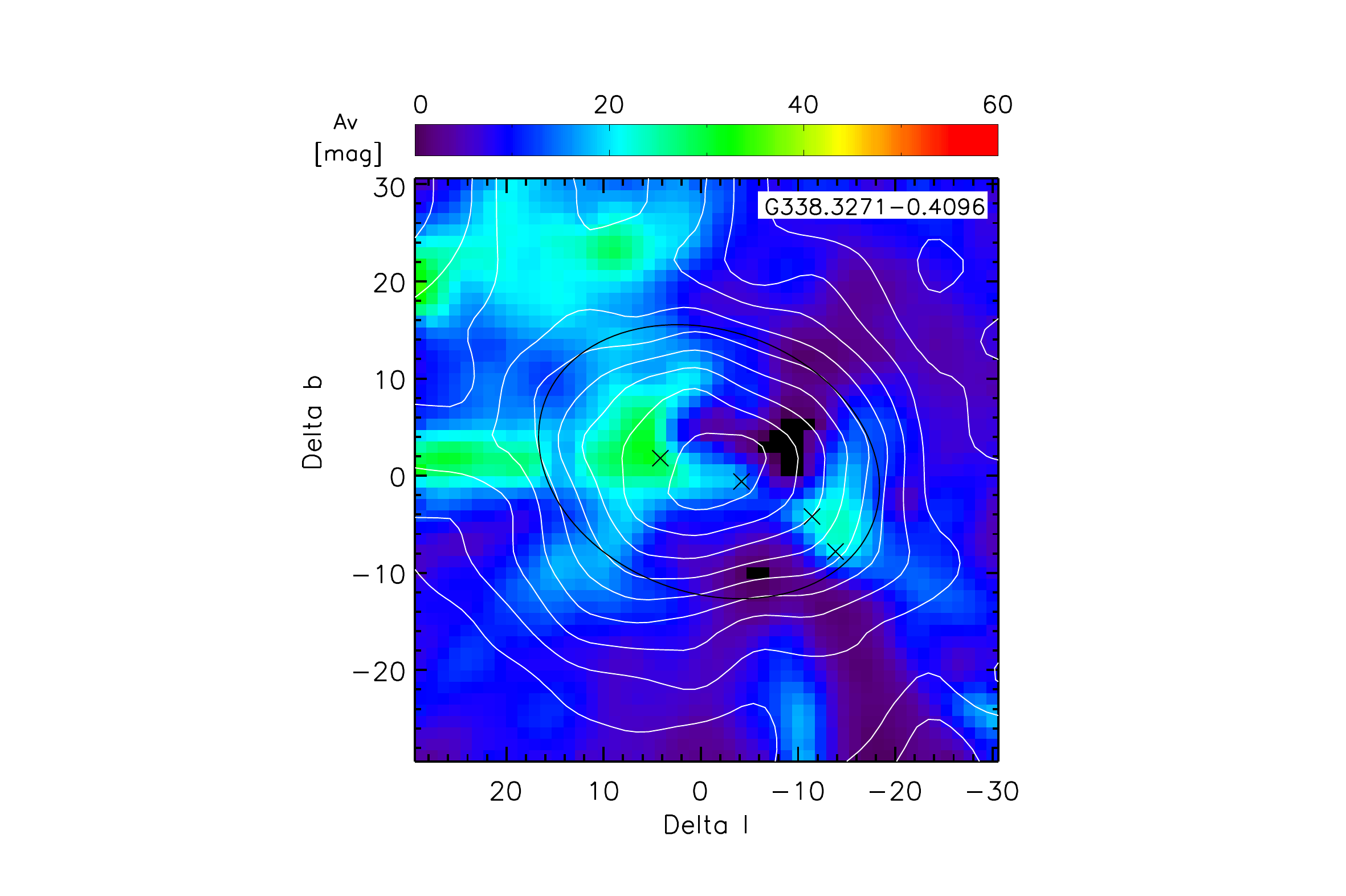}
\end{minipage}
\begin{minipage}{0.23\textwidth}
\includegraphics[width=\textwidth, clip=true, trim= 7.4cm 2.3cm 5.8cm 3cm]{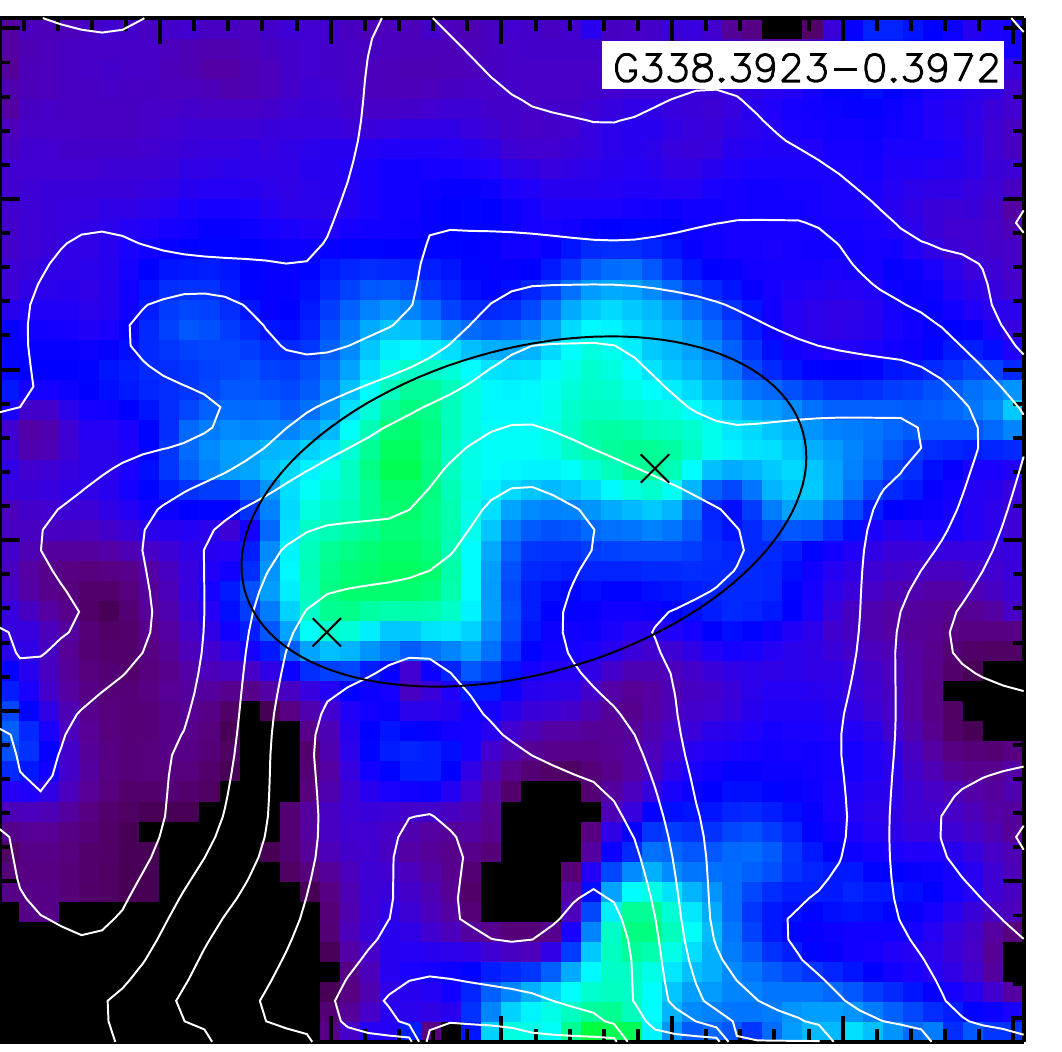}
\end{minipage}
\begin{minipage}{0.23\textwidth}
\includegraphics[width=\textwidth, clip=true, trim= 7.4cm 2.3cm 5.8cm 3cm]{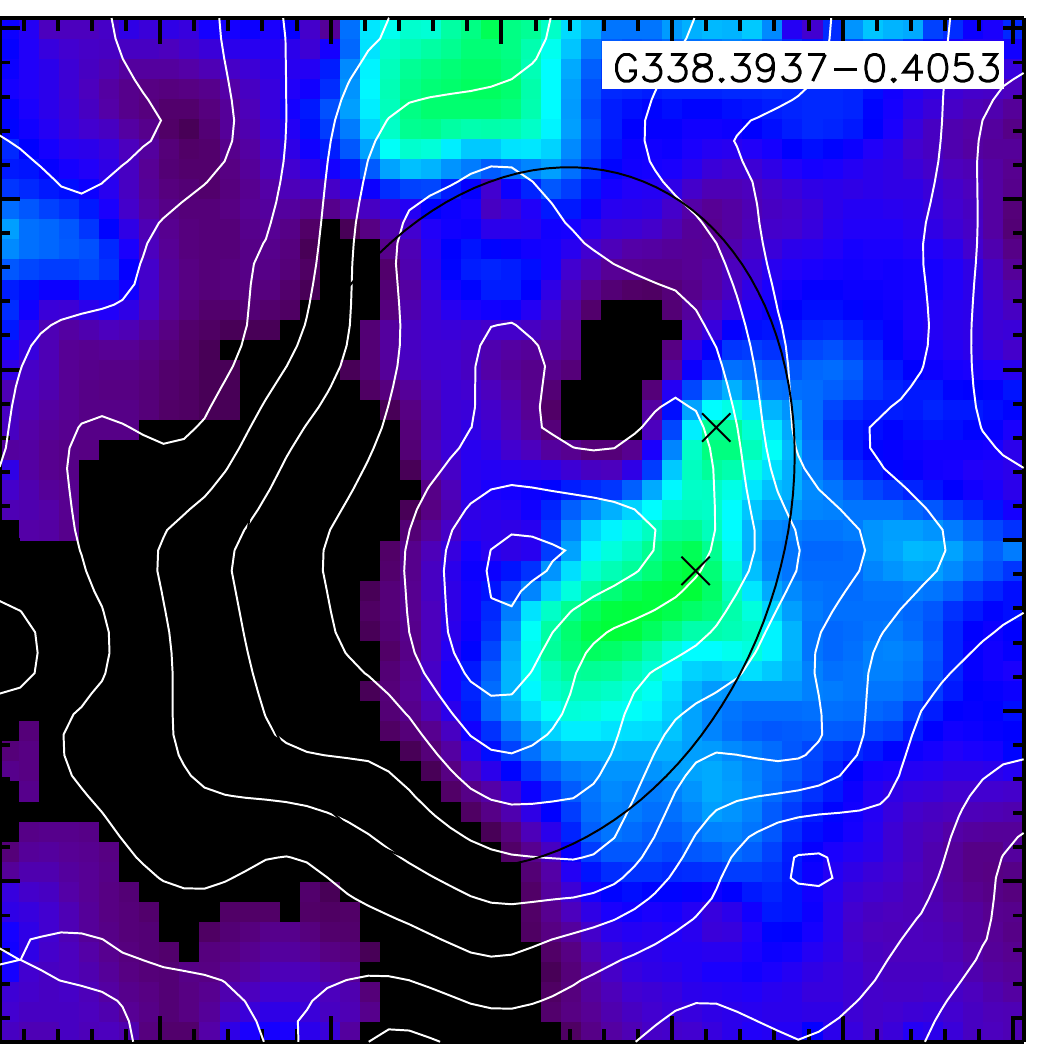}
\end{minipage}
\begin{minipage}{0.23\textwidth}
\includegraphics[width=\textwidth, clip=true, trim= 7.4cm 2.3cm 5.8cm 3cm]{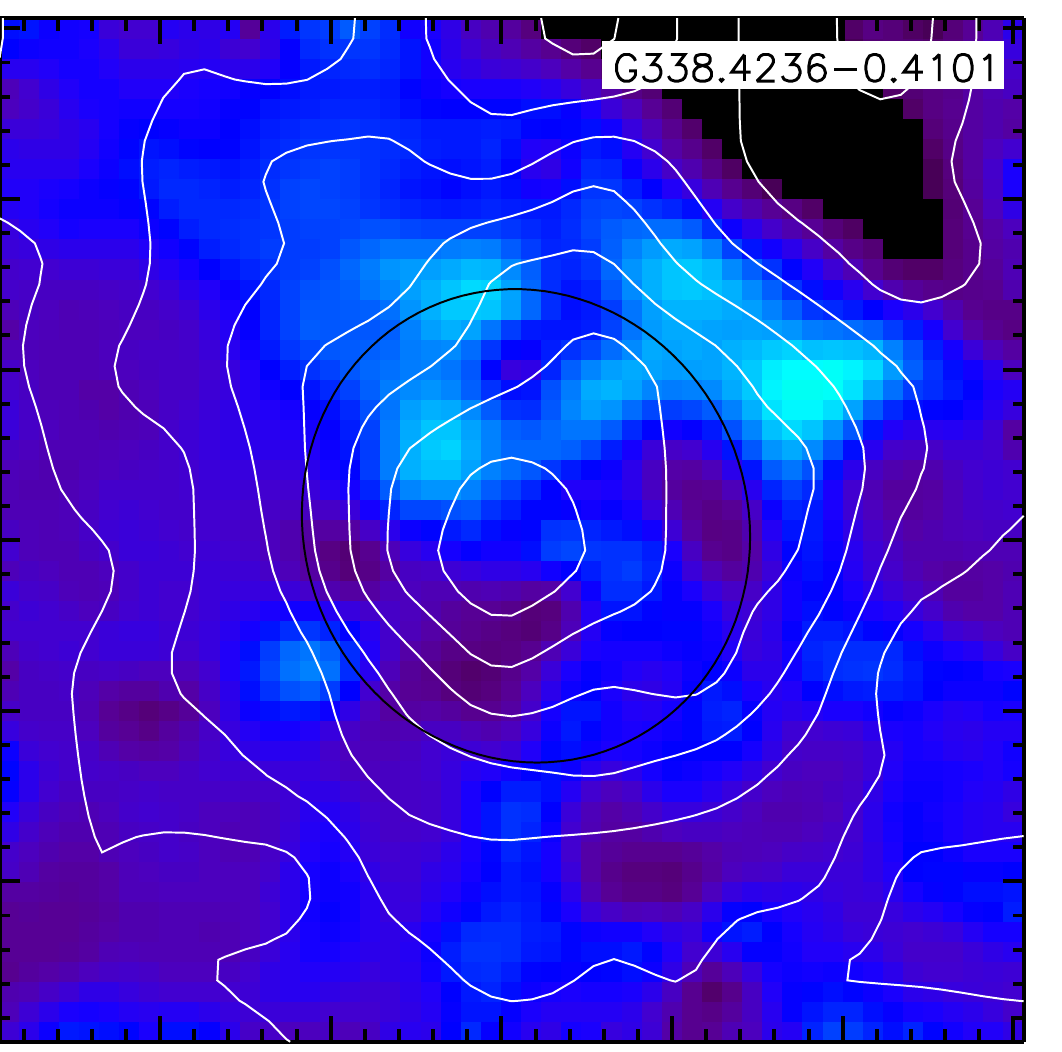}
\end{minipage}

\begin{minipage}{0.28\textwidth}
\includegraphics[width=\textwidth, clip=true, trim= 5.0cm 2.3cm 5.8cm 3cm]{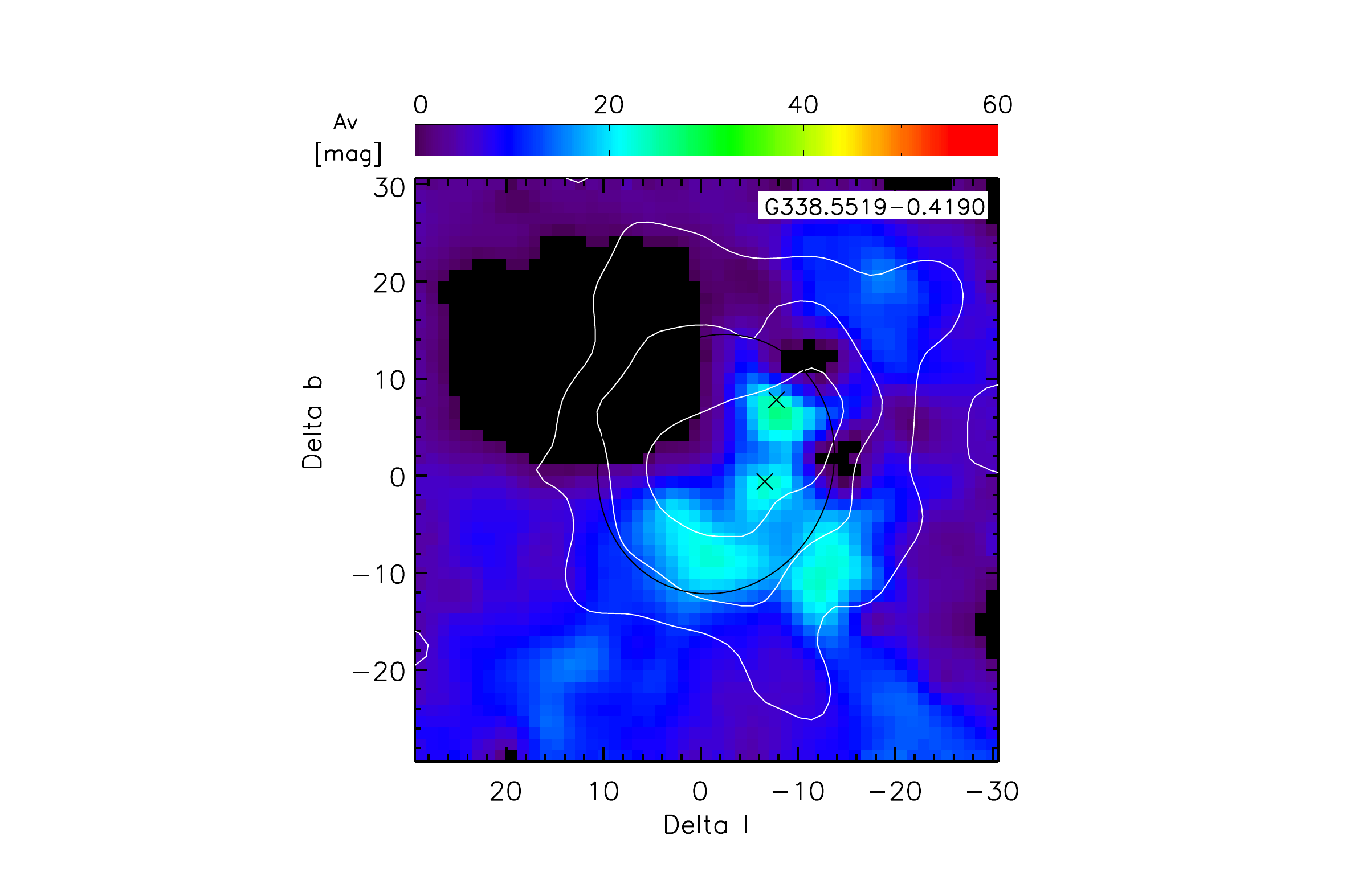}
\end{minipage}
\begin{minipage}{0.23\textwidth}
\includegraphics[width=\textwidth, clip=true, trim= 7.4cm 2.3cm 5.8cm 3cm]{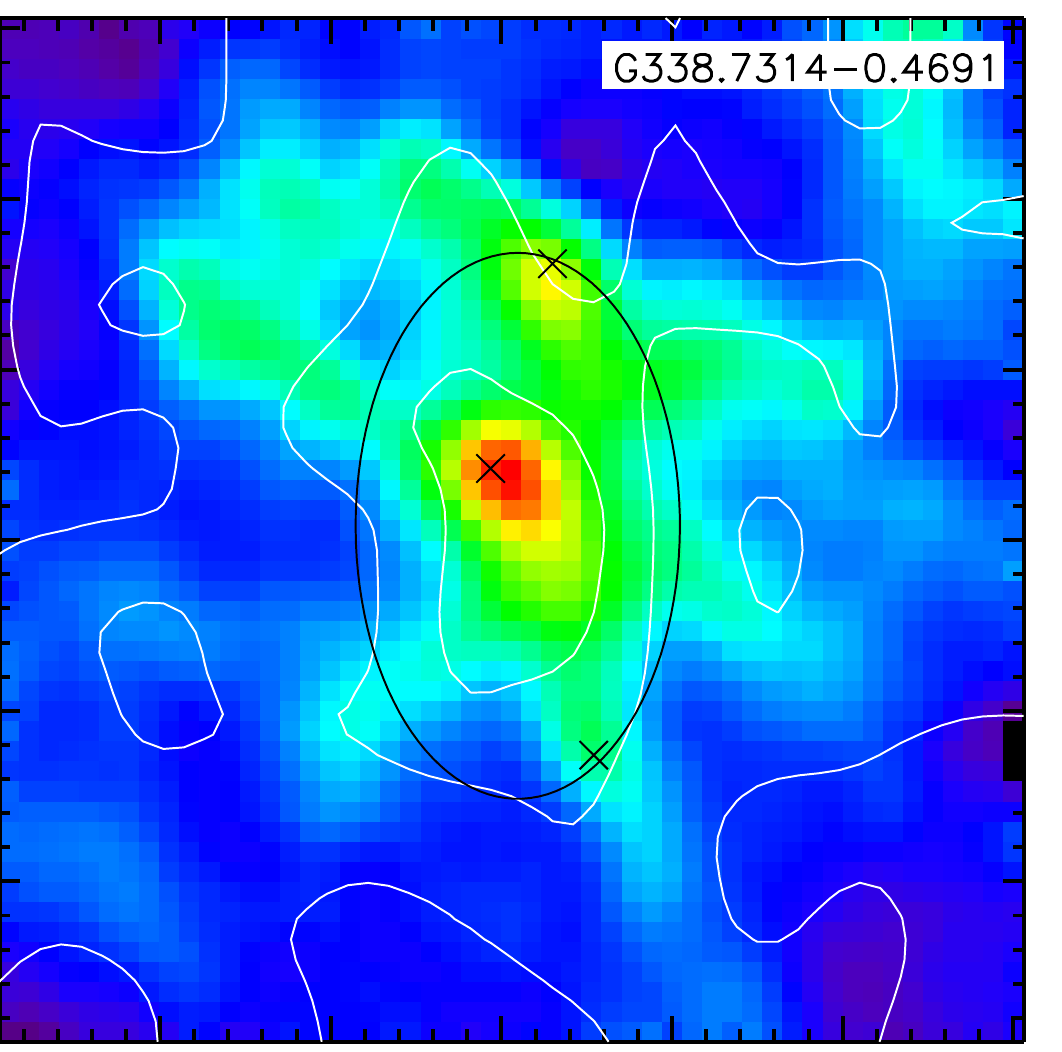}
\end{minipage}
\begin{minipage}{0.23\textwidth}
\includegraphics[width=\textwidth, clip=true, trim= 7.4cm 2.3cm 5.8cm 3cm]{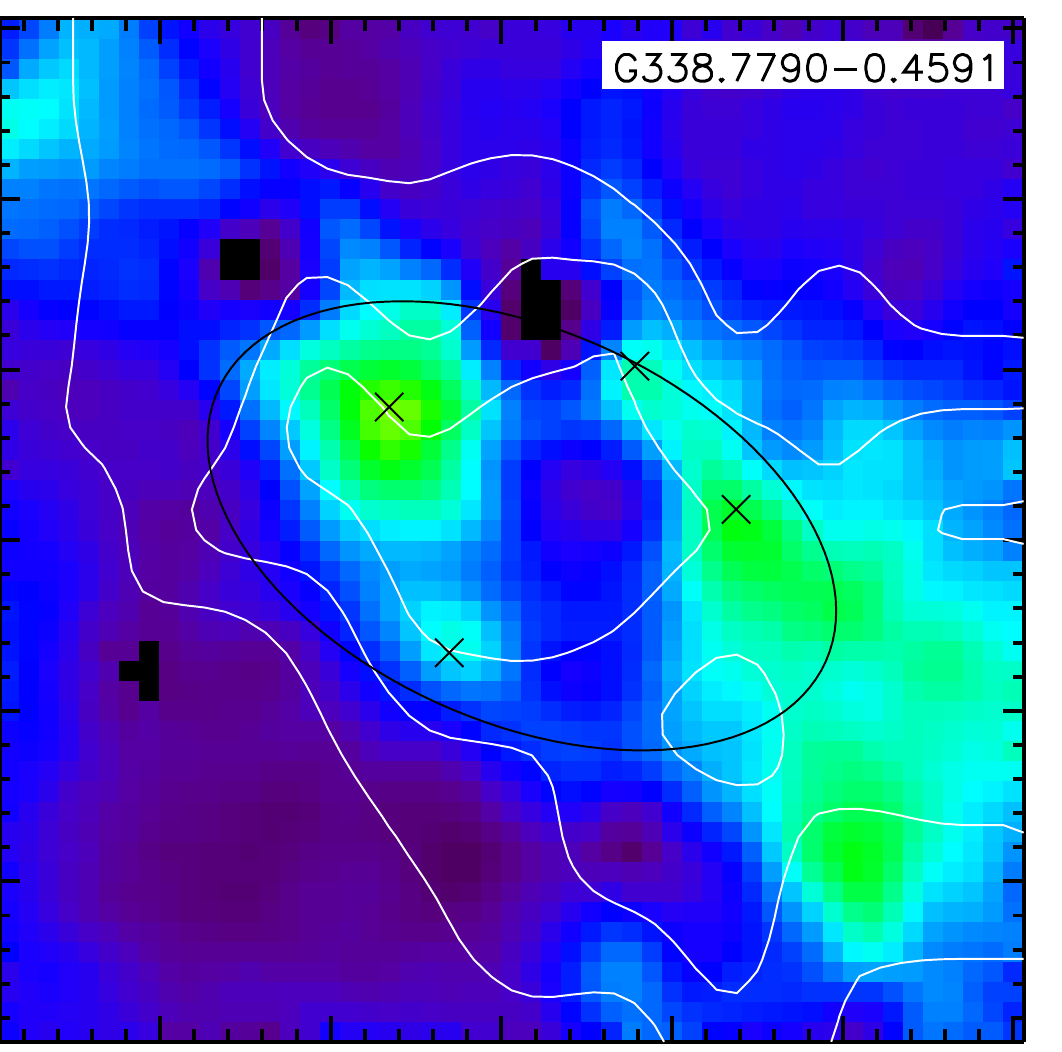}
\end{minipage}
\begin{minipage}{0.23\textwidth}
\includegraphics[width=\textwidth, clip=true, trim= 7.4cm 2.3cm 5.8cm 3cm]{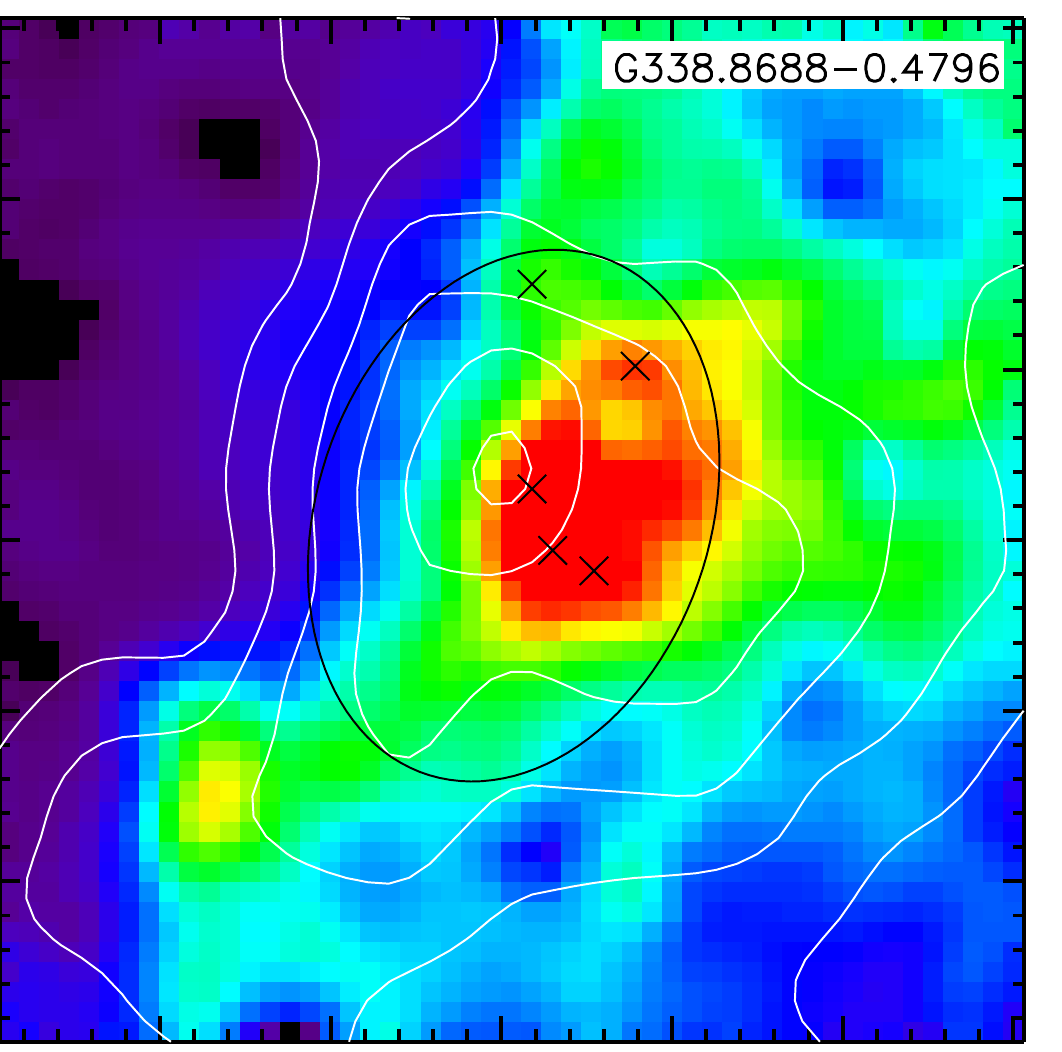}
\end{minipage}

\begin{minipage}{0.28\textwidth}
\includegraphics[width=\textwidth, clip=true, trim= 5.0cm 1.0cm 5.8cm 3cm]{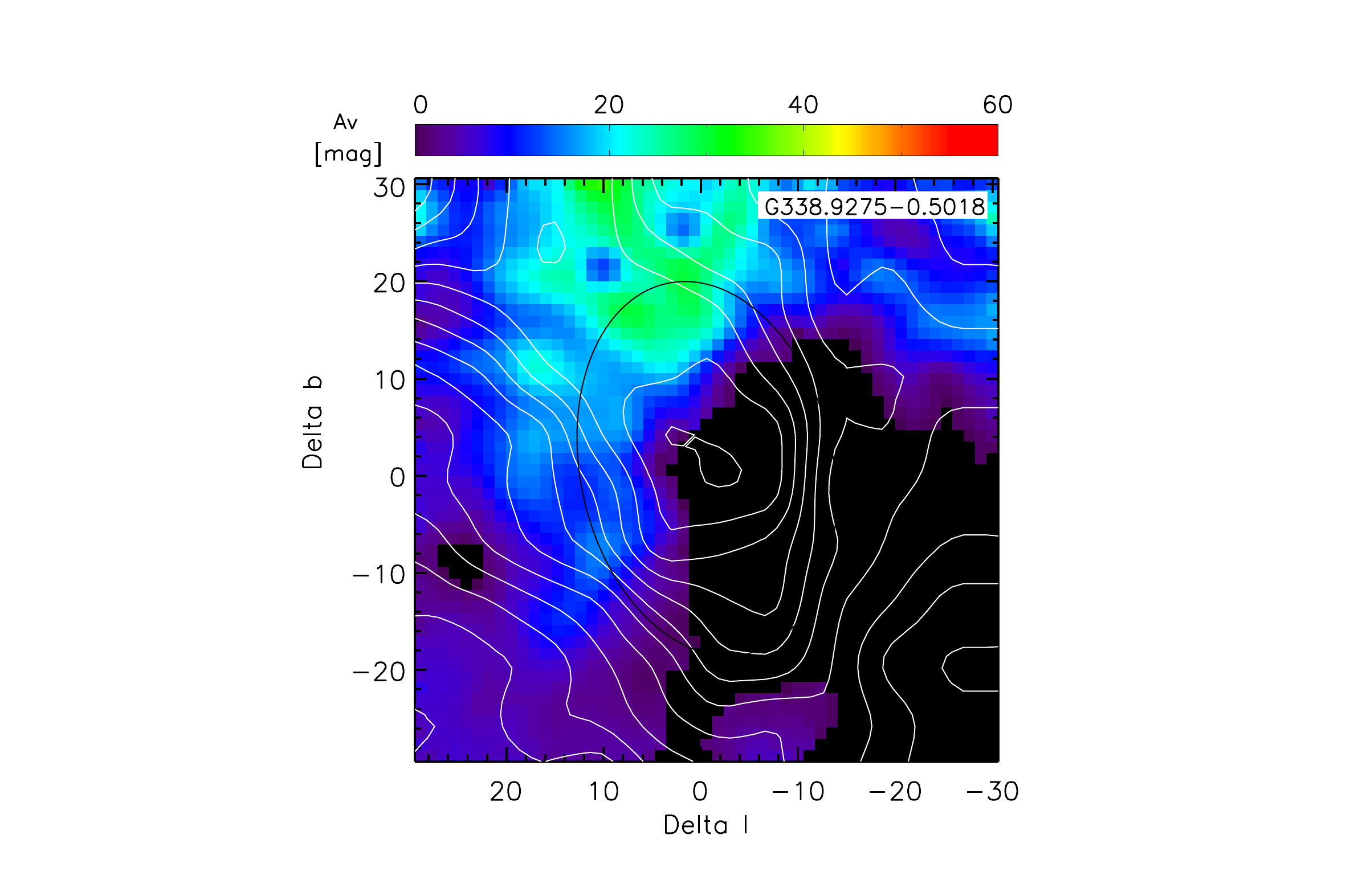}
\end{minipage}
\begin{minipage}{0.23\textwidth}
\includegraphics[width=\textwidth, clip=true, trim= 7.4cm 1.0cm 5.8cm 3cm]{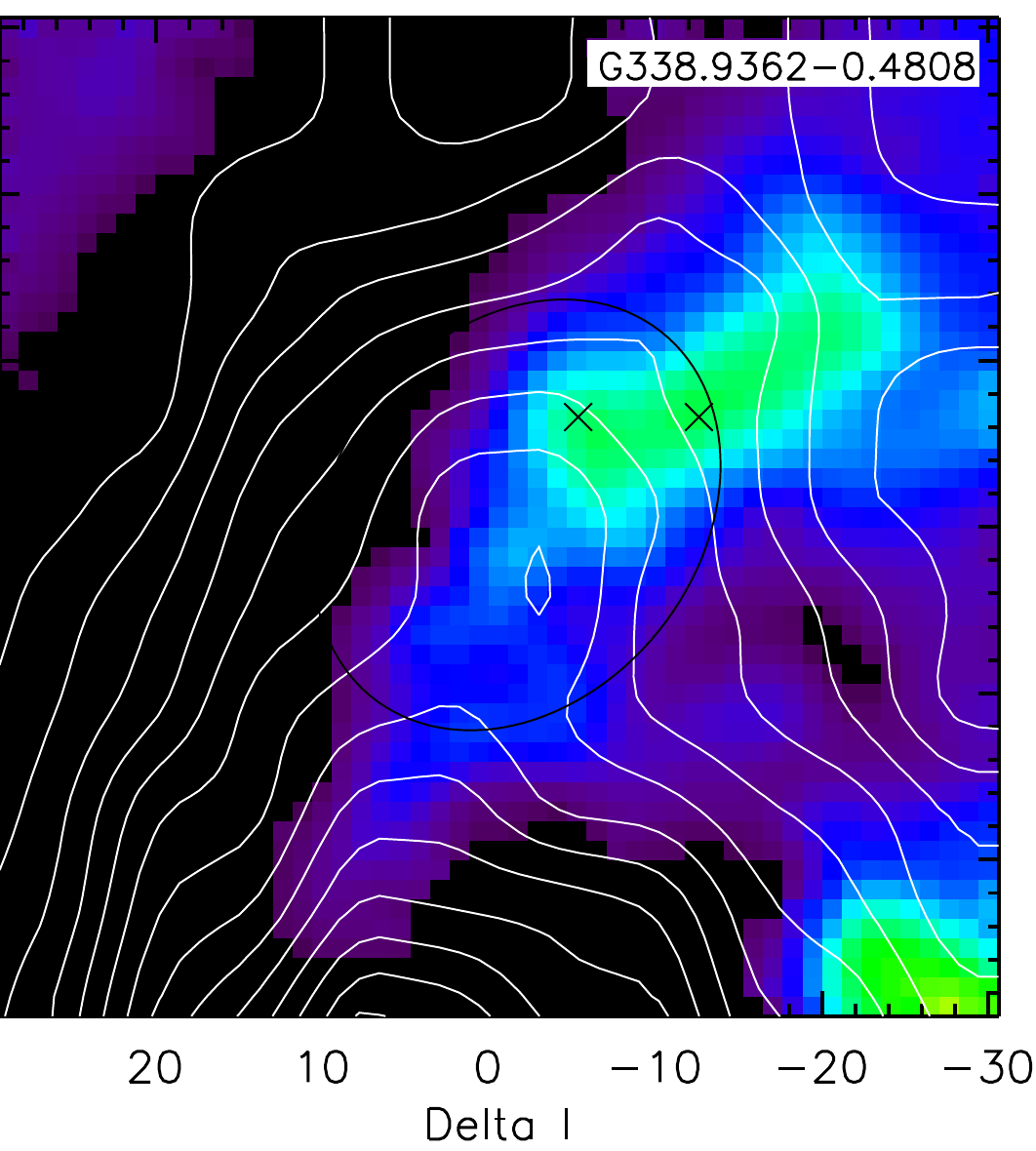}
\end{minipage}
\begin{minipage}{0.23\textwidth}
\includegraphics[width=\textwidth, clip=true, trim= 7.4cm 1.0cm 5.8cm 3cm]{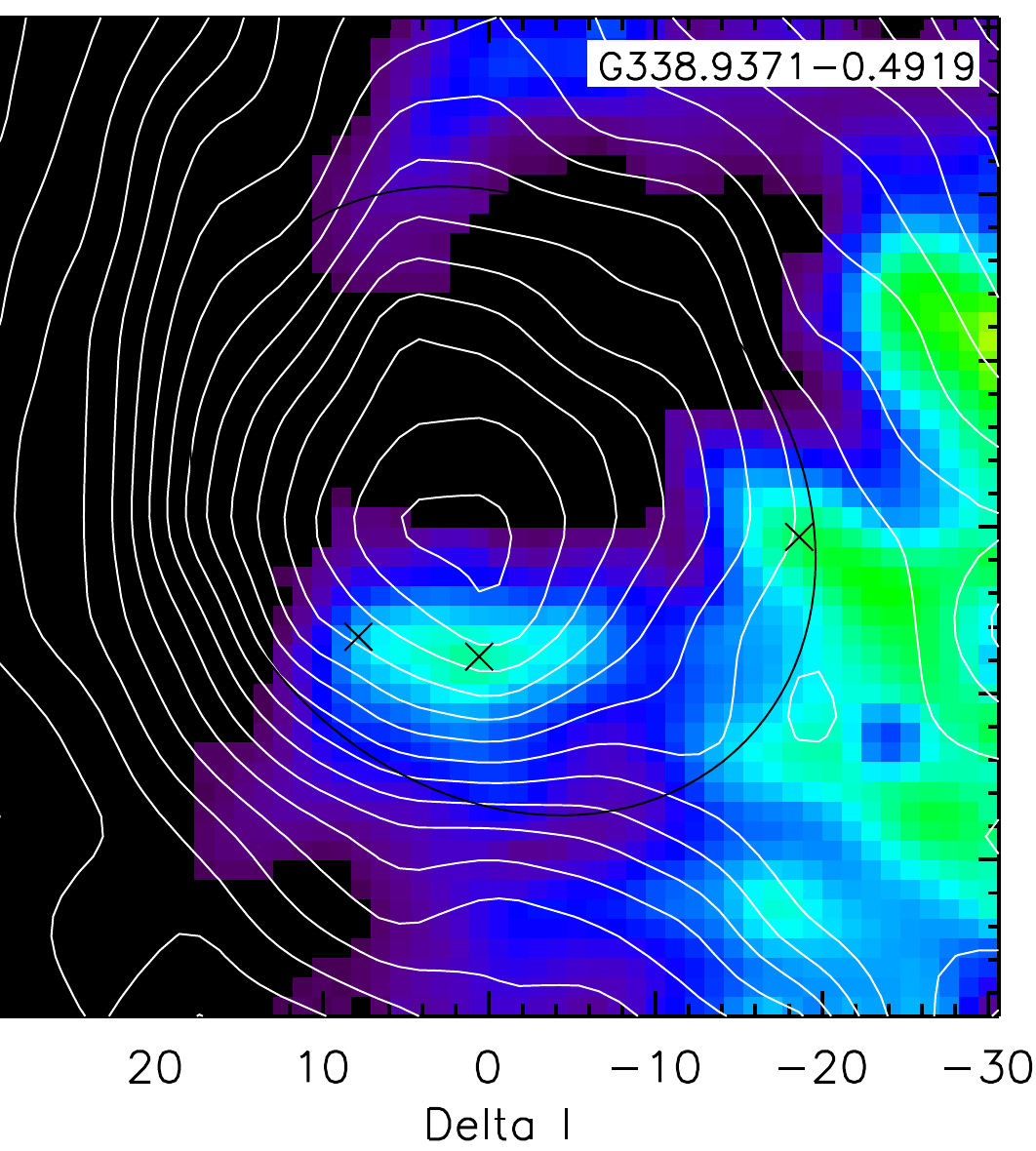}
\end{minipage}
\begin{minipage}{0.23\textwidth}
\includegraphics[width=\textwidth, clip=true, trim= 7.4cm 1.0cm 5.8cm 3cm]{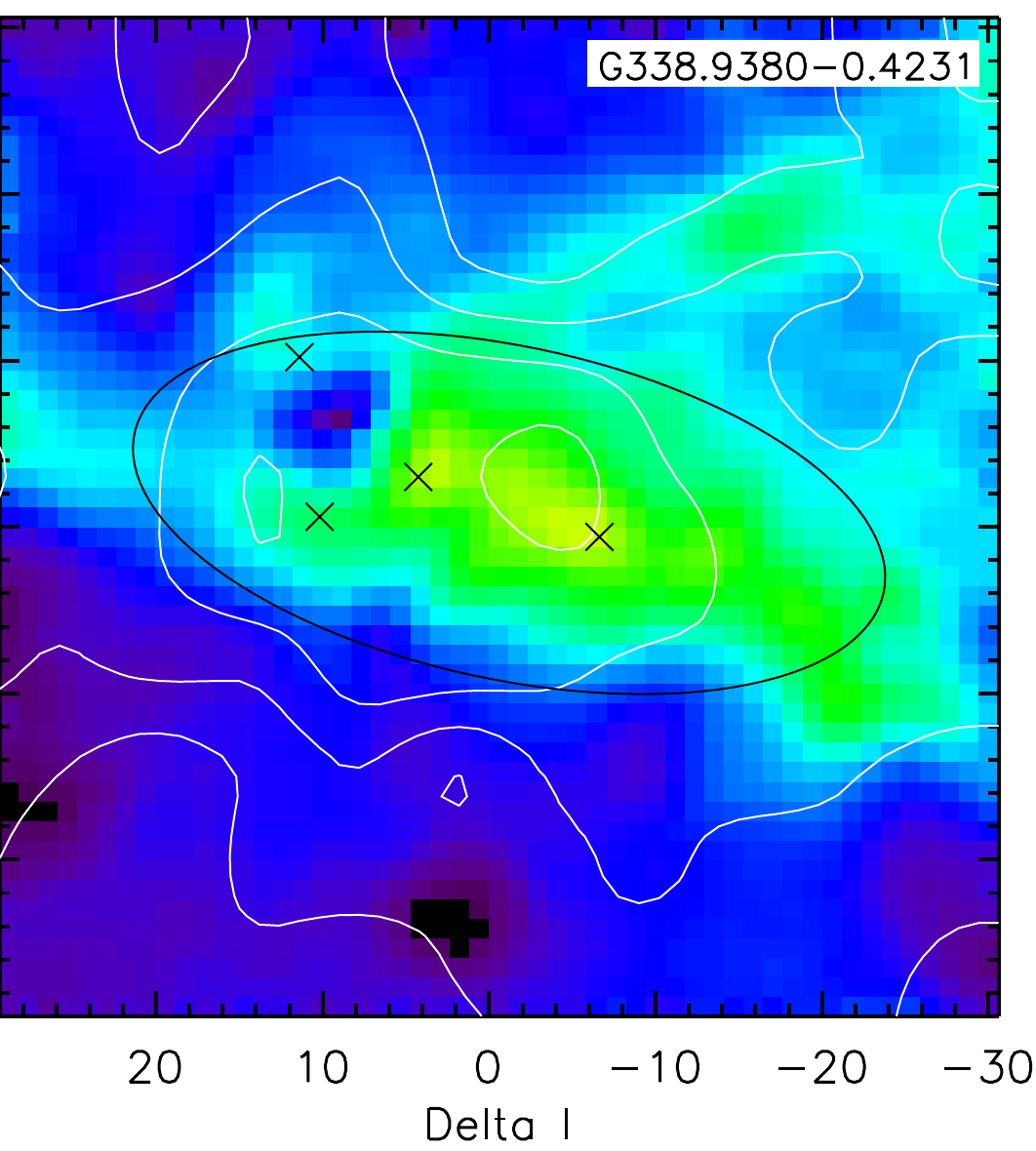}
\end{minipage}

\caption{Half power ellipses of $16$ GCSC ATLASGAL sources (black) overlaid on combined near- and mid-infrared extinction maps. The crosses mark the position of substructures detected on a scale-map ($s=2$) by the clumpfind-2D algorithm within the ATLASGAL sources. The white lines indicate the contours of the ATLASGAL emission.}
\label{ATLAS clumps}
\end{figure*}

\section{Properties of the identified structures}
Here we show the properties of the identified small-scale structures, which are likely to become star formation sites. The shown properties are the results of the clumpfind-2D algorithm applied to the column density map of scale ($i=2$).
\longtab[1]{
\begin{longtable}{ccccccccc}
\caption{structures identified on the $i=2$ scale-map}\\
\hline 
\noalign{\vspace{0.5mm}}
ID & $\textit{l}$ & $\textit{b}$ & $N(\text{H}_2)_{\text{peak}}$ & FWHM$_{\text{x}}$ & FWHM$_{\text{y}}$ & $R$ & $N(\text{H}_2)_{\text{tot}}$ & $N_\text{pix}$ \\ 
 & [$^\circ$] & [$^\circ$] & [$\rm \frac{1}{cm^2}$] & pix & pix & pix & [$\rm \frac{1}{cm^2}$] & \\
\hline
\endfirsthead

\multicolumn{9}{c}%
{{\bfseries \tablename\ \thetable{} -- Continued from previous page}} \\
\hline 
\noalign{\vspace{0.5mm}}
ID & $\textit{l}$ & $\textit{b}$ & $N(\text{H}_2)_{\text{peak}}$ & FWHM$_{\text{x}}$ & FWHM$_{\text{y}}$ & $R$ & $N(\text{H}_2)_{\text{tot}}$ & $N_\text{pix}$ \\ 
 & [$^\circ$] & [$^\circ$] & [$\rm \frac{1}{cm^2}$] & pix & pix & pix & [$\rm \frac{1}{cm^2}$] & \\
\hline
\endhead

\hline \hline \multicolumn{9}{r}{{Continued on next page}} \\
\endfoot

\hline \hline
\endlastfoot

\noalign{\vspace{0.5mm}}
9 & 338.11 & -0.47 & 10.38 & 4.38 & 2.93 & 3.39 & 194.42 & 36 \\
10 & 338.08 & -0.45 & 9.78 & 4.96 & 6.44 & 3.95 & 224.46 & 49 \\
11 & 338.09 & -0.45 & 9.66 & 2.87 & 2.67 & 2.65 & 127.39 & 22 \\
16 & 338.86 & -0.47 & 8.71 & 10.33 & 3.52 & 4.62 & 232.36 & 67 \\
17 & 338.70 & -0.46 & 8.63 & 7.02 & 2.49 & 3.87 & 173.85 & 47 \\
20 & 338.73 & -0.47 & 8.38 & 3.43 & 3.19 & 3.14 & 126.11 & 31 \\
24 & 338.09 & -0.45 & 8.13 & 4.51 & 6.27 & 3.74 & 160.13 & 44 \\
26 & 338.09 & -0.45 & 8.01 & 3.21 & 2.67 & 2.82 & 113.50 & 25 \\
29 & 338.65 & -0.46 & 7.77 & 5.00 & 3.77 & 3.19 & 118.37 & 32 \\
33 & 338.71 & -0.46 & 7.55 & 4.21 & 2.61 & 3.09 & 108.18 & 30 \\
36 & 338.11 & -0.46 & 7.38 & 2.27 & 4.12 & 2.82 & 85.24 & 25 \\
43 & 338.64 & -0.46 & 7.14 & 2.93 & 2.09 & 2.33 & 82.64 & 17 \\
44 & 338.73 & -0.46 & 7.09 & 3.01 & 3.62 & 2.93 & 102.15 & 27 \\
46 & 338.11 & -0.47 & 6.95 & 4.01 & 4.30 & 3.52 & 134.24 & 39 \\
48 & 338.08 & -0.45 & 6.92 & 3.21 & 2.83 & 2.52 & 72.30 & 20 \\
49 & 338.69 & -0.46 & 6.89 & 5.20 & 3.13 & 3.61 & 125.90 & 41 \\
50 & 338.87 & -0.48 & 6.87 & 2.74 & 5.54 & 3.61 & 153.59 & 41 \\
51 & 338.34 & -0.50 & 6.84 & 3.37 & 2.89 & 2.88 & 89.20 & 26 \\
53 & 338.09 & -0.43 & 6.79 & 1.80 & 4.40 & 2.65 & 83.95 & 22 \\
54 & 338.10 & -0.45 & 6.73 & 3.88 & 1.90 & 2.52 & 80.81 & 20 \\
56 & 338.08 & -0.44 & 6.69 & 5.41 & 2.96 & 3.24 & 111.74 & 33 \\
58 & 338.31 & -0.51 & 6.66 & 3.01 & 2.54 & 2.71 & 76.39 & 23 \\
59 & 338.11 & -0.47 & 6.61 & 1.91 & 3.39 & 2.39 & 63.31 & 18 \\
60 & 338.55 & -0.42 & 6.45 & 6.23 & 4.03 & 3.61 & 133.05 & 41 \\
67 & 338.09 & -0.44 & 6.27 & 1.92 & 2.17 & 1.95 & 43.95 & 12 \\
71 & 338.87 & -0.48 & 6.15 & 2.36 & 3.23 & 2.46 & 67.43 & 19 \\
77 & 338.08 & -0.43 & 6.04 & 2.23 & 2.42 & 2.26 & 52.61 & 16 \\
78 & 338.87 & -0.47 & 6.01 & 5.25 & 2.75 & 3.19 & 95.52 & 32 \\
79 & 338.08 & -0.43 & 6.00 & 2.70 & 1.72 & 2.11 & 45.62 & 14 \\
83 & 338.78 & -0.46 & 5.92 & 2.34 & 3.13 & 2.33 & 50.45 & 17 \\
85 & 338.87 & -0.48 & 5.84 & 2.25 & 3.44 & 2.46 & 72.41 & 19 \\
87 & 338.39 & -0.40 & 5.79 & 2.66 & 3.43 & 2.65 & 65.60 & 22 \\
88 & 338.62 & -0.44 & 5.74 & 3.83 & 5.21 & 3.09 & 78.61 & 30 \\
89 & 338.09 & -0.45 & 5.74 & 2.60 & 2.47 & 2.33 & 53.72 & 17 \\
90 & 338.32 & -0.41 & 5.73 & 4.68 & 3.34 & 3.19 & 105.23 & 32 \\
95 & 338.32 & -0.51 & 5.67 & 8.75 & 2.10 & 3.57 & 115.17 & 40 \\
96 & 338.19 & -0.48 & 5.64 & 2.84 & 3.06 & 2.65 & 67.07 & 22 \\
97 & 338.24 & -0.44 & 5.63 & 2.63 & 4.44 & 3.04 & 88.73 & 29 \\
100 & 338.13 & -0.49 & 5.62 & 3.73 & 3.68 & 3.14 & 85.00 & 31 \\
102 & 338.11 & -0.45 & 5.60 & 2.71 & 2.89 & 2.65 & 66.99 & 22 \\
103 & 338.10 & -0.45 & 5.59 & 6.57 & 3.81 & 3.39 & 107.10 & 36 \\
106 & 338.46 & -0.43 & 5.50 & 5.86 & 2.02 & 3.04 & 85.06 & 29 \\
108 & 338.10 & -0.45 & 5.49 & 3.55 & 2.47 & 2.39 & 58.37 & 18 \\
110 & 338.85 & -0.47 & 5.47 & 2.49 & 2.68 & 2.46 & 59.34 & 19 \\
111 & 338.33 & -0.51 & 5.46 & 2.03 & 2.91 & 2.33 & 50.69 & 17 \\
113 & 338.09 & -0.44 & 5.45 & 3.26 & 2.60 & 2.65 & 70.43 & 22 \\
116 & 338.20 & -0.46 & 5.44 & 2.69 & 3.57 & 2.71 & 75.55 & 23 \\
117 & 338.60 & -0.44 & 5.44 & 1.83 & 5.45 & 2.82 & 79.80 & 25 \\
118 & 338.64 & -0.46 & 5.43 & 3.59 & 1.88 & 2.39 & 58.47 & 18 \\
120 & 338.67 & -0.45 & 5.40 & 4.36 & 2.46 & 2.71 & 63.15 & 23 \\
121 & 338.08 & -0.44 & 5.35 & 2.96 & 2.73 & 2.52 & 64.19 & 20 \\
122 & 338.34 & -0.40 & 5.33 & 6.25 & 3.01 & 3.24 & 89.47 & 33 \\
125 & 338.82 & -0.45 & 5.32 & 2.92 & 3.32 & 2.52 & 59.44 & 20 \\
126 & 338.81 & -0.48 & 5.31 & 3.19 & 2.31 & 2.59 & 60.68 & 21 \\
129 & 338.69 & -0.46 & 5.28 & 3.21 & 2.11 & 2.39 & 52.69 & 18 \\
130 & 338.60 & -0.44 & 5.27 & 3.64 & 2.43 & 2.71 & 75.71 & 23 \\
134 & 338.82 & -0.47 & 5.23 & 3.11 & 3.57 & 2.65 & 61.68 & 22 \\
135 & 338.94 & -0.42 & 5.22 & 2.34 & 4.97 & 2.99 & 83.16 & 28 \\
136 & 338.27 & -0.43 & 5.22 & 3.72 & 2.35 & 2.82 & 70.46 & 25 \\
138 & 338.50 & -0.42 & 5.21 & 3.78 & 3.16 & 2.71 & 69.47 & 23 \\
139 & 338.33 & -0.41 & 5.21 & 3.61 & 1.76 & 2.33 & 55.73 & 17 \\
141 & 338.09 & -0.46 & 5.19 & 3.75 & 6.70 & 2.99 & 73.09 & 28 \\
145 & 338.61 & -0.44 & 5.16 & 3.26 & 4.76 & 3.09 & 95.21 & 30 \\
147 & 338.60 & -0.44 & 5.12 & 3.34 & 2.18 & 2.46 & 65.92 & 19 \\
148 & 338.78 & -0.46 & 5.12 & 4.06 & 4.37 & 3.43 & 94.78 & 37 \\
149 & 338.87 & -0.48 & 5.12 & 3.81 & 1.96 & 2.39 & 48.98 & 18 \\
154 & 338.85 & -0.47 & 5.03 & 2.84 & 2.09 & 2.33 & 47.11 & 17 \\
157 & 338.48 & -0.43 & 4.95 & 4.17 & 2.94 & 3.14 & 87.01 & 31 \\
159 & 338.29 & -0.43 & 4.93 & 3.24 & 3.61 & 3.09 & 81.85 & 30 \\
160 & 338.30 & -0.52 & 4.92 & 3.00 & 4.62 & 3.29 & 98.33 & 34 \\
161 & 338.62 & -0.44 & 4.91 & 3.04 & 5.80 & 3.09 & 79.39 & 30 \\
164 & 338.52 & -0.43 & 4.89 & 2.45 & 2.04 & 2.11 & 38.29 & 14 \\
168 & 338.62 & -0.44 & 4.85 & 2.20 & 2.72 & 2.33 & 47.48 & 17 \\
170 & 338.60 & -0.44 & 4.83 & 3.11 & 2.49 & 2.46 & 56.86 & 19 \\
173 & 338.30 & -0.52 & 4.82 & 4.16 & 2.31 & 2.82 & 72.29 & 25 \\
178 & 338.09 & -0.44 & 4.78 & 4.77 & 2.31 & 2.88 & 77.22 & 26 \\
179 & 338.50 & -0.42 & 4.77 & 3.09 & 4.57 & 2.88 & 78.79 & 26 \\
181 & 338.84 & -0.45 & 4.76 & 7.52 & 3.05 & 3.57 & 100.89 & 40 \\
182 & 338.75 & -0.46 & 4.76 & 6.73 & 5.13 & 3.74 & 107.58 & 44 \\
183 & 338.34 & -0.51 & 4.75 & 2.98 & 2.28 & 2.33 & 51.20 & 17 \\
184 & 338.76 & -0.48 & 4.74 & 2.20 & 4.51 & 2.71 & 57.15 & 23 \\
185 & 338.61 & -0.44 & 4.73 & 6.08 & 3.39 & 3.57 & 104.06 & 40 \\
189 & 338.70 & -0.48 & 4.71 & 4.08 & 3.40 & 3.19 & 81.20 & 32 \\
190 & 338.11 & -0.46 & 4.71 & 3.54 & 1.95 & 2.33 & 42.64 & 17 \\
191 & 338.61 & -0.44 & 4.70 & 3.15 & 2.90 & 2.71 & 62.21 & 23 \\
195 & 338.30 & -0.52 & 4.66 & 2.44 & 6.96 & 3.14 & 80.88 & 31 \\
196 & 338.09 & -0.45 & 4.66 & 6.14 & 2.50 & 3.14 & 80.21 & 31 \\
197 & 338.10 & -0.46 & 4.66 & 2.83 & 2.30 & 2.26 & 45.35 & 16 \\
199 & 338.73 & -0.47 & 4.62 & 3.08 & 6.62 & 2.76 & 57.63 & 24 \\
201 & 338.57 & -0.44 & 4.61 & 3.60 & 2.26 & 2.46 & 50.63 & 19 \\
202 & 338.65 & -0.45 & 4.61 & 2.44 & 1.79 & 1.95 & 33.23 & 12 \\
204 & 338.17 & -0.47 & 4.59 & 2.19 & 2.51 & 2.19 & 39.57 & 15 \\
208 & 338.18 & -0.46 & 4.56 & 2.20 & 1.96 & 1.95 & 35.85 & 12 \\
211 & 338.31 & -0.52 & 4.56 & 5.37 & 2.54 & 2.71 & 57.95 & 23 \\
213 & 338.10 & -0.46 & 4.55 & 2.73 & 3.91 & 2.76 & 66.28 & 24 \\
217 & 339.04 & -0.39 & 4.54 & 4.51 & 2.87 & 3.14 & 85.78 & 31 \\
218 & 338.33 & -0.41 & 4.52 & 3.38 & 1.75 & 2.11 & 41.53 & 14 \\
219 & 338.93 & -0.49 & 4.52 & 2.79 & 4.27 & 3.14 & 84.45 & 31 \\
220 & 338.31 & -0.51 & 4.50 & 1.74 & 2.43 & 1.95 & 32.52 & 12 \\
227 & 338.86 & -0.47 & 4.47 & 4.71 & 3.95 & 3.48 & 86.50 & 38 \\
228 & 338.08 & -0.44 & 4.46 & 3.36 & 1.79 & 2.33 & 43.44 & 17 \\
229 & 338.33 & -0.41 & 4.46 & 2.90 & 1.84 & 2.11 & 38.98 & 14 \\
231 & 338.81 & -0.46 & 4.45 & 2.25 & 2.34 & 2.03 & 38.57 & 13 \\
236 & 338.30 & -0.52 & 4.43 & 3.76 & 3.27 & 2.82 & 62.30 & 25 \\
239 & 338.20 & -0.46 & 4.42 & 4.80 & 3.31 & 3.19 & 76.03 & 32 \\
240 & 338.93 & -0.43 & 4.42 & 2.64 & 2.19 & 2.26 & 42.37 & 16 \\
242 & 338.77 & -0.46 & 4.40 & 3.02 & 2.80 & 2.46 & 51.70 & 19 \\
246 & 338.19 & -0.46 & 4.37 & 3.46 & 2.83 & 2.71 & 64.63 & 23 \\
248 & 338.24 & -0.44 & 4.36 & 1.55 & 3.30 & 2.03 & 37.13 & 13 \\
249 & 338.29 & -0.43 & 4.36 & 2.86 & 2.19 & 2.26 & 41.10 & 16 \\
252 & 338.31 & -0.43 & 4.35 & 1.68 & 3.33 & 2.19 & 38.06 & 15 \\
253 & 338.45 & -0.42 & 4.35 & 7.54 & 2.76 & 3.29 & 78.07 & 34 \\
256 & 338.20 & -0.48 & 4.32 & 2.47 & 4.63 & 2.88 & 67.15 & 26 \\
257 & 338.62 & -0.44 & 4.31 & 2.13 & 2.44 & 1.95 & 34.91 & 12 \\
259 & 338.09 & -0.42 & 4.31 & 3.15 & 2.10 & 2.39 & 46.75 & 18 \\
267 & 338.77 & -0.46 & 4.27 & 3.16 & 2.20 & 2.26 & 45.37 & 16 \\
269 & 338.21 & -0.48 & 4.27 & 2.08 & 1.87 & 1.87 & 30.97 & 11 \\
271 & 338.18 & -0.48 & 4.26 & 4.38 & 3.80 & 2.93 & 69.68 & 27 \\
272 & 338.33 & -0.41 & 4.26 & 5.35 & 4.27 & 3.39 & 99.61 & 36 \\
273 & 338.18 & -0.46 & 4.25 & 3.00 & 2.82 & 2.52 & 56.71 & 20 \\
276 & 338.87 & -0.49 & 4.23 & 3.27 & 4.13 & 2.88 & 62.19 & 26 \\
277 & 338.87 & -0.48 & 4.22 & 4.13 & 2.40 & 2.46 & 48.62 & 19 \\
279 & 338.63 & -0.45 & 4.21 & 2.34 & 3.10 & 2.33 & 42.57 & 17 \\
282 & 338.12 & -0.47 & 4.20 & 3.42 & 3.57 & 3.09 & 72.84 & 30 \\
285 & 338.90 & -0.43 & 4.20 & 4.70 & 3.30 & 2.93 & 66.23 & 27 \\
287 & 338.46 & -0.43 & 4.18 & 1.89 & 2.47 & 1.95 & 31.58 & 12 \\
288 & 338.47 & -0.43 & 4.18 & 3.32 & 2.58 & 2.26 & 46.97 & 16 \\
289 & 338.90 & -0.43 & 4.18 & 2.82 & 2.09 & 2.19 & 39.80 & 15 \\
290 & 338.51 & -0.42 & 4.17 & 5.28 & 2.82 & 3.09 & 69.12 & 30 \\
291 & 338.85 & -0.47 & 4.17 & 2.92 & 3.41 & 2.82 & 66.00 & 25 \\
293 & 338.87 & -0.49 & 4.15 & 2.24 & 2.26 & 2.03 & 32.41 & 13 \\
296 & 338.33 & -0.51 & 4.13 & 2.07 & 2.52 & 2.11 & 38.05 & 14 \\
298 & 338.26 & -0.38 & 4.12 & 2.75 & 2.28 & 2.26 & 39.48 & 16 \\
300 & 338.08 & -0.45 & 4.12 & 2.29 & 5.21 & 2.19 & 34.42 & 15 \\
303 & 338.19 & -0.47 & 4.11 & 7.73 & 3.58 & 2.93 & 60.83 & 27 \\
306 & 338.71 & -0.47 & 4.09 & 2.51 & 2.26 & 2.11 & 38.51 & 14 \\
307 & 338.09 & -0.45 & 4.09 & 3.07 & 2.07 & 2.33 & 41.94 & 17 \\
308 & 338.75 & -0.47 & 4.09 & 2.99 & 3.47 & 2.88 & 67.70 & 26 \\
310 & 338.81 & -0.49 & 4.09 & 2.98 & 3.04 & 2.76 & 60.01 & 24 \\
312 & 338.56 & -0.44 & 4.08 & 5.54 & 2.29 & 3.04 & 74.99 & 29 \\
315 & 338.34 & -0.41 & 4.08 & 3.59 & 1.91 & 2.11 & 34.87 & 14 \\
316 & 338.16 & -0.48 & 4.08 & 1.99 & 3.17 & 2.26 & 41.14 & 16 \\
317 & 338.93 & -0.49 & 4.08 & 3.41 & 2.55 & 2.46 & 45.47 & 19 \\
319 & 338.78 & -0.46 & 4.07 & 2.91 & 4.07 & 2.76 & 58.32 & 24 \\
326 & 338.32 & -0.42 & 4.05 & 3.33 & 3.87 & 2.59 & 45.35 & 21 \\
329 & 338.46 & -0.43 & 4.05 & 3.33 & 1.96 & 2.39 & 43.65 & 18 \\
330 & 338.93 & -0.49 & 4.04 & 6.58 & 4.80 & 3.61 & 111.99 & 41 \\
333 & 338.65 & -0.45 & 4.03 & 4.66 & 2.04 & 2.65 & 54.46 & 22 \\
337 & 338.86 & -0.48 & 4.02 & 4.28 & 2.64 & 2.59 & 55.51 & 21 \\
338 & 338.65 & -0.45 & 4.02 & 3.83 & 4.01 & 2.88 & 60.45 & 26 \\
339 & 338.58 & -0.44 & 4.02 & 1.73 & 2.48 & 1.95 & 29.73 & 12 \\
340 & 338.58 & -0.43 & 4.02 & 1.65 & 5.11 & 2.59 & 52.32 & 21 \\
342 & 338.24 & -0.45 & 4.01 & 1.79 & 2.55 & 1.95 & 30.15 & 12 \\
343 & 338.87 & -0.49 & 4.01 & 1.77 & 2.60 & 2.03 & 31.81 & 13 \\
344 & 338.54 & -0.43 & 4.00 & 2.98 & 2.09 & 2.33 & 38.68 & 17 \\
345 & 339.09 & -0.41 & 3.99 & 3.68 & 2.49 & 2.59 & 53.84 & 21 \\
346 & 338.71 & -0.48 & 3.99 & 2.20 & 2.50 & 2.03 & 33.55 & 13 \\
350 & 338.31 & -0.42 & 3.97 & 3.38 & 2.09 & 2.19 & 39.75 & 15 \\
353 & 338.72 & -0.47 & 3.96 & 3.01 & 2.18 & 2.26 & 42.21 & 16 \\
358 & 338.32 & -0.41 & 3.94 & 3.42 & 3.18 & 2.46 & 47.99 & 19 \\
360 & 338.13 & -0.49 & 3.93 & 5.36 & 2.08 & 2.65 & 51.31 & 22 \\
361 & 338.99 & -0.40 & 3.93 & 5.93 & 3.47 & 3.39 & 90.02 & 36 \\
363 & 338.87 & -0.49 & 3.92 & 1.99 & 3.28 & 2.19 & 35.27 & 15 \\
367 & 338.92 & -0.50 & 3.91 & 1.74 & 3.51 & 2.33 & 43.16 & 17 \\
368 & 338.62 & -0.45 & 3.91 & 3.24 & 2.77 & 2.39 & 43.53 & 18 \\
373 & 338.88 & -0.49 & 3.88 & 3.06 & 2.38 & 2.33 & 39.56 & 17 \\
374 & 338.95 & -0.42 & 3.88 & 3.84 & 4.19 & 2.88 & 67.02 & 26 \\
376 & 338.57 & -0.44 & 3.87 & 2.86 & 2.99 & 2.46 & 44.64 & 19 \\
379 & 338.13 & -0.47 & 3.86 & 2.24 & 2.57 & 2.19 & 35.89 & 15 \\
380 & 338.85 & -0.47 & 3.85 & 3.28 & 2.52 & 2.39 & 42.99 & 18 \\
386 & 338.86 & -0.48 & 3.82 & 3.18 & 2.00 & 2.19 & 36.43 & 15 \\
387 & 338.32 & -0.41 & 3.82 & 2.75 & 2.72 & 2.39 & 46.55 & 18 \\
389 & 338.11 & -0.46 & 3.81 & 3.14 & 7.97 & 3.48 & 90.03 & 38 \\
391 & 339.02 & -0.40 & 3.81 & 5.29 & 2.79 & 3.14 & 79.56 & 31 \\
393 & 338.68 & -0.46 & 3.81 & 5.74 & 5.04 & 3.09 & 61.83 & 30 \\
394 & 338.10 & -0.46 & 3.81 & 2.87 & 3.85 & 2.93 & 70.69 & 27 \\
397 & 338.85 & -0.45 & 3.80 & 3.34 & 3.10 & 2.39 & 46.74 & 18 \\
400 & 338.87 & -0.47 & 3.79 & 2.38 & 2.87 & 2.33 & 40.43 & 17 \\
402 & 338.78 & -0.46 & 3.79 & 1.79 & 2.52 & 1.87 & 27.91 & 11 \\
407 & 338.55 & -0.42 & 3.77 & 3.22 & 2.70 & 2.39 & 44.22 & 18 \\
408 & 338.78 & -0.46 & 3.77 & 2.81 & 2.28 & 2.26 & 39.44 & 16 \\
409 & 338.40 & -0.40 & 3.76 & 2.31 & 2.37 & 2.03 & 34.17 & 13 \\
410 & 338.50 & -0.42 & 3.76 & 3.23 & 2.94 & 2.39 & 44.86 & 18 \\
416 & 338.25 & -0.44 & 3.74 & 4.34 & 3.55 & 2.65 & 51.41 & 22 \\
418 & 338.94 & -0.49 & 3.74 & 2.97 & 2.28 & 2.33 & 41.33 & 17 \\
419 & 338.33 & -0.40 & 3.74 & 2.52 & 1.93 & 2.03 & 31.91 & 13 \\
424 & 338.11 & -0.46 & 3.73 & 2.23 & 3.18 & 2.52 & 47.40 & 20 \\
427 & 338.81 & -0.46 & 3.73 & 2.21 & 2.09 & 1.95 & 29.11 & 12 \\
430 & 338.41 & -0.41 & 3.70 & 4.55 & 2.54 & 2.59 & 50.33 & 21 \\
432 & 338.55 & -0.43 & 3.70 & 1.62 & 3.16 & 2.11 & 30.81 & 14 \\
435 & 338.94 & -0.42 & 3.69 & 3.72 & 2.56 & 2.52 & 46.00 & 20 \\
436 & 338.30 & -0.48 & 3.69 & 8.93 & 3.74 & 3.24 & 67.69 & 33 \\
437 & 338.42 & -0.42 & 3.69 & 3.96 & 2.96 & 2.33 & 37.20 & 17 \\
441 & 338.47 & -0.43 & 3.68 & 4.43 & 6.44 & 3.19 & 70.14 & 32 \\
443 & 338.94 & -0.49 & 3.67 & 3.32 & 2.28 & 2.46 & 46.19 & 19 \\
445 & 338.34 & -0.51 & 3.66 & 3.00 & 1.76 & 2.11 & 33.98 & 14 \\
447 & 338.92 & -0.49 & 3.66 & 2.54 & 2.26 & 2.19 & 35.85 & 15 \\
448 & 339.10 & -0.40 & 3.66 & 5.58 & 2.93 & 2.76 & 51.35 & 24 \\
455 & 338.87 & -0.47 & 3.65 & 3.50 & 2.63 & 2.59 & 46.82 & 21 \\
456 & 338.17 & -0.47 & 3.64 & 2.70 & 2.84 & 2.33 & 42.97 & 17 \\
458 & 338.11 & -0.46 & 3.64 & 4.67 & 3.00 & 3.09 & 65.79 & 30 \\
459 & 338.93 & -0.42 & 3.63 & 4.40 & 2.19 & 2.65 & 51.21 & 22 \\
464 & 338.41 & -0.40 & 3.62 & 3.43 & 2.33 & 2.39 & 43.00 & 18 \\
466 & 338.86 & -0.50 & 3.62 & 4.56 & 2.46 & 2.82 & 59.96 & 25 \\
467 & 338.64 & -0.45 & 3.62 & 2.01 & 3.10 & 2.03 & 29.45 & 13 \\
469 & 338.62 & -0.44 & 3.62 & 2.80 & 2.32 & 2.33 & 40.72 & 17 \\
472 & 338.38 & -0.41 & 3.61 & 3.10 & 2.22 & 2.39 & 40.07 & 18 \\
474 & 338.38 & -0.40 & 3.60 & 8.46 & 1.77 & 3.52 & 85.07 & 39 \\
476 & 338.43 & -0.41 & 3.59 & 2.82 & 2.19 & 2.33 & 38.70 & 17 \\
481 & 338.25 & -0.42 & 3.59 & 6.31 & 4.20 & 2.88 & 51.53 & 26 \\
482 & 338.73 & -0.47 & 3.59 & 3.65 & 2.43 & 2.03 & 30.09 & 13 \\
483 & 338.18 & -0.46 & 3.58 & 3.29 & 2.87 & 2.26 & 39.31 & 16 \\
484 & 339.09 & -0.41 & 3.58 & 6.07 & 2.06 & 2.99 & 66.74 & 28 \\
485 & 338.85 & -0.47 & 3.58 & 1.73 & 2.72 & 1.87 & 27.13 & 11 \\
487 & 338.28 & -0.43 & 3.58 & 4.09 & 1.68 & 2.39 & 41.53 & 18 \\
488 & 338.25 & -0.44 & 3.58 & 7.14 & 5.04 & 3.19 & 78.22 & 32 \\
494 & 338.63 & -0.46 & 3.56 & 2.64 & 3.38 & 2.52 & 43.19 & 20 \\
497 & 338.88 & -0.54 & 3.56 & 2.31 & 1.74 & 1.87 & 25.55 & 11 \\
499 & 338.32 & -0.41 & 3.55 & 3.74 & 2.00 & 2.39 & 45.00 & 18 \\
503 & 339.09 & -0.41 & 3.53 & 3.84 & 1.94 & 2.46 & 44.86 & 19 \\
504 & 338.16 & -0.48 & 3.53 & 6.17 & 3.02 & 3.19 & 69.67 & 32 \\
505 & 338.62 & -0.45 & 3.53 & 4.27 & 6.16 & 3.19 & 64.71 & 32 \\
508 & 338.26 & -0.44 & 3.52 & 2.63 & 5.68 & 2.99 & 60.92 & 28 \\
509 & 338.93 & -0.42 & 3.52 & 3.87 & 2.86 & 2.52 & 41.61 & 20 \\
510 & 338.39 & -0.40 & 3.52 & 4.35 & 2.13 & 2.39 & 38.16 & 18 \\
511 & 338.32 & -0.42 & 3.52 & 3.86 & 2.65 & 2.19 & 33.45 & 15 \\
512 & 338.15 & -0.48 & 3.51 & 2.03 & 2.86 & 2.03 & 29.11 & 13 \\
515 & 338.72 & -0.46 & 3.51 & 2.39 & 3.91 & 2.52 & 47.61 & 20 \\
516 & 338.84 & -0.45 & 3.50 & 7.58 & 3.22 & 3.57 & 92.03 & 40 \\
519 & 339.02 & -0.40 & 3.50 & 2.39 & 1.71 & 1.87 & 26.38 & 11 \\
525 & 338.31 & -0.50 & 3.48 & 3.09 & 3.07 & 2.26 & 33.52 & 16 \\
526 & 338.09 & -0.46 & 3.48 & 3.13 & 1.74 & 2.11 & 32.44 & 14 \\
527 & 338.21 & -0.47 & 3.48 & 3.10 & 4.38 & 3.04 & 63.24 & 29 \\
530 & 338.93 & -0.49 & 3.47 & 3.38 & 2.63 & 2.46 & 43.78 & 19 \\
534 & 338.15 & -0.49 & 3.46 & 2.34 & 1.94 & 1.87 & 27.83 & 11 \\
541 & 338.23 & -0.49 & 3.45 & 2.19 & 2.79 & 2.26 & 35.14 & 16 \\
548 & 338.32 & -0.42 & 3.44 & 6.93 & 2.16 & 2.88 & 62.58 & 26 \\
549 & 338.93 & -0.49 & 3.44 & 2.35 & 3.64 & 2.19 & 34.08 & 15 \\
551 & 339.08 & -0.41 & 3.43 & 2.53 & 2.37 & 2.19 & 33.81 & 15 \\
552 & 338.11 & -0.48 & 3.43 & 2.75 & 2.52 & 2.39 & 41.90 & 18 \\
555 & 338.42 & -0.42 & 3.43 & 3.61 & 2.05 & 2.19 & 34.05 & 15 \\
556 & 338.55 & -0.42 & 3.43 & 5.50 & 6.00 & 2.93 & 49.09 & 27 \\
558 & 338.93 & -0.50 & 3.41 & 3.18 & 2.20 & 2.19 & 34.37 & 15 \\
559 & 338.87 & -0.48 & 3.41 & 1.92 & 3.23 & 2.19 & 33.82 & 15 \\
563 & 338.41 & -0.41 & 3.41 & 1.97 & 2.45 & 2.03 & 29.67 & 13 \\
565 & 338.42 & -0.41 & 3.40 & 4.76 & 6.13 & 3.34 & 71.32 & 35 \\
569 & 338.15 & -0.48 & 3.39 & 5.04 & 2.66 & 2.71 & 50.21 & 23 \\
573 & 338.10 & -0.46 & 3.38 & 2.86 & 1.95 & 2.19 & 35.27 & 15 \\
575 & 338.96 & -0.43 & 3.37 & 4.13 & 2.74 & 2.65 & 52.43 & 22 \\
576 & 338.75 & -0.46 & 3.37 & 2.31 & 3.18 & 2.26 & 38.80 & 16 \\
581 & 338.28 & -0.43 & 3.36 & 3.05 & 2.02 & 2.26 & 37.93 & 16 \\
583 & 338.85 & -0.45 & 3.36 & 3.12 & 1.62 & 1.87 & 25.73 & 11 \\
584 & 338.93 & -0.42 & 3.35 & 2.38 & 2.02 & 1.87 & 25.95 & 11 \\
588 & 338.94 & -0.48 & 3.35 & 3.08 & 3.71 & 2.82 & 55.20 & 25 \\
593 & 339.03 & -0.40 & 3.34 & 2.37 & 2.50 & 2.11 & 32.78 & 14 \\
594 & 338.26 & -0.43 & 3.34 & 3.57 & 2.11 & 2.39 & 41.79 & 18 \\
595 & 338.24 & -0.37 & 3.34 & 3.32 & 3.32 & 2.52 & 45.93 & 20 \\
598 & 338.78 & -0.45 & 3.34 & 2.18 & 2.80 & 2.03 & 29.06 & 13 \\
605 & 338.25 & -0.42 & 3.31 & 3.13 & 2.21 & 1.95 & 25.39 & 12 \\
606 & 338.73 & -0.46 & 3.30 & 2.62 & 2.71 & 2.26 & 35.77 & 16 \\
608 & 338.65 & -0.45 & 3.30 & 3.57 & 2.30 & 2.26 & 38.13 & 16 \\
613 & 338.87 & -0.48 & 3.30 & 2.60 & 1.69 & 1.87 & 24.71 & 11 \\
616 & 338.65 & -0.45 & 3.29 & 3.46 & 3.37 & 2.46 & 47.44 & 19 \\
618 & 338.20 & -0.47 & 3.29 & 2.53 & 3.00 & 2.46 & 39.80 & 19 \\
620 & 338.09 & -0.46 & 3.29 & 1.82 & 2.79 & 1.95 & 26.62 & 12 \\
622 & 338.64 & -0.45 & 3.28 & 2.54 & 3.15 & 2.52 & 46.68 & 20 \\
623 & 338.30 & -0.48 & 3.28 & 2.60 & 2.90 & 2.33 & 36.33 & 17 \\
626 & 338.32 & -0.42 & 3.27 & 1.61 & 4.58 & 2.33 & 42.92 & 17 \\
627 & 338.09 & -0.43 & 3.27 & 2.32 & 2.30 & 1.95 & 28.44 & 12 \\
628 & 338.31 & -0.42 & 3.27 & 2.62 & 2.46 & 2.11 & 32.06 & 14 \\
632 & 338.67 & -0.46 & 3.26 & 3.29 & 2.19 & 2.26 & 32.28 & 16 \\
634 & 338.77 & -0.46 & 3.26 & 3.01 & 4.47 & 2.93 & 61.89 & 27 \\
636 & 338.13 & -0.45 & 3.26 & 1.85 & 3.80 & 2.26 & 36.59 & 16 \\
639 & 338.74 & -0.47 & 3.25 & 3.64 & 4.05 & 2.76 & 55.48 & 24 \\
641 & 338.82 & -0.47 & 3.25 & 1.83 & 4.12 & 1.95 & 24.01 & 12 \\
650 & 338.33 & -0.41 & 3.23 & 3.26 & 2.43 & 2.11 & 31.07 & 14 \\
654 & 338.76 & -0.46 & 3.23 & 3.67 & 5.64 & 2.65 & 42.56 & 22 \\
657 & 338.82 & -0.46 & 3.22 & 2.36 & 3.43 & 2.19 & 33.18 & 15 \\
662 & 338.89 & -0.44 & 3.21 & 2.99 & 2.63 & 2.26 & 35.37 & 16 \\
664 & 339.04 & -0.43 & 3.21 & 3.53 & 1.70 & 2.19 & 29.72 & 15 \\
670 & 338.19 & -0.47 & 3.20 & 4.98 & 2.81 & 2.82 & 50.91 & 25 \\
671 & 338.77 & -0.46 & 3.20 & 4.11 & 3.46 & 2.93 & 53.53 & 27 \\
672 & 338.92 & -0.50 & 3.20 & 2.43 & 1.75 & 1.87 & 24.94 & 11 \\
676 & 338.31 & -0.42 & 3.19 & 2.05 & 2.11 & 1.87 & 24.23 & 11 \\
677 & 338.30 & -0.52 & 3.19 & 3.17 & 1.82 & 2.19 & 32.27 & 15 \\
681 & 338.93 & -0.42 & 3.18 & 2.06 & 2.65 & 1.95 & 26.58 & 12 \\
689 & 338.94 & -0.42 & 3.17 & 2.46 & 3.88 & 1.95 & 24.56 & 12 \\
695 & 338.15 & -0.49 & 3.17 & 1.87 & 2.40 & 1.87 & 23.86 & 11 \\
700 & 338.81 & -0.46 & 3.15 & 5.73 & 4.56 & 2.93 & 50.88 & 27 \\
705 & 338.25 & -0.44 & 3.14 & 2.55 & 3.53 & 2.11 & 27.85 & 14 \\
706 & 338.63 & -0.46 & 3.14 & 4.05 & 2.33 & 2.71 & 55.49 & 23 \\
709 & 338.15 & -0.48 & 3.14 & 2.35 & 2.85 & 1.87 & 26.52 & 11 \\
714 & 338.11 & -0.45 & 3.13 & 3.05 & 2.41 & 2.33 & 35.36 & 17 \\
724 & 338.38 & -0.40 & 3.10 & 3.01 & 2.46 & 2.26 & 34.84 & 16 \\
730 & 338.94 & -0.42 & 3.10 & 3.94 & 2.95 & 2.76 & 49.44 & 24 \\
736 & 338.31 & -0.42 & 3.09 & 4.75 & 1.54 & 2.19 & 36.73 & 15 \\
738 & 338.09 & -0.44 & 3.08 & 2.19 & 2.36 & 2.03 & 27.83 & 13 \\
743 & 338.09 & -0.43 & 3.04 & 1.58 & 4.01 & 2.19 & 31.44 & 15 \\
747 & 338.32 & -0.50 & 3.04 & 3.21 & 2.91 & 2.59 & 45.57 & 21 \\
756 & 338.87 & -0.49 & 3.03 & 1.79 & 3.62 & 2.26 & 33.49 & 16 \\
757 & 339.10 & -0.41 & 3.02 & 3.32 & 2.94 & 2.19 & 30.79 & 15 \\
761 & 339.00 & -0.41 & 3.01 & 2.80 & 2.12 & 2.11 & 28.16 & 14 \\
762 & 338.39 & -0.41 & 3.01 & 3.81 & 4.05 & 2.59 & 44.03 & 21 \\
766 & 338.92 & -0.51 & 3.00 & 2.51 & 4.78 & 2.93 & 58.06 & 27 \\
767 & 338.32 & -0.44 & 3.00 & 2.55 & 3.28 & 2.19 & 31.35 & 15 \\
771 & 338.45 & -0.42 & 2.99 & 4.15 & 4.00 & 2.71 & 43.32 & 23 \\
780 & 338.93 & -0.49 & 2.98 & 4.18 & 8.68 & 3.14 & 67.19 & 31 \\
786 & 338.26 & -0.43 & 2.97 & 4.00 & 4.10 & 2.71 & 48.26 & 23 \\
787 & 338.91 & -0.53 & 2.97 & 3.53 & 2.00 & 2.19 & 30.76 & 15 \\
789 & 338.12 & -0.50 & 2.96 & 2.54 & 2.13 & 2.11 & 29.34 & 14 \\
790 & 338.37 & -0.40 & 2.96 & 3.62 & 3.54 & 2.71 & 47.90 & 23 \\
793 & 339.01 & -0.40 & 2.96 & 2.62 & 7.07 & 2.82 & 48.46 & 25 \\
794 & 338.14 & -0.48 & 2.96 & 5.21 & 3.54 & 2.71 & 41.30 & 23 \\
795 & 338.18 & -0.46 & 2.96 & 3.52 & 1.74 & 2.11 & 27.88 & 14 \\
796 & 338.64 & -0.45 & 2.96 & 2.71 & 2.31 & 1.95 & 26.54 & 12 \\
800 & 338.71 & -0.47 & 2.95 & 2.10 & 2.36 & 1.95 & 26.85 & 12 \\
801 & 338.95 & -0.42 & 2.95 & 4.03 & 3.81 & 3.29 & 73.55 & 34 \\
803 & 338.28 & -0.43 & 2.95 & 4.34 & 2.31 & 2.52 & 38.05 & 20 \\
805 & 338.52 & -0.43 & 2.94 & 2.01 & 3.06 & 2.19 & 31.25 & 15 \\
806 & 338.42 & -0.41 & 2.94 & 2.43 & 2.17 & 2.03 & 25.85 & 13 \\
809 & 338.63 & -0.45 & 2.94 & 3.87 & 3.45 & 2.65 & 42.43 & 22 \\
812 & 338.94 & -0.49 & 2.94 & 2.39 & 2.75 & 2.26 & 33.87 & 16 \\
819 & 338.11 & -0.45 & 2.92 & 3.38 & 2.40 & 2.33 & 32.89 & 17 \\
821 & 338.32 & -0.41 & 2.92 & 2.78 & 4.96 & 2.52 & 37.38 & 20 \\
822 & 338.55 & -0.42 & 2.92 & 2.01 & 3.45 & 2.33 & 35.02 & 17 \\
823 & 339.00 & -0.40 & 2.92 & 2.54 & 3.53 & 2.52 & 39.96 & 20 \\
824 & 338.09 & -0.46 & 2.92 & 2.71 & 2.71 & 2.39 & 38.23 & 18 \\
825 & 339.00 & -0.41 & 2.92 & 2.37 & 2.88 & 1.95 & 25.02 & 12 \\
826 & 338.86 & -0.50 & 2.92 & 2.80 & 2.03 & 2.19 & 30.56 & 15 \\
833 & 338.09 & -0.44 & 2.91 & 2.07 & 3.41 & 2.03 & 25.15 & 13 \\
835 & 338.85 & -0.47 & 2.90 & 3.57 & 3.08 & 2.39 & 36.98 & 18 \\
841 & 338.39 & -0.40 & 2.90 & 3.31 & 3.20 & 2.52 & 41.88 & 20 \\
843 & 338.41 & -0.42 & 2.89 & 2.30 & 1.81 & 1.87 & 23.09 & 11 \\
848 & 338.34 & -0.41 & 2.89 & 3.28 & 1.57 & 2.03 & 26.28 & 13 \\
851 & 338.09 & -0.43 & 2.88 & 1.56 & 3.19 & 1.87 & 22.86 & 11 \\
855 & 338.47 & -0.42 & 2.88 & 3.92 & 2.29 & 2.39 & 39.05 & 18 \\
860 & 338.91 & -0.44 & 2.87 & 2.76 & 1.97 & 2.03 & 25.09 & 13 \\
862 & 338.48 & -0.43 & 2.87 & 1.82 & 3.44 & 2.03 & 24.75 & 13 \\
868 & 339.00 & -0.41 & 2.86 & 2.94 & 2.47 & 2.19 & 30.87 & 15 \\
873 & 338.77 & -0.46 & 2.86 & 4.29 & 2.22 & 2.59 & 44.53 & 21 \\
875 & 338.78 & -0.45 & 2.86 & 2.38 & 3.50 & 2.11 & 27.94 & 14 \\
881 & 338.39 & -0.41 & 2.85 & 4.44 & 3.62 & 2.99 & 56.27 & 28 \\
882 & 338.41 & -0.40 & 2.85 & 2.22 & 3.04 & 2.33 & 35.67 & 17 \\
886 & 338.96 & -0.41 & 2.84 & 2.74 & 2.40 & 2.03 & 26.13 & 13 \\
889 & 339.07 & -0.41 & 2.84 & 3.20 & 2.86 & 2.19 & 30.98 & 15 \\
894 & 338.86 & -0.50 & 2.83 & 2.99 & 1.41 & 1.87 & 23.21 & 11 \\
896 & 338.13 & -0.46 & 2.83 & 3.60 & 2.03 & 2.03 & 25.17 & 13 \\
897 & 338.73 & -0.47 & 2.82 & 2.46 & 3.07 & 2.26 & 30.70 & 16 \\
898 & 338.38 & -0.39 & 2.82 & 6.66 & 3.16 & 3.34 & 65.60 & 35 \\
900 & 338.93 & -0.48 & 2.82 & 2.71 & 2.23 & 2.11 & 29.19 & 14 \\
909 & 338.35 & -0.41 & 2.80 & 5.46 & 1.97 & 2.76 & 47.88 & 24 \\
912 & 338.09 & -0.44 & 2.80 & 1.64 & 3.50 & 2.03 & 25.88 & 13 \\
921 & 338.33 & -0.51 & 2.78 & 5.40 & 2.97 & 2.59 & 37.55 & 21 \\
922 & 338.14 & -0.48 & 2.78 & 2.09 & 3.20 & 2.19 & 29.89 & 15 \\
923 & 339.04 & -0.42 & 2.78 & 4.01 & 1.68 & 2.11 & 28.54 & 14 \\
924 & 338.82 & -0.45 & 2.78 & 2.70 & 3.66 & 2.59 & 39.57 & 21 \\
927 & 338.31 & -0.43 & 2.78 & 2.18 & 3.32 & 2.26 & 33.61 & 16 \\
930 & 338.08 & -0.42 & 2.77 & 2.29 & 2.81 & 2.03 & 26.13 & 13 \\
931 & 338.93 & -0.50 & 2.77 & 4.96 & 1.52 & 2.26 & 30.70 & 16 \\
933 & 338.83 & -0.46 & 2.77 & 2.47 & 3.37 & 2.59 & 41.85 & 21 \\
934 & 338.32 & -0.42 & 2.77 & 1.97 & 4.12 & 2.33 & 36.09 & 17 \\
936 & 338.08 & -0.45 & 2.76 & 1.66 & 2.60 & 1.87 & 21.71 & 11 \\
939 & 338.19 & -0.48 & 2.76 & 1.74 & 2.69 & 1.87 & 23.72 & 11 \\
940 & 338.85 & -0.47 & 2.76 & 2.98 & 3.87 & 2.39 & 37.16 & 18 \\
941 & 338.18 & -0.47 & 2.76 & 2.46 & 2.66 & 2.26 & 30.54 & 16 \\
944 & 338.83 & -0.45 & 2.75 & 6.34 & 7.60 & 3.74 & 85.64 & 44 \\
949 & 338.82 & -0.46 & 2.74 & 3.20 & 2.57 & 2.39 & 37.41 & 18 \\
950 & 338.10 & -0.46 & 2.74 & 3.28 & 2.50 & 2.33 & 36.22 & 17 \\
953 & 338.78 & -0.46 & 2.74 & 4.17 & 2.04 & 2.46 & 38.05 & 19 \\
954 & 338.40 & -0.40 & 2.74 & 6.99 & 3.35 & 2.99 & 47.57 & 28 \\
956 & 338.81 & -0.48 & 2.73 & 5.81 & 5.17 & 2.93 & 52.59 & 27 \\
963 & 338.41 & -0.41 & 2.73 & 2.17 & 2.39 & 1.87 & 21.75 & 11 \\
970 & 338.69 & -0.49 & 2.72 & 1.49 & 4.69 & 2.11 & 26.56 & 14 \\
973 & 338.41 & -0.40 & 2.72 & 1.57 & 3.93 & 2.11 & 26.85 & 14 \\
980 & 338.37 & -0.40 & 2.70 & 2.67 & 3.47 & 2.03 & 25.73 & 13 \\
981 & 338.20 & -0.47 & 2.70 & 3.26 & 5.64 & 2.65 & 44.85 & 22 \\
987 & 338.31 & -0.50 & 2.69 & 2.55 & 2.70 & 2.26 & 30.31 & 16 \\
994 & 338.37 & -0.40 & 2.69 & 2.56 & 4.04 & 2.19 & 30.63 & 15 \\
1002 & 338.09 & -0.46 & 2.68 & 4.93 & 2.33 & 2.39 & 35.03 & 18 \\
1005 & 338.93 & -0.48 & 2.68 & 2.58 & 2.44 & 2.19 & 29.21 & 15 \\
1009 & 338.86 & -0.47 & 2.68 & 2.71 & 3.94 & 2.26 & 29.48 & 16 \\
1011 & 339.03 & -0.40 & 2.67 & 2.44 & 3.07 & 2.26 & 31.96 & 16 \\
1016 & 338.94 & -0.42 & 2.66 & 3.65 & 3.96 & 2.82 & 49.61 & 25 \\
1022 & 338.11 & -0.47 & 2.65 & 1.75 & 3.64 & 2.26 & 32.96 & 16 \\
1025 & 338.32 & -0.40 & 2.65 & 5.67 & 3.46 & 2.59 & 40.37 & 21 \\
1027 & 338.76 & -0.46 & 2.64 & 2.44 & 2.77 & 2.19 & 28.37 & 15 \\
1031 & 338.08 & -0.42 & 2.63 & 4.11 & 3.86 & 2.03 & 24.55 & 13 \\
1032 & 338.31 & -0.52 & 2.63 & 2.77 & 1.89 & 2.03 & 26.27 & 13 \\
1033 & 338.93 & -0.47 & 2.63 & 3.04 & 3.09 & 2.26 & 28.61 & 16 \\
1037 & 338.92 & -0.47 & 2.63 & 3.53 & 2.26 & 2.11 & 25.46 & 14 \\
1040 & 338.65 & -0.46 & 2.62 & 1.73 & 3.97 & 2.11 & 25.92 & 14 \\
1046 & 338.31 & -0.40 & 2.62 & 4.60 & 2.31 & 2.65 & 39.79 & 22 \\
1049 & 338.23 & -0.46 & 2.61 & 2.09 & 2.92 & 2.19 & 27.37 & 15 \\
1051 & 338.39 & -0.40 & 2.61 & 2.43 & 2.66 & 1.95 & 21.65 & 12 \\
1053 & 338.42 & -0.41 & 2.60 & 2.34 & 2.28 & 1.95 & 22.77 & 12 \\
1056 & 338.78 & -0.50 & 2.60 & 4.00 & 2.24 & 2.46 & 36.40 & 19 \\
1058 & 338.73 & -0.48 & 2.60 & 2.31 & 2.08 & 1.87 & 20.14 & 11 \\
1059 & 338.86 & -0.49 & 2.60 & 2.60 & 3.83 & 2.33 & 32.30 & 17 \\
1061 & 339.03 & -0.42 & 2.59 & 2.68 & 3.08 & 2.26 & 30.22 & 16 \\
1062 & 338.82 & -0.45 & 2.59 & 3.55 & 2.87 & 2.39 & 36.06 & 18 \\
1072 & 338.32 & -0.51 & 2.57 & 6.73 & 3.80 & 2.65 & 39.47 & 22 \\
1074 & 338.82 & -0.48 & 2.57 & 2.42 & 3.00 & 2.33 & 33.41 & 17 \\
1075 & 338.18 & -0.46 & 2.57 & 2.87 & 2.45 & 2.03 & 23.19 & 13 \\
1077 & 338.18 & -0.47 & 2.57 & 1.88 & 3.66 & 2.03 & 22.67 & 13 \\
1079 & 338.92 & -0.52 & 2.56 & 2.09 & 6.18 & 2.59 & 38.36 & 21 \\
1089 & 338.93 & -0.50 & 2.54 & 2.79 & 1.70 & 1.95 & 22.25 & 12 \\
1098 & 338.08 & -0.42 & 2.53 & 2.58 & 4.52 & 2.52 & 37.33 & 20 \\
1099 & 338.32 & -0.51 & 2.53 & 6.36 & 2.29 & 2.39 & 31.33 & 18 \\
1100 & 338.92 & -0.49 & 2.53 & 4.13 & 2.46 & 2.65 & 40.62 & 22 \\
1103 & 338.78 & -0.47 & 2.53 & 1.82 & 2.66 & 1.95 & 23.11 & 12 \\
1107 & 338.19 & -0.47 & 2.52 & 4.23 & 4.07 & 2.46 & 35.50 & 19 \\
1109 & 338.68 & -0.46 & 2.52 & 3.00 & 3.41 & 2.59 & 37.29 & 21 \\
1110 & 338.66 & -0.46 & 2.52 & 2.36 & 3.03 & 2.33 & 30.75 & 17 \\
1111 & 338.23 & -0.54 & 2.51 & 2.72 & 2.88 & 2.33 & 31.96 & 17 \\
1112 & 338.92 & -0.49 & 2.51 & 3.51 & 1.76 & 2.19 & 29.06 & 15 \\
1115 & 338.75 & -0.47 & 2.51 & 2.49 & 3.75 & 2.19 & 27.81 & 15 \\
1118 & 338.96 & -0.44 & 2.50 & 2.76 & 2.17 & 1.87 & 20.08 & 11 \\
1122 & 338.35 & -0.41 & 2.50 & 3.72 & 3.95 & 2.33 & 31.33 & 17 \\
1132 & 338.11 & -0.47 & 2.48 & 3.15 & 1.93 & 1.95 & 22.37 & 12 \\
1140 & 339.03 & -0.40 & 2.47 & 2.94 & 4.73 & 2.33 & 28.47 & 17 \\
1144 & 338.68 & -0.48 & 2.46 & 2.10 & 2.74 & 2.03 & 23.67 & 13 \\
1154 & 338.76 & -0.48 & 2.45 & 3.15 & 2.07 & 2.03 & 23.77 & 13 \\
1155 & 339.01 & -0.41 & 2.45 & 3.40 & 2.87 & 2.03 & 23.04 & 13 \\
1157 & 338.42 & -0.41 & 2.45 & 3.02 & 2.06 & 2.11 & 24.91 & 14 \\
1163 & 338.86 & -0.48 & 2.44 & 3.54 & 3.71 & 2.52 & 34.61 & 20 \\
1164 & 338.33 & -0.51 & 2.44 & 2.79 & 1.77 & 1.95 & 21.37 & 12 \\
1166 & 338.79 & -0.45 & 2.44 & 3.51 & 2.96 & 2.26 & 29.53 & 16 \\
1167 & 338.65 & -0.45 & 2.44 & 2.83 & 4.72 & 2.52 & 37.71 & 20 \\
1170 & 338.32 & -0.40 & 2.44 & 3.62 & 2.71 & 2.46 & 34.45 & 19 \\
1174 & 338.31 & -0.41 & 2.43 & 1.76 & 3.56 & 2.11 & 25.21 & 14 \\
1186 & 338.93 & -0.44 & 2.41 & 2.16 & 2.09 & 1.87 & 20.10 & 11 \\
1193 & 338.41 & -0.43 & 2.40 & 2.91 & 4.72 & 2.76 & 43.26 & 24 \\
1199 & 338.43 & -0.41 & 2.38 & 3.00 & 2.42 & 2.26 & 28.97 & 16 \\
1201 & 338.13 & -0.47 & 2.38 & 2.13 & 3.51 & 2.26 & 29.05 & 16 \\
1203 & 338.29 & -0.46 & 2.38 & 2.15 & 2.94 & 1.87 & 19.95 & 11 \\
1204 & 338.75 & -0.50 & 2.38 & 2.94 & 1.66 & 1.95 & 22.68 & 12 \\
1205 & 338.81 & -0.47 & 2.38 & 4.46 & 2.38 & 2.33 & 28.46 & 17 \\
1206 & 338.80 & -0.46 & 2.38 & 4.55 & 1.80 & 2.11 & 23.51 & 14 \\
1210 & 338.39 & -0.41 & 2.37 & 2.82 & 1.63 & 1.87 & 19.66 & 11 \\
1211 & 338.31 & -0.44 & 2.37 & 2.29 & 3.39 & 2.19 & 28.20 & 15 \\
1216 & 338.34 & -0.50 & 2.36 & 1.40 & 3.81 & 1.95 & 23.55 & 12 \\
1226 & 338.64 & -0.44 & 2.34 & 2.71 & 1.73 & 1.95 & 21.87 & 12 \\
1242 & 338.07 & -0.43 & 2.33 & 4.49 & 4.30 & 2.59 & 37.37 & 21 \\
1246 & 338.84 & -0.48 & 2.32 & 1.62 & 2.74 & 1.87 & 19.24 & 11 \\
1248 & 339.04 & -0.41 & 2.32 & 1.74 & 2.79 & 1.95 & 21.91 & 12 \\
1249 & 338.25 & -0.44 & 2.32 & 2.62 & 3.13 & 2.03 & 22.78 & 13 \\
1251 & 339.02 & -0.40 & 2.32 & 3.15 & 4.01 & 2.39 & 31.33 & 18 \\
1252 & 339.00 & -0.41 & 2.32 & 3.96 & 1.93 & 2.11 & 25.44 & 14 \\
1258 & 338.12 & -0.47 & 2.31 & 2.82 & 2.50 & 2.19 & 25.60 & 15 \\
1259 & 338.69 & -0.48 & 2.31 & 1.80 & 2.68 & 1.87 & 18.46 & 11 \\
1261 & 338.20 & -0.48 & 2.31 & 2.76 & 7.84 & 2.39 & 30.11 & 18 \\
1262 & 338.70 & -0.46 & 2.31 & 4.21 & 3.62 & 1.95 & 20.75 & 12 \\
1266 & 338.69 & -0.46 & 2.30 & 3.16 & 1.71 & 1.95 & 22.74 & 12 \\
1271 & 338.11 & -0.46 & 2.30 & 3.51 & 1.85 & 2.03 & 23.63 & 13 \\
1275 & 338.16 & -0.47 & 2.29 & 1.88 & 2.94 & 2.03 & 23.12 & 13 \\
1278 & 338.75 & -0.50 & 2.29 & 2.67 & 9.65 & 3.04 & 48.86 & 29 \\
1281 & 338.75 & -0.46 & 2.29 & 2.75 & 2.76 & 2.26 & 29.68 & 16 \\
1288 & 338.68 & -0.46 & 2.28 & 3.35 & 1.92 & 2.19 & 26.11 & 15 \\
1289 & 338.94 & -0.48 & 2.28 & 5.99 & 1.85 & 2.19 & 25.31 & 15 \\
1290 & 339.03 & -0.40 & 2.27 & 3.28 & 2.19 & 2.11 & 23.38 & 14 \\
1293 & 338.92 & -0.43 & 2.27 & 6.72 & 4.35 & 2.76 & 42.14 & 24 \\
1307 & 338.08 & -0.44 & 2.24 & 5.37 & 1.44 & 2.03 & 23.45 & 13 \\
1309 & 338.28 & -0.43 & 2.24 & 3.77 & 1.77 & 2.03 & 22.34 & 13 \\
1312 & 338.21 & -0.49 & 2.24 & 2.75 & 2.46 & 1.87 & 20.47 & 11 \\
1314 & 338.59 & -0.45 & 2.23 & 3.06 & 4.20 & 2.26 & 27.29 & 16 \\
1315 & 338.39 & -0.40 & 2.23 & 3.21 & 2.08 & 2.03 & 22.14 & 13 \\
1318 & 338.77 & -0.47 & 2.23 & 4.02 & 2.83 & 2.03 & 20.97 & 13 \\
1322 & 338.80 & -0.48 & 2.22 & 4.08 & 1.91 & 2.19 & 26.50 & 15 \\
1324 & 339.06 & -0.42 & 2.22 & 1.77 & 2.95 & 2.03 & 22.86 & 13 \\
1325 & 338.82 & -0.47 & 2.22 & 4.72 & 1.58 & 2.03 & 23.03 & 13 \\
1336 & 338.24 & -0.46 & 2.21 & 3.37 & 2.03 & 2.11 & 23.68 & 14 \\
1341 & 338.24 & -0.45 & 2.20 & 2.75 & 6.28 & 2.59 & 35.06 & 21 \\
1343 & 338.33 & -0.54 & 2.19 & 2.78 & 3.55 & 1.95 & 21.35 & 12 \\
1346 & 338.61 & -0.44 & 2.19 & 2.50 & 2.28 & 1.87 & 18.92 & 11 \\
1348 & 338.31 & -0.49 & 2.19 & 2.14 & 5.71 & 2.39 & 29.30 & 18 \\
1351 & 338.90 & -0.43 & 2.18 & 2.04 & 2.78 & 1.87 & 18.29 & 11 \\
1357 & 339.06 & -0.41 & 2.18 & 2.29 & 2.16 & 1.95 & 21.11 & 12 \\
1359 & 338.58 & -0.45 & 2.17 & 1.87 & 4.60 & 2.46 & 31.17 & 19 \\
1361 & 339.05 & -0.41 & 2.17 & 2.60 & 2.69 & 2.03 & 22.80 & 13 \\
1363 & 338.55 & -0.42 & 2.17 & 1.68 & 2.79 & 1.87 & 19.69 & 11 \\
1364 & 338.97 & -0.41 & 2.17 & 2.00 & 3.08 & 1.87 & 19.79 & 11 \\
1365 & 338.93 & -0.49 & 2.17 & 5.81 & 11.05 & 2.88 & 44.12 & 26 \\
1366 & 338.52 & -0.43 & 2.17 & 1.89 & 2.52 & 1.95 & 21.52 & 12 \\
1375 & 338.11 & -0.46 & 2.16 & 3.46 & 4.06 & 2.19 & 26.28 & 15 \\
1380 & 338.51 & -0.41 & 2.15 & 5.56 & 2.41 & 2.52 & 32.87 & 20 \\
1382 & 338.21 & -0.48 & 2.15 & 2.33 & 3.05 & 1.95 & 20.87 & 12 \\
1384 & 338.31 & -0.46 & 2.14 & 2.61 & 2.04 & 1.87 & 18.32 & 11 \\
1386 & 338.61 & -0.45 & 2.14 & 1.82 & 2.42 & 1.87 & 19.15 & 11 \\
1392 & 338.21 & -0.46 & 2.13 & 7.33 & 1.18 & 2.52 & 33.94 & 20 \\
1396 & 338.18 & -0.49 & 2.12 & 2.00 & 2.43 & 1.87 & 18.83 & 11 \\
1399 & 338.39 & -0.40 & 2.12 & 2.20 & 2.23 & 1.87 & 18.85 & 11 \\
1400 & 338.93 & -0.48 & 2.12 & 3.15 & 2.78 & 1.87 & 18.96 & 11 \\
1405 & 338.12 & -0.46 & 2.11 & 2.77 & 2.00 & 1.87 & 19.07 & 11 \\
1408 & 338.13 & -0.49 & 2.10 & 7.10 & 5.57 & 2.46 & 31.00 & 19 \\
1411 & 339.02 & -0.40 & 2.10 & 8.15 & 3.09 & 2.93 & 44.73 & 27 \\
1419 & 338.72 & -0.46 & 2.08 & 2.14 & 3.05 & 2.11 & 24.18 & 14 \\
1421 & 339.09 & -0.41 & 2.08 & 6.14 & 1.73 & 2.03 & 22.01 & 13 \\
1426 & 338.66 & -0.45 & 2.07 & 3.70 & 1.57 & 1.95 & 18.95 & 12 \\
1428 & 339.05 & -0.41 & 2.07 & 4.41 & 1.95 & 2.33 & 28.24 & 17 \\
1438 & 338.76 & -0.46 & 2.05 & 2.26 & 2.87 & 1.87 & 19.42 & 11 \\
1442 & 338.12 & -0.50 & 2.05 & 2.24 & 2.15 & 1.87 & 18.54 & 11 \\
1444 & 338.80 & -0.48 & 2.04 & 3.64 & 2.13 & 1.87 & 18.07 & 11 \\
1454 & 338.93 & -0.50 & 2.03 & 4.36 & 5.94 & 2.76 & 39.36 & 24 \\
1460 & 338.93 & -0.48 & 2.02 & 2.78 & 1.75 & 1.87 & 18.35 & 11 \\
1464 & 338.95 & -0.42 & 2.01 & 5.24 & 3.37 & 2.59 & 34.32 & 21 \\
1466 & 339.10 & -0.41 & 2.01 & 2.80 & 3.20 & 1.87 & 17.33 & 11 \\
1473 & 338.24 & -0.45 & 2.00 & 2.75 & 2.79 & 1.95 & 20.27 & 12 \\
1474 & 338.92 & -0.50 & 1.99 & 3.29 & 1.73 & 1.87 & 18.02 & 11 \\
1475 & 338.63 & -0.46 & 1.99 & 1.68 & 3.27 & 1.87 & 19.41 & 11 \\
1478 & 338.75 & -0.48 & 1.98 & 3.83 & 4.53 & 2.19 & 25.00 & 15 \\
1485 & 338.92 & -0.47 & 1.97 & 2.21 & 2.50 & 1.95 & 19.61 & 12 \\
1492 & 338.25 & -0.41 & 1.95 & 1.55 & 4.60 & 1.95 & 18.79 & 12 \\
1495 & 338.91 & -0.45 & 1.94 & 5.13 & 1.67 & 2.11 & 22.28 & 14 \\
1513 & 338.79 & -0.50 & 1.90 & 3.47 & 3.03 & 1.95 & 18.84 & 12 \\
1515 & 338.68 & -0.46 & 1.89 & 3.22 & 2.39 & 1.95 & 18.85 & 12 \\
1517 & 338.16 & -0.48 & 1.89 & 3.66 & 2.43 & 1.87 & 18.60 & 11 \\
1522 & 338.97 & -0.41 & 1.86 & 3.08 & 2.27 & 1.87 & 17.50 & 11 \\
1527 & 338.83 & -0.46 & 1.85 & 3.54 & 4.50 & 1.95 & 19.37 & 12 \\
1535 & 338.61 & -0.45 & 1.79 & 2.29 & 3.22 & 1.87 & 17.12 & 11 \\
1538 & 338.80 & -0.48 & 1.77 & 3.13 & 2.48 & 1.87 & 17.05 & 11 \\
1539 & 338.79 & -0.47 & 1.76 & 4.52 & 1.12 & 1.87 & 17.07 & 11 \\
1540 & 338.09 & -0.43 & 1.76 & 4.58 & 3.86 & 1.95 & 17.82 & 12 \\
1546 & 338.80 & -0.49 & 1.73 & 5.75 & 1.51 & 2.11 & 20.86 & 14 \\
\hline
\label{strc2-tbl}
\end{longtable}
}

\end{appendix}

\end{document}